\documentclass[aps,prb,twocolumn,floatfix]{revtex4-1}
\usepackage{amsfonts,bm}
\usepackage{amsmath}
\usepackage{amssymb}
\usepackage{amsthm}
\usepackage{graphicx}
\usepackage{graphics}
\usepackage{subfigure}
\usepackage{color}
\usepackage{bm}
\usepackage{hyperref}
\usepackage{rotating}
\usepackage{comment}
\usepackage{mathrsfs}

\renewcommand{\>}{\rangle}
\newcommand{\be}{\begin{equation} }
\newcommand{\ee}{\end{equation} }
\newcommand{\ba}{\begin{eqnarray} }
\newcommand{\ea}{\end{eqnarray} }

\newcommand{\bpm}{\begin{pmatrix}}
\newcommand{\epm}{\end{pmatrix}}
\newcommand{\bmm}{\begin{matrix}}
\newcommand{\emm}{\end{matrix}}

\begin{document}

\title{Loop braiding statistics in exactly soluble 3D lattice models}
\author{Chien-Hung Lin and Michael Levin}
\affiliation{James Franck Institute and Department of Physics, University of Chicago, Chicago, Illinois 60637, USA}

\begin{abstract}
We construct two exactly soluble lattice spin models that demonstrate the importance of three-loop braiding statistics for the classification of 3D gapped quantum phases. The two models are superficially similar: both are gapped and both support particle-like and loop-like excitations similar to that of charges and vortex lines in a $\mathbb{Z}_2 \times \mathbb{Z}_2$ gauge theory. Furthermore, in both models the particle excitations are bosons, and in both models the particle and loop excitations have the same mutual braiding statistics. The difference between the two models is only apparent when one considers the recently proposed three-loop braiding process in which one loop is braided around another while both are linked to a third loop. We find that the statistical phase associated with this process is different in the two models, thus proving that they belong to two distinct phases. An important feature of this work is that we derive our results using a concrete approach: we construct string and membrane operators that create and move the particle and loop excitations and then we extract the braiding statistics from the commutation algebra of these operators.
\end{abstract}

\maketitle

\section{Introduction}

The discovery of quantum Hall states and more recently, topological insulators\cite{HasanKaneRMP,QiZhangRMP}, has taught us that there are many different types of 
gapped quantum many body systems. In order to understand the relationship between these 
systems, it is useful to divide them into classes in such a way that the members of each class share the same qualitative 
properties. Typically these classes are defined as follows: two gapped Hamiltonians $H$ and $H'$ are assigned to the same class if 
they can be adiabatically connected to one another ---  that is, if there exists a one parameter family of interpolating 
Hamiltonians $H(s)$ with (1) $H(0) = H$, $H(1) = H'$ and (2) a finite energy gap for all $s$, $0 \leq s \leq 1$. The different 
classes of Hamiltonians are then called ``gapped phases.''

The precise definition of gapped phases depends on what type of systems we wish to consider. For example, if we are interested 
in systems with particular symmetries, then it is natural to assign $H$ and $H'$ to the same phase if there exists an 
interpolating Hamiltonian $H(s)$ which is \emph{both} gapped and invariant under the relevant symmetries. Including such symmetry constraints typically 
leads to a finer classification of gapped phases, as illustrated by the example of topological and conventional insulators.\cite{HasanKaneRMP,QiZhangRMP}

In this paper, we consider the coarsest possible classification scheme: that is, we do not impose any symmetry constraints and we 
say that two gapped Hamiltonians $H$ and $H'$ belong to the same phase if they can be adiabatically connected by \emph{any} 
interpolating Hamiltonian $H(s)$ with local interactions. Our starting point is a basic question: how can we tell whether or not 
two gapped Hamiltonians belong to the same phase?

This question has an appealing answer in the case of two dimensional (2D) systems. \footnote {The 1D case is less interesting, since it is 
known that in the absence of symmetry all 1D bosonic systems belong to the same 
phase\cite{PollmannSPT2,XieSPT1,XieSPT2,NorbertSPT,LukaszfSPT2,PollmannSPT1} while all 1D fermionic systems belong to one of two phases.\cite{LukaszfSPT2}} 
To determine whether two gapped 2D Hamiltonians belong to the same phase, one can simply compare the braiding statistics
\footnote{Here, when we say braiding statistics, we mean the complete set of algebraic data for anyon systems, including quantum dimensions 
and fusion rules. For more details, see e.g. Appendix E of Ref. \onlinecite{KitaevHoneycomb}.} 
of their quasiparticle excitations. If the braiding statistics data do not match, then the two Hamiltonians must belong to different phases 
since braiding statistics cannot change under an adiabatic deformation. Conversely, if the braiding statistics do match, then we can 
\emph{almost} conclude that the two Hamiltonians belong to the same phase. To reach this conclusion, we simply need to compare one 
other quantity, namely the thermal Hall conductance\cite{KaneFisherThermal}. Indeed, according to a plausible (but unproven) 
conjecture, if two Hamiltonians have the 
same braiding statistics and the same thermal Hall conductance, then they must be adiabatically connected to one another in the 
absence of any symmetry constraints. 

\begin{figure}[ptb]
\begin{center}
\includegraphics[
height=0.8in,
width=1.5in
]{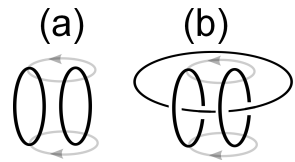}
\end{center}
\caption{(a) Two-loop braiding process. (b) Three-loop braiding process. The gray curves show the paths of two points on the moving loop.}
\label{loopbraiding}
\end{figure}

In the 3D case, our understanding is much more limited. One way to attack the classification problem is to 
try to generalize the concept of quasiparticle braiding statistics to the 3D case. The simplest generalization begins with the 
observation that many 3D Hamiltonians support \emph{loop-like} excitations in addition to particles. 
Given this observation, we can consider several different types of braiding statistics. 
First, we can look at the exchange statistics of particle-like excitations. These exchange statistics can take only 
one of two values for each particle: every particle must be either bosonic or fermionic.\footnote{Here, we implicitly exclude 3D 
layered systems, like a stack of fractional quantum Hall states, from our discussion.} Second, we can consider the statistical phase 
associated with braiding a particle around a loop.\cite{AharonovBohm,AlfordWilczek,KraussWilczek,PreskillKrauss} Finally, we can look at the 
statistical phase associated with braiding one 
loop excitation around another (Fig. \ref{loopbraiding}a).\cite{Anezirisloop,Alfordloop,Baezloop} If we combine all of these types of braiding statistics, we can indeed distinguish many 
different 3D gapped phases.

Interestingly, however, this data is incomplete: Refs. \onlinecite{WangLevin,JiangMesarosRan} argued that we also need to consider the statistical phase 
associated with a \emph{three-loop} braiding process. In this process, one loop is braided around another loop while both are 
linked to a third loop (Fig. \ref{loopbraiding}b). It is unclear whether the three-loop braiding data is the last piece of the puzzle or 
whether there exist further distinctions between 3D gapped phases that can only be seen if we consider other braiding processes 
or other types of probes. However, either way, three-loop braiding statistics has already proven to be useful in certain cases. \cite{WangWen14,JianQi,WangLevin2,SurfaceTOChen}

One weakness of previous studies of three-loop braiding statistics is that this quantity has only been calculated using indirect
and abstract arguments. For example, Ref. \onlinecite{WangLevin} computed the three-loop braiding statistics of 3D Dijkgraaf-Witten gauge theories 
using a dimensional reduction argument which relates the braiding statistics of loops in 3D Dijkgraaf-Witten models to the braiding statistics of particles in 
2D Dijkgraaf-Witten models\cite{DijkgraafWitten}. The approaches of Refs. \onlinecite{JiangMesarosRan,WangWen14} were also indirect: Refs. \onlinecite{JiangMesarosRan,WangWen14} computed loop braiding statistics by relating this quantity to modular transformations on a three dimensional torus.

In this paper, we study three-loop braiding statistics using a more concrete approach. We focus on two exactly soluble lattice models
and we compute their three-loop braiding statistics by explicitly implementing the loop braiding process on the lattice. Our approach is 
similar to how quasiparticle braiding statistics is commonly calculated in 2D lattice models\cite{KitaevToric, LevinWenHop, LevinWenstrnet}: we construct
membrane operators that create and move loop-like excitations, and then we extract the three-loop statistics from the commutation
algebra of these operators. 

The two spin models that we analyze provide an explicit demonstration of the importance of three-loop braiding statistics for 
distinguishing 3D gapped phases. Indeed, we show that the models share the same particle exchange statistics as well as the 
same particle-loop and loop-loop braiding statistics. The only difference between the models is that they have different
three-loop braiding statistics. Thus, it is only this quantity that reveals that the two models belong to different phases.

The models that we study are not completely new and have appeared previously in the literature in different forms. In particular,
the first model is essentially identical to the 3D generalized toric code model\cite{KitaevToric,HammaZanardiWen,LevinWenstrnet,CastelnovoChamon} 
corresponding to the group $\mathbb{Z}_2 \times \mathbb{Z}_2$. Thus, 
the low energy properties of this model are similar to that of conventional $\mathbb{Z}_2 \times \mathbb{Z}_2$ gauge theory\cite{Kogut}. The second model can be 
thought of as a different type of $\mathbb{Z}_2 \times \mathbb{Z}_2$ gauge theory. More specifically, this model can be obtained by starting with the
spin model in Ref. \onlinecite{ChenLuVishwanath} which describes a nontrivial symmetry-protected topological phase\cite{GuSPT,PollmannSPT2,LukaszfSPT2,XieSPT1,XieSPT2,NorbertSPT,ChenGuWenSPT} with $\mathbb{Z}_2 \times \mathbb{Z}_2$ symmetry, and then 
coupling this system to a $\mathbb{Z}_2 \times \mathbb{Z}_2$ lattice gauge field\cite{Kogut}. We will explain this connection in more detail in a
separate publication\cite{LinLevinprep}. In addition to its connection with the spin model
in Ref. \onlinecite{ChenLuVishwanath}, we believe that the second model belongs to the same phase as one of the
exactly soluble $\mathbb{Z}_2 \times \mathbb{Z}_2$ Dijkgraaf-Witten models.\cite{DijkgraafWitten,Wan3dmodel} This conjecture is based on the fact that
the braiding statistics in the two systems seem to match one another.

The connection between our models and symmetry-protected topological phases is not accidental: we specifically designed our 
models to be equivalent to gauged symmetry-protected topological phases, because according to the results of Ref. \onlinecite{WangLevin,JiangMesarosRan}, such 
gauge theories can support different types of three-loop braiding statistics. The reason that we chose to gauge the spin model 
from Ref. \onlinecite{ChenLuVishwanath} is that this is one of the simplest known models for a 3D symmetry-protected topological phase with unitary symmetry 
group.

The rest of the paper is organized as follows. In section \ref{section:membrane} we introduce the two exactly soluble 3D spin 
models that we will analyze. In section \ref{section:excitations}, we study the particle-like and loop-like excitations of these models 
by explicitly constructing the string and membrane operators that create and move these excitations. We then compute the braiding 
statistics of these particles and loops in section \ref{section:statistics}. Some technical details can be found in 
the appendices. 

\section{Models \label{section:membrane}}

\subsection{Hilbert space for the models}

\begin{figure}[ptb]
\begin{center}
\includegraphics[
height=0.9in,
width=1.5in
]{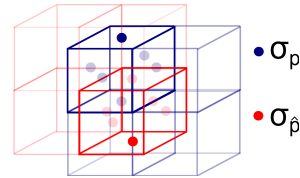}
\end{center}
\caption{Both models are built out of two species of spins. The blue spins $\sigma_p$ live on the plaquettes $p$ of the cubic lattice while the red spins $\sigma_{\hat{p}}$ live on the plaquettes $\hat{p}$ of the dual cubic lattice.
}
\label{fig:lattice1}
\end{figure}

The models that we will discuss are spin-$1/2$ systems made up of two species of spins: ``blue spins'' and ``red spins.'' The blue spins live on the plaquettes $p$ of the cubic lattice and will be denoted by $\sigma_p$ while the red spins live on the plaquettes $\hat{p}$ of the dual cubic lattice and will be denoted by $\sigma_{\hat{p}}$ (Fig. \ref{fig:lattice1}). In this notation, the $S^z$ eigenstates $|\{\sigma^z_p, \sigma^z_{\hat{p}}\}\>$ provide a complete basis for the Hilbert space.

We will often find it convenient to describe spin states using an alternative language based on ``membranes.'' In the membrane language, each $S^z$ eigenstate $|\{\sigma^z_p, \sigma^z_{\hat{p}}\}\>$ corresponds to a spatial configuration of red and blue membranes on the cubic lattice and dual cubic lattice. The dictionary between spin states and membrane configurations is as follows: if $\sigma_p^z = -1$ we say that the plaquette $p$ is occupied by a blue membrane, while if $\sigma_p^z = +1$ we say that the plaquette $p$ is empty. Similarly, if $\sigma_{\hat{p}}^z = -1$ then $\hat{p}$ is occupied by a red membrane while if $\sigma_{\hat{p}}^z = +1$, then $\hat{p}$ is empty. In this way, each spin state can be equivalently described as a membrane state. We will label our membrane states as $|X_b, X_r\>$ where $X_b$ denotes the subset of plaquettes that are occupied by blue membranes, and $X_r$ denotes the subset of plaquettes occupied by red membranes.  

\subsection{Ground state wave functions}
The two models that we will discuss have been engineered to have particular ground states. These ground states are easiest to describe if we assume an infinite (non-periodic) geometry. In such a geometry, the ground state of the first model is
\begin{equation}
|\Psi_0\> = \sum_{\text{closed } X_b, X_r} |X_b, X_r\>
\label{psi0}
\end{equation}
where the sum runs over all \emph{closed} membrane states $|X_b, X_r\>$. Here, by a closed membrane state, we mean a membrane configuration $(X_b, X_r)$ in which all the blue and red membranes form closed surfaces, i.e. surfaces without boundaries. More precisely, a closed membrane state is defined to be a state in which every edge $l$ in the cubic lattice and every edge $\hat{l}$ in the dual cubic lattice is adjacent to an \emph{even} number of occupied plaquettes.

The ground state of the second model has a similar form
\begin{equation}
|\Psi_1\> = \sum_{\text{closed } X_b, X_r} (-1)^{N_g(X_b, X_r)} |X_b, X_r\>
\label{psi1}
\end{equation}
where again the sum runs over all possible closed membrane states $|X_b, X_r\>$. The quantity $N_g(X_b, X_r)$ is defined as follows: for each closed 
membrane configuration $(X_b, X_r)$, the intersections between the red and blue membranes $X_b \cap X_r$ form a collection of disconnected closed 
curves, which we will call ``green loops.'' The quantity $N_g(X_b, X_r)$ is defined to be the number of the green loops in $X_b \cap X_r$. As 
discussed in the introduction, the ground state $|\Psi_1\>$ is closely related to the ground state of the spin model in Ref. \onlinecite{ChenLuVishwanath} which describes a nontrivial symmetry-protected topological phase with $\mathbb{Z}_2 \times \mathbb{Z}_2$ symmetry.

\subsection{Sheared cubic lattice \label{section:skew}}
The reader may notice that there is a technical problem with the above definition of $N_g(X_b, X_r)$: the problem is that the closed membrane condition allows for membrane configurations in which an edge $l$ is adjacent to four occupied plaquettes. Geometrically such configurations correspond to the case where two blue membranes touch one another along the edge $l$. This membrane touching is problematic because it means that the ``green loops'' defined by the intersections of red and blue membranes can also touch one another at corners. As a result, there is some ambiguity in determining the number of disconnected green loops corresponding to a membrane configuration $(X_b, X_r)$.

\begin{figure}[ptb]
\begin{center}
\includegraphics[
height=1.4in,
width=1.4in
]{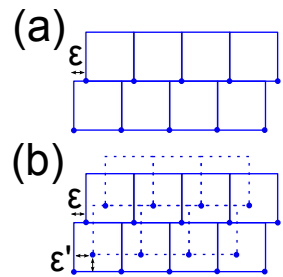}
\end{center}
\caption{(a) The sheared square lattice is formed by shifting the corners of the square plaquettes (the blue dots) according to $(i,j) \rightarrow (i+\epsilon j,j)$ with $\epsilon>0$. 
	(b) Top view of the sheared cubic lattice. The solid/dashed squares denote cubes on two neighboring layers. The sheared cubic lattice is formed by shifting the corners of the cubes according to $(i,j,k) \rightarrow (i+\epsilon j+\epsilon' k,j+\epsilon' k,k)$ with $\epsilon'>\epsilon'>0$. 
} 
\label{fig:skewedlattice}
\end{figure}

To deal with this issue, we now describe a way to infinitesimally deform the cubic lattice so as to eliminate membrane touching. Before describing this deformation, we first warm up with an analogous deformation of the square lattice. The basic idea is to think of the square plaquettes that make up the square lattice as rigid blocks that can be shifted around. The deformation we have in mind corresponds to shifting the position of the square plaquettes so that their corners move from $(i,j)  \rightarrow (i + \epsilon j, j)$. The resulting lattice, which we call the \emph{sheared} square lattice is shown in Fig. \ref{fig:skewedlattice}(a). 

We are now ready to consider the cubic lattice. In this case, we think of the cubes that make up the cubic lattice as rigid blocks and we shift these 
cubes so that their corners move from
\begin{equation}
(i, j, k) \rightarrow (i+\epsilon j + \epsilon'k, j+\epsilon' k, k)
\label{splconv}
\end{equation}
where $\epsilon' > \epsilon > 0$. The resulting \emph{sheared} cubic lattice is shown in Fig. \ref{fig:skewedlattice}(b). 

Now let us imagine performing the same shearing deformation to both the original cubic lattice and the dual cubic lattice. We can then deform an arbitrary closed membrane configuration $(X_b, X_r)$ on the cubic lattice (and dual cubic lattice) to a membrane configuration on the sheared cubic lattice (and sheared dual cubic lattice). The result is a closed membrane configuration without any membrane touching of any kind. In what follows, we will always think of membrane configurations as living on the sheared cubic lattice rather than the cubic lattice so we don't have to worry about membrane touching. 

\subsection{Hamiltonians}
\subsubsection{\texorpdfstring{$H_0$}{H0} \label{section:h0}}
We are now ready to write down the Hamiltonians for the two models. The Hamiltonian for the first membrane model is a sum of four terms,
\begin{equation}
H_0 = - \sum_{l} A_l - \sum_{\hat{l}} A_{\hat{l}} - \sum_{c} B^0_c - \sum_{\hat{c}} B^0_{\hat{c}}
\label{h0}
\end{equation}
where the indices $l,\hat{l}$ run over the links of the cubic lattice and dual cubic lattice respectively, while $c,\hat{c}$ run over the ``cubes'' 
of the cubic lattice and dual cubic lattice. To define the operators $A_l, A_{\hat{l}}, B_c, B_{\hat{c}}$, it suffices to explain how they act on 
the membrane basis states $|X_b, X_r\>$. The $A_l,A_{\hat{l}}$ operators are given by
\begin{align}
A_l = \frac{1}{2}(1+ \mathscr{A}_l)  \ \ \ , \ \ \ A_{\hat{l}} = \frac{1}{2}(1+\mathscr{A}_{\hat{l}})
\end{align}
where 
\begin{align}
\mathscr{A}_l |X_b, X_r\> &= (-1)^{N_l} |X_b, X_r\>, \nonumber\\
\mathscr{A}_{\hat{l}} |X_b, X_r\> &= (-1)^{N_{\hat{l}}} |X_b, X_r\>.
\end{align}
Here $N_l$ and $N_{\hat{l}}$ are defined to be the number of occupied plaquettes adjacent to $l$ and $\hat{l}$ respectively. The $B^0_c, B^0_{\hat{c}}$ 
operators are defined by
\begin{align}
B^0_c = \frac{1}{2}(1+\mathcal{B}^0_c)  \ \ \ , \ \ \ B^0_{\hat{c}} = \frac{1}{2}(1+\mathcal{B}^0_{\hat{c}}) 
\end{align}
where
\begin{align}
\mathcal{B}^0_{c} |X_b, X_r\> &= |X_b + c, X_r\>, \nonumber \\
\mathcal{B}^0_{\hat{c}} |X_b, X_r\> &= |X_b, X_r + \hat{c}\>.
\label{b0}
\end{align}
Here the notation $X_b+c$ is meant to denote a kind of $\mathbb{Z}_2$ addition on membrane configurations. More specifically, given the membrane 
configuration $X_b$, the configuration $X_b + c$ is obtained by flipping the occupation numbers of the six plaquettes of the cube $c$: that is, 
one changes the plaquettes from unoccupied to occupied and vice versa.

There is a simple physical picture for the $A$ and $B$ in terms of the membrane language: the $A$ terms favor \emph{closed} membrane configurations, 
while the $B$ terms provide an amplitude for the membranes to fluctuate. Together these terms ensure that the ground state is a superposition of many different closed membrane configurations. 

Alternatively, we can express the $A$ and $B$ operators in the spin language:
\begin{align}
A_l = \frac{1}{2}(1+\prod_{p \in l} \sigma^z_p) \quad , \quad A_{\hat{l}} = \frac{1}{2}(1+\prod_{\hat{p} \in \hat{l}} \sigma^z_{\hat{p}})
\label{ql}
\end{align}
where these products run over the four plaquettes adjacent to $l, \hat{l}$ respectively. Similarly,
\begin{align}
B_c = \frac{1}{2}(1+\prod_{p \in c} \sigma^x_{p}) \quad , \quad B_{\hat{c}} = \frac{1}{2} (1+ \prod_{\hat{p} \in \hat{c}} \sigma^x_{\hat{p}})
\label{qbspin0}
\end{align}
where these products run over the six plaquettes adjacent to $c,\hat{c}$ respectively.

\begin{figure}[ptb]
\begin{center}
\includegraphics[
height=1.2in,
width=3in
]{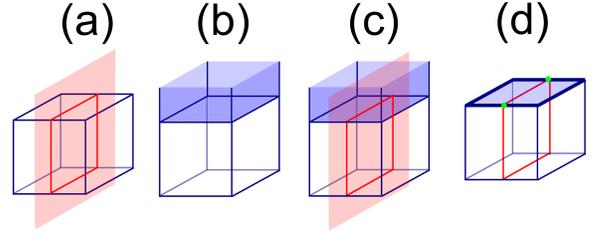}
\end{center}
\caption{
	(a) A blue cube intersects a red membrane; their intersection consists of a single red loop.
	(b) A blue cube intersects/overlaps with a blue membrane; their intersection consists of a 
single blue region.
	(c) A blue cube intersects with both a red membrane and a blue membrane; their intersection consists 
of one red loop and one blue region. 
	(d) For the membrane configuration shown in (c), the 
integer $m_c = 1$ because there is \emph{one} red loop while $n_c =2$ because there are \emph{two} intersections (denoted by green 
dots) between the red loop and the boundary of the blue region (thick blue line).
} 
\label{fig:intersection}
\end{figure}

\subsubsection{\texorpdfstring{$H_1$}{H1} \label{section:h1}}
The Hamiltonian for the second membrane model is similar:
\begin{equation}
H_1 = - \sum_{l} A_l - \sum_{\hat{l}} A_{\hat{l}} - \sum_{c} B^1_c - \sum_{\hat{c}} B^1_{\hat{c}}.
\label{h1}
\end{equation}
The first two terms are defined as above, while the last two terms are defined by
\begin{align}
B^1_c = \frac{1}{2}(1+\mathcal{B}^1_c) \cdot P_c \ \ \ , \ \ \ B^1_{\hat{c}} = \frac{1}{2}(1+ \mathcal{B}^1_{\hat{c}}) \cdot P_{\hat{c}}.
\label{bc1}
\end{align}
Here $P_c$ is a projector that projects onto states obeying the closed membrane constraint in the neighborhood of the cube $c$. More specifically,
\begin{align}
P_{c} =  \prod_{l \in c} A_l \cdot \prod_{\hat{l} \perp c} A_{\hat{l}}
\end{align}
where the first product runs over the twelve edges $l$ of $c$ and the second product runs over the six edges $\hat{l}$ that are perpendicular to the six plaquettes of the cube $c$.
Similarly,
\begin{align} 
P_{\hat{c}} = \prod_{\hat{l} \in \hat{c}} A_{\hat{l}}  \cdot \prod_{l \perp \hat{c}} A_{l}. 
\end{align}
The $\mathcal{B}^1_c$ and $\mathcal{B}^1_{\hat{c}}$ operators have a similar structure as (\ref{b0}), but their matrix elements have some additional phase factors:
\begin{align}
\mathcal{B}^1_{c} |X_b, X_r\> &= (-1)^{m_c} i^{n_c} |X_b+c, X_r\>, \nonumber \\
\mathcal{B}^1_{\hat{c}} |X_b, X_r\> &= (-1)^{m_{\hat{c}}} i^{n_{\hat{c}}} |X_b, X_r + \hat{c}\>.
\label{matrixB}
\end{align}
Here $m_c,m_{\hat{c}}, n_c, n_{\hat{c}}$ are integer-valued functions of $(X_b, X_r)$, which we will now define. In fact, we will only define $m_c, n_c$ for membrane states $(X_b, X_r)$ that satisfy $P_c = 1$, and we will only define $m_{\hat{c}}, n_{\hat{c}}$ for states with $P_{\hat{c}} = 1$; it suffices to discuss these subsets of states since the projectors $P_c, P_{\hat{c}}$ ensure that the matrix elements of $B^1_c$ and $B^1_{\hat{c}}$ vanish for all other states.

We begin with $m_c$. To define the value of $m_c$ for some $(X_b, X_r)$, consider the intersection between the set of red membranes $X_r$ and the cube $c$ . This intersection defines a collection of closed loops that live on the surface of the cube $c$ (Fig. \ref{fig:intersection}a). (The fact that the loops are closed follows from the closed membrane constraint $P_c = 1$). We will call these loops ``red loops.'' The integer $m_c$ is defined to be the number of these red loops, i.e.,
\begin{equation}
m_c(X_b, X_r) = \text{\#\{red loops on $c$\}}
\end{equation}
(See Fig. \ref{fig:intersection}d for an example). 

To define the value of $n_c$ for some $(X_b, X_r)$, consider the intersection between the set of blue membranes $X_b$ 
and the cube $c$. 
Because the blue membranes $X_b$ and the cube $c$ both live on the same cubic lattice, the intersection 
$X_b \cap c$ consists of 2D \emph{regions} instead of loops. These 2D regions live on the surface of the cube $c$ 
(Fig. \ref{fig:intersection}b). Consider the boundary of these 2D
regions. This boundary consists of a collection of closed loops. We will call these loops ``blue loops.'' The integer 
$n_c$ is defined to be the
number of intersections between the blue loops and the red loops defined above:
\begin{equation}
n_c(X_b, X_r) = \text{\#\{blue-red intersections on $c$\}}
\end{equation}
(See Fig. \ref{fig:intersection}d for an example). Notice that $n_c$ is always even. 

The integers $m_{\hat{c}}$ and $n_{\hat{c}}$ are defined in an identical way but with the colors reversed:
\begin{equation}
m_{\hat{c}}(X_b, X_r) = \text{\#\{blue loops on $\hat{c}$\}} 
\end{equation}
and
\begin{equation}
n_{\hat{c}}(X_b, X_r) = \text{\#\{red-blue intersections on $\hat{c}$\}}
\end{equation}
Here, the blue loops are defined by the intersection between the blue membranes $X_b$ and the red cube $\hat{c}$, while
the red loops are the boundaries of the intersections between the red membranes $X_r$ and the cube $\hat{c}$. 

\begin{figure}[ptb]
\begin{center}
\includegraphics[
height=1.2in,
width=2.8in
]{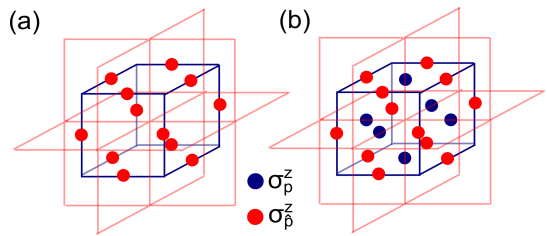}
\end{center}
\caption{(a) The integer $m_c$ is a function of the $12$ spins $\sigma_{\hat{p}}^z$ that are closest to the cube $c$.
(b) The integer $n_c$ is a function of the $6$ blue spins $\sigma_p^z$ and $12$ red spins $\sigma_{\hat{p}}^z$ around the cube $c$.}
\label{cube}
\end{figure}

Alternatively, we can express $m_c, m_{\hat{c}}, n_c, n_{\hat{c}}$ in the spin language. In this language, the operator $m_c$ can be written as a function of the twelve spins $\sigma_{\hat{p}_1}^z,...,\sigma_{\hat{p}_{12}}^z$ that are closest to $c$ (Fig. \ref{cube}a), and similarly for $m_{\hat{c}}$:
\begin{align}
m_c &= f(\sigma_{\hat{p}_1}^z, \sigma_{\hat{p}_2}^z,\dots, \sigma_{\hat{p}_{12}}^z), \nonumber \\
m_{\hat{c}} &= f(\sigma^z_{p_1}, \sigma^z_{p_2},\dots, \sigma^z_{p_{12}}). 
\label{mfun}
\end{align}
Here $f$ is a finite polynomial, but the explicit form of $f$ is not illuminating so we do not show it here. Likewise, we can write $n_c$ as a function of the $6$ blue spins $\sigma_{p_1}^z,...,\sigma_{p_6}^z$ and $12$ red spins, $\sigma_{\hat{p}_1}^z,...,\sigma_{\hat{p}_{12}}^z$ surrounding the cube $c$ (Fig. \ref{cube}b), and similarly for $n_{\hat{c}}$:
\begin{align}
n_c &=  g(\{\sigma^z_{p}\},\{\sigma^z_{\hat{p}}\}), \nonumber \\
n_{\hat{c}} &=  g(\{\sigma^z_{\hat{p}}\},\{\sigma^z_{p}\}).
\label{nfun}
\end{align}
As above, the expression for $g$ is complicated so we do not show it here.

While the Hamiltonian $H_0$ is essentially identical to the well-known 3D toric code model\cite{KitaevToric,HammaZanardiWen,LevinWenstrnet,CastelnovoChamon}, some readers may be curious about the origin
of the Hamiltonian $H_1$. As we mentioned previously, this Hamiltonian has been designed to have a particular ground state --- namely, $|\Psi_1\>$. The state $|\Psi_1\>$ is in turn
motivated by the ground state of the spin model in Ref. \onlinecite{ChenLuVishwanath} which describes a nontrivial symmetry-protected topological phase with $\mathbb{Z}_2 \times \mathbb{Z}_2$ 
symmetry. Thus, the main question is how to design an exactly soluble Hamiltonian with a particular ground state, $|\Psi_1\>$. One way to do this is to follow a similar approach to the string-net
construction of Ref. \onlinecite{LevinWenstrnet}. We recall that the string-net models of Ref. \onlinecite{LevinWenstrnet} can written as a sum of two 
types of operators --- a $Q$ operator and a 
$B$ operator. The $Q$ operator prefers closed string configurations and the $B$ operator creates a closed loop. While the form of the $Q$ operator is simple and intuitive, the $B$ operator is
more complicated, and its matrix elements are obtained by fusing a closed loop onto the lattice using 2D local rules. The Hamiltonian $H_1$ can be constructed using a similar approach: in this case the Hamiltonian is built out of $A$ operators which prefer closed membrane configurations and $B$ operators which create a cube. Similarly to the string-net models, the precise form of the $B$ operator can be obtained by writing down local rules obeyed by $|\Psi_1\>$, and then fusing a cube onto the lattice using 3D local rules. This is one way to obtain $H_1$. Another way is to use membrane operators: as we will see in section \ref{otherM}, the $B_c,B_{\hat{c}}$ are examples of spherical membrane operators (\ref{wb},\ref{sphere}).

\subsection{Properties of the Hamiltonians}
The two Hamiltonians $H_0,H_1$ have many nice properties. One property is that all the operators $\{A_l, A_{\hat{l}}, B_c^0, B_{\hat{c}}^0, B_c^1, B_{\hat{c}}^1\}$ are Hermitian so that $H_0$ and $H_1$ are also Hermitian. Indeed, the Hermiticity of $\{A_l, A_{\hat{l}}, B_c^0, B_{\hat{c}}^0\}$ is clear from the definitions (\ref{ql},\ref{qbspin0}), while the Hermiticity of $B_c^1$ and $B_{\hat{c}}^1$ can be seen by noting that the matrix elements of $B_c^1$ and $B_{\hat{c}}^1$, as defined in (\ref{matrixB}), are both symmetric and real.

Another important property is that all the terms in $H_0, H_1$ commute with one another, so that the two models are exactly soluble. For the case of $H_0$, the fact that $\{A_l, A_{\hat{l}}, B_c^0, B_{\hat{c}}^0\}$ all commute with one another follows easily from the definitions (\ref{ql},\ref{qbspin0}) of these operators. For the case of $H_1$, more work is required to verify this commutativity: while simple algebra shows that
\begin{align}
	[A_l,A_{l'}]&=[A_{\hat{l}},A_{\hat{l}'}]=0
\end{align}
and
\begin{align}
	[A_l,B_c^1]&=[A_{\hat{l}},B_c^1]=[A_l,B_{\hat{c}}^1]=[A_{\hat{l}},B_{\hat{c}}^1]=0, \nonumber \\
\end{align}
the fact that
\begin{align}
	[B_c^1,B_{c'}^1]&=[B_{\hat{c}}^1,B_{\hat{c}'}^1] = [B_c^1,B_{\hat{c}'}^1]=0
	\label{commutingbc}
\end{align}
is not obvious. We leave the derivation of the latter identity (\ref{commutingbc}) to Appendix \ref{app:show}.

A third property of $H_0, H_1$ is that the operators $\{A_l, A_{\hat{l}}, B_c^0, B_{\hat{c}}^0, B_c^1, B_{\hat{c}}^1\}$ have eigenvalues $0$ or $1$:
\begin{equation}
a_l, a_{\hat{l}}, b_c^0, b_{\hat{c}}^0, b_c^1, b_{\hat{c}}^1 = 0 , 1.
\end{equation}
(The first two eigenvalue spectra can be derived from $A_l^2 = A_l$, etc.). A final property of these models is that $|\Psi_0\>$ is an simultaneous eigenstate of $\{A_l, A_{\hat{l}}, B_c^0, B_{\hat{c}}^0\}$ with 
\begin{align}
a_l = a_{\hat{l}} = b_c^0 = b_{\hat{c}}^0 = 1.
\label{psi0prop}
\end{align}
Similarly, $|\Psi_1\>$ is a simultaneous eigenstate of $\{A_l, A_{\hat{l}}, B_c^1, B_{\hat{c}}^1\}$ with
\begin{align} 
a_l = a_{\hat{l}} = b_c^1 = b_{\hat{c}}^1 = 1.
\label{psi1prop}
\end{align}
A derivation of the relations (\ref{psi0prop}) and (\ref{psi1prop}) is given in Appendix \ref{app:show}. 

\subsection{Solving the models}
In this section, we show that $H_0$ and $H_1$ are gapped, and that $|\Psi_0\>$ and $|\Psi_1\>$ are ground states of these Hamiltonians. We begin with $H_0$. To find the energy spectrum of $H_0$, recall that the operators $\{A_l, A_{\hat{l}}, B^0_c, B^0_{\hat{c}}\}$ commute with one another and can therefore be simultaneously diagonalized. Let us label these simultaneous eigenstates by $|a_l, a_{\hat{l}}, b^0_c, b^0_{\hat{c}}\>$ where $a_l, a_{\hat{l}}, b^0_c, b^0_{\hat{c}}$ denote the eigenvalues. It is clear that these states are energy eigenstates with energy
\begin{equation}
E = - \sum_{l} a_l - \sum_{\hat{l}} a_{\hat{l}} - \sum_{c} b^0_c - \sum_{\hat{c}} b^0_{\hat{c}}.
\label{energy}
\end{equation}
Now, since the eigenvalues $\{a_l, a_{\hat{l}}, b^0_c, b^0_{\hat{c}}\}$ take values in $0,1$, it follows that the ground state(s) of $H_0$ have 
$a_l = a_{\hat{l}} = b^0_c = b^0_{\hat{c}} = 1$, while the excited states have at least one $a_l, a_{\hat{l}},b^0_c, b^0_{\hat{c}}$ equal to $0$. We 
conclude that there is a finite energy gap, $\Delta = 1$, separating the ground state(s) and excited states. Furthermore, we can see that $|\Psi_0\>$ is a ground state of $H_0$ since it obeys (\ref{psi0prop}). All that remains is to determine the ground state degeneracy of $H_0$. This degeneracy depends on the global topology of the system on which $H_0$ is defined. In an infinite non-periodic geometry, one can show that $H_0$ has a unique ground state --- namely, $|\Psi_0\>$. On the other hand, in periodic (3D torus) geometry, it can be shown that the ground state degeneracy is $4^3 = 64$.
These degenerate ground states are characterized by different parities of noncontractible red and blue membranes along the $x,y$ and $z$ directions of the 3D torus.  

Now let us consider Hamiltonian $H_1$. Similarly to $H_0$, we can simultaneously diagonalize $\{A_l, A_{\hat{l}}, B^1_c, B^1_{\hat{c}}\}$ and label their eigenstates as $|a_l, a_{\hat{l}}, b^1_c, b^1_{\hat{c}}\>$. Following the same reasoning as above, we conclude that there is a finite energy gap, $\Delta = 1$, separating the ground state(s) and excited states. Furthermore, we can see that $|\Psi_1\>$ is a ground state of $H_1$ since it obeys (\ref{psi1prop}). Like $H_0$, it can be shown that $H_1$ has a unique ground state in an infinite non-periodic geometry ($|\Psi_1\>$), and the ground state degeneracy is $64$ in a 3D torus geometry.

\subsection{Particle-like and loop-like excitations}
If we examine equation (\ref{energy}), we can see that $H_0$ supports both particle-like and loop-like excitations. An example of a 
particle-like excitation is a point defect where $b^0_c$ or $b^0_{\hat{c}}$ is equal to $0$ instead of the ground state value of $1$. Likewise, 
an example of a loop-like excitation is a line defect along which $a_l$ or $a_{\hat{l}}$ is equal to $0$ instead of the ground state value of $1$. These line-like defects always form closed loops: to see this, note that the quantum numbers $a_l$ and $a_{\hat{l}}$ obey the local constraints
\begin{equation}
\prod_{l \in s} (1-2a_l) = 1 \ \ \ , \ \ \ \prod_{\hat{l} \in \hat{s}} (1-2a_{\hat{l}}) = 1
\end{equation}
for every site $s$ in the cubic lattice and $\hat{s}$ in the dual cubic lattice. 
Here, the first product runs over the six edges $l$ that are adjacent to $s$ and 
similarly for the second product. The above constraints guarantee that each site 
$s$ is adjacent to an even number of line-line defects, so that the defects always 
form closed loops. (To derive these constraints, note that they follow from the 
corresponding operator identities, $\prod_{l \in s} (1-2 A_l) = \prod_{\hat{l} \in \hat{s}} (1-2 A_{\hat{l}}) = 1$ 
which in turn follow from the definition (\ref{ql})). In exactly the same way, 
one can see that the Hamiltonian $H_1$ supports particle-like excitations with 
$b^1_c$ or $b^1_{\hat{c}}$ is equal to $0$, and loop-like excitations where $a_l$ 
or $a_{\hat{l}}$ is equal to $0$ along some closed loop. 

Below we will see that in both models, these particle-like and loop-like excitations 
have nontrivial braiding statistics with one another and have similar properties to 
the charges and vortex loops in $\mathbb{Z}_2 \times \mathbb{Z}_2$ gauge theories. In view of this connection, 
we will refer to the particle-like excitations with $b^0_c$ or $b^1_c$ equal to $0$ as ``blue charges'' 
and the excitations with $b^0_{\hat{c}}$ or $b^1_{\hat{c}}$ equal to $0$ as ``red charges.'' 
Similarly, we will refer to the loop-like excitations with $a_l = 0$ as ``blue vortex loops'' 
and the excitations with $a_{\hat{l}} = 0$ as ``red vortex loops.'' In the following, we 
will study these charge and vortex-loop excitations in more detail, with a focus on their topological properties.

\section{Excitations and the associated creation operators \label{section:excitations}}
In this section, we find operators that create the charge and vortex loop excitations of $H_0$ and $H_1$. These 
operators are useful because their commutation algebra contains information about the braiding statistics of the associated particles and loops.

\subsection{General picture for string and membrane operators}
In general, topologically nontrivial particle excitations cannot be created using local operators. Instead, the easiest way to create these excitations 
is to use string-like operators. In the following sections we will find string-like creation operators for each topologically distinct charge 
excitation $\alpha$ in $H_0$ and $H_1$. We will denote these operators by $W_\alpha(P)$ where $P$ is the path along which the string operator acts. These 
operators satisfy two key properties. First, if $P$ is an \emph{open} path, then when $W_\alpha(P)$ is applied to the ground state $|\Psi\>$, it creates 
an excited state $|\Psi_{ex}\>$ with two charge excitations $\alpha$ at the two ends of $P$:
\begin{equation*}
	W_\alpha(P)|\Psi\>=|\Psi_{ex}\>.
\end{equation*}
(Here the excited state $|\Psi_{ex}\>$ only depends on the endpoints of $P$ and not on the choice of path). Second, if $P$ is a \emph{closed} path, then $W_\alpha(P)$ does not create any excitation at all: $W_\alpha(P) |\Psi\> \propto |\Psi\>$. 

In addition to these string operators, which create charge excitations, we will also find membrane operators that create vortex loop excitations. That is, 
for each topologically distinct vortex loop excitation $\alpha$ in $H_0$ and $H_1$, we will find a corresponding membrane creation operator, which we 
will denote by $M_{\alpha}(S)$ with $S$ being the surface where the membrane operator acts.
These membrane operators satisfy similar properties to the string operators. First, if $S$ is a \emph{cylindrical} surface, then when $M_{\alpha}(S)$ is 
applied to the ground state $|\Psi\>$ it creates an excited state with two loop excitations at the ends of the cylinder: 
\begin{equation*}
	M_{\alpha}(S)|\Psi\>=|\Psi_{ex}\>.
\end{equation*}
(Similarly to before, the excited state $|\Psi_{ex}\>$ only depends on the two boundaries of $S$ and not on the choice of surface that joins them). 
Second, if $S$ is a \emph{toroidal} surface, then $M_\alpha(S)$ does not create any loop excitations at all: $M_\alpha(S) |\Psi\> \propto |\Psi\>$.

We can also consider membrane operators with other topologies
beyond the cylinder and torus case:
for example, later we will discuss \emph{spherical} membrane operators. However,  
the structure of membrane operators is different for different topologies\footnote{For example, 
toroidal operators can be decorated by 
string operators that encircle the torus while spherical membrane operators cannot be decorated in this way. 
See section \ref{otherM} for more details.} and thus each case has to be treated separately. 
Here we will focus on cylindrical and toroidal cases as they are sufficient for our purposes.

The string and membrane operators have simple physical interpretations: the string operator $W_\alpha(P)$ describes a process in which 
a pair of charge excitations is created and then moved to the two ends of the path $P$. Likewise, the membrane operator $M_\alpha(S)$ 
describes a process in which a pair of loops is created and then moved to the two ends of the cylinder $S$. On the other hand, if $P$ 
is a closed path then $W_\alpha(P)$ describes a process in which a pair of charge excitations is created and then moved around $P$ and 
annihilated with each other. Similarly, if $S$ is a toroidal surface then $M_\alpha(S)$ describes a process in which a loop-antiloop pair 
is created, moved around $S$ and annihilated with one another.

\subsection{String and membrane operators for \texorpdfstring{$H_0$}{H0}}

\subsubsection{String operators}
It is easy to find string operators $W_b^0$, $W_r^0$ that create the blue and red charge excitations of $H_0$. These operators are given by
\begin{equation}	
	W_b^0(P)=\prod_{p \perp P} \sigma_p^z, \quad W_r^0(P)=\prod_{\hat{p} \perp P} \sigma_{\hat{p}}^z
	\label{string0}
\end{equation}
where $P$ is a path on the dual cubic lattice in the first expression, and a path in the cubic lattice in the second expression. The two products run over plaquettes that are perpendicular to these two paths. Equivalently, in the membrane representation, $W_b^0$ and $W_r^0$ are given by:
\begin{align}
	W_b^0(P)|X_b,X_r\> &=(-1)^{N_b}|X_b,X_r\>, \nonumber \\
	W_r^0(P)|X_b,X_r\> &= (-1)^{N_r}|X_b,X_r\>
	\label{string1a}
\end{align}
where $N_b$ and $N_r$ are the number of blue and red membranes that cross the path $P$.

Let us now verify that when $P$ is an open path, $W_b^0(P)$ creates blue charge excitations at the two endpoints of $P$. From the definitions 
(\ref{string0}), it is easy to see that $W_b^0(P)$ commutes with all the terms in the Hamiltonian $H_0$ except for $\mathcal{B}^0_{c_1}$ and $\mathcal{B}^0_{c_2}$ where $c_1$ and $c_2$ are the two cubes at the endpoints of $P$. These two operators \emph{anticommute} with $W_b^0(P)$ rather than commute. We conclude that $W_b^0(P) |\Psi_0\>$ is a simultaneous eigenstate of $A_l, A_{\hat{l}}, B^0_p, B^0_{\hat{p}}$ with eigenvalues $b^0_{c_1} = b^0_{c_2} = 0$ and all other eigenvalues equal to $1$. Hence, $W_b^0(P) |\Psi_0\>$ contains two blue charge excitations at the endpoints of $P$. A similar argument shows that $W_r^0$ also creates red charge excitations at the endpoints of $P$.

\subsubsection{Membrane operators \label{memH0}}
It is also easy to find membrane operators $M_b^0$, $M_r^0$ that create the blue and red vortex excitations of $H_0$. These operators are given by
\begin{equation}
	M_{b}^0(S) = \prod_{p\in S}  \sigma_p^x, \quad
	M_{r}^0(S) = \prod_{\hat{p}\in S}  \sigma_{\hat{p}}^x 
\label{membrane0}
\end{equation}
where $S$ is a surface made up of plaquettes $p$ living in the cubic lattice in the first expression, and a surface consisting of plaquettes $\hat{p}$ in the dual cubic lattice in the second expression. Equivalently, we can express $M_b^0$ and $M_r^0$ in the membrane representation as:
\begin{align}
M_b^0(S) |X_b, X_r\> &= |X_b+S, X_r\>, \nonumber \\
M_r^0(S) |X_b, X_r\> &= |X_b, X_r + S \>
\label{membrane0m}
\end{align}
where $X_b + S$ denotes the $\mathbb{Z}_2$ addition operation defined below equation (\ref{b0}).

We now check that when $S$ is a cylindrical surface, $M_b^0(S)$ creates blue loop excitations at the two ends of $S$. To establish this fact, we note that $M_b^0(S)$ commutes with all the terms in the 
Hamiltonian $H_0$ except for $\mathscr{A}_l$ when $l$ lies along the two boundaries of $S$. These $\mathscr{A}_l$ operators anticommute with $M_b^0(S)$ 
rather than commute. It then follows that $M_b^0(S) |\Psi_0\>$ is an eigenstate of $A_l, A_{\hat{l}}, B^0_p, B^0_{\hat{p}}$ with eigenvalues 
$a_l = 0$ along the boundaries of $S$, and all other eigenvalues equal to $1$. This establishes the claim. The same argument applies to $M_r^0(S)$. 

It should be noted that the excitations created by $M_b^0$ and $M_r^0$ are \emph{not} the most general possible vortex loop excitations. In fact, the 
$H_0$ model supports three other types of blue vortex loops and three other types of red vortex loops. These excitations can be obtained by attaching either a red charge, a blue charge, or both a red charge and a blue charge, to the vortex loops created by $M_b^0$ and $M_r^0$. We will label the four types of blue vortex loops by $(b, q_b, q_r)$, and the four types of red vortex loops by $(r, q_b, q_r)$ where $q_b, q_r$ take values in $0,1$. In this labeling scheme, the excitations created by $M_b^0$ and $M_r^0$ are denoted by $(b,0,0)$ and $(r,0,0)$, with the other excitations labeled according to the amount of charge attached to them. As we will see later, these loop excitations are all topologically distinct from one another in the sense that they have different braiding statistics.

It is easy to find membrane operators that create these more general vortex loop excitations. For example, consider the operator $M_b^0(S)$ in the case 
where $S$ is a cylinder. When this operator is applied to the ground state $|\Psi_0\>$, it creates two vortex loops of type $(b,0,0)$ at the two ends of 
the cylinder. We can modify this operator to create other vortex loops $(b, q_b, q_r)$ by multiplying $M_b^0(S)$ by the string operators 
$W_b^0(P)$ or $W_r^0(P')$ or $W_b^0(P) \cdot W_r^0(P')$ where the paths $P,P'$ run along the length of the cylinder. In the same way, we can
construct membrane creation operators for the red vortex loops $(r, q_b, q_r)$ by multiplying $M_r^0(S)$ by appropriate string operators.

Before concluding, we should mention that there is yet another type of vortex loop excitation which we have not discussed, namely a \emph{composite} of a red and blue vortex loop. This type of loop excitation can be
obtained by fusing together a red and blue loop, and again it comes in four subtypes which we can denote by $(rb,q_b, q_r)$. In what follows, we will generally ignore this additional kind of vortex loop excitation since its braiding statistics properties are completely determined by the properties of the individual red and blue loops.

\subsection{String and membrane operators for \texorpdfstring{$H_1$}{H1} \label{membraneop}}
In this section, we construct string and membrane operators for the Hamiltonian $H_1$. The string operators are relatively easy: following the same analysis as above, it is simple to check that the
same string operators that create the blue and red charge excitations in the $H_0$ model can also be used to create the charge excitations in the $H_1$ model. That is, just as in (\ref{string1a}), we can set 
\begin{align}
	W_b^1(P)|X_b,X_r\> &=(-1)^{N_b}|X_b,X_r\>, \nonumber \\
	W_r^1(P)|X_b,X_r\> &= (-1)^{N_r}|X_b,X_r\>
\end{align}
where $N_b$ and $N_r$ are the number of blue and red membranes that cross the path $P$. 

We will need to do more work to construct membrane operators for $H_1$. The rest of this section is devoted to this problem.

\subsubsection{Cylindrical membrane operators for blue vortex loops}
We begin by finding a membrane operator $M_b^1(S)$ that creates blue vortex loops. To proceed, it is convenient to work in the membrane basis. Let $S$ be a cylindrical surface made up of plaquettes $p$ living in the cubic lattice. We need to define the action of $M_b^1(S)$ on a general membrane state $|X_b, X_r\>$. A natural guess, inspired by (\ref{membrane0m}), is that $M_b^1(S)$ should act as
\begin{align}
M_b^1(S) |X_b, X_r\> = f_b(X_b, X_r,S)|X_b + S, X_r\>
\label{wb}
\end{align}
for some complex-valued function $f_b(X_b, X_r,S)$. Indeed, it is easy to see that any operator of this type has the property that it 
anticommutes with the $\mathscr{A}_l$ terms that lie along the two boundaries of $S$ and it commutes with every other $\mathscr{A}_l$ and 
$\mathscr{A}_{\hat{l}}$ term. Hence, any operator of this type has the property that if we apply it to the ground state $|\Psi_1\>$, it will create blue 
vortex loop excitations at the boundaries of $S$. The problem is that most of these operators also create many other excitations along the surface $S$ since most of these operators do \emph{not} commute with $B_{c}^1$ and $B_{\hat{c}}^1$. Thus, our task is to choose $f_b$ appropriately so that $M_b^1(S)$ does not create any excitations except at the boundaries of $S$.

\begin{figure}[ptb]
\begin{center}
\includegraphics[
height=0.7in,
width=2.4in
]{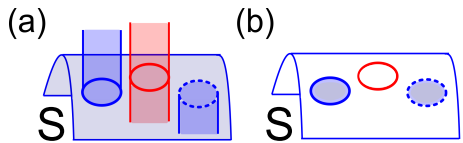}
\end{center}
\caption{(a) A blue cylinder $S$ intersects with a red membrane and two blue membranes (only part of $S$ is shown for clarity). One of the blue membranes is incident from above $S$ and the other from below.
	(b) We represent the intersections as a picture drawn on the cylinder $S$. In this case, the picture consists of a red loop, and two blue regions --- one with a solid boundary and one with a dotted boundary.
} 
\label{intersection2}
\end{figure}

We now describe one choice of $f_b$ that does the job. Below, we will just present the definition of $f_b$ without any motivation. Later, in section \ref{showpathindep}, we will explain why this choice works and we will provide some motivation as to where it comes from. 

\begin{figure}[ptb]
\begin{center}
\includegraphics[
height=0.7in,
width=0.8in
]{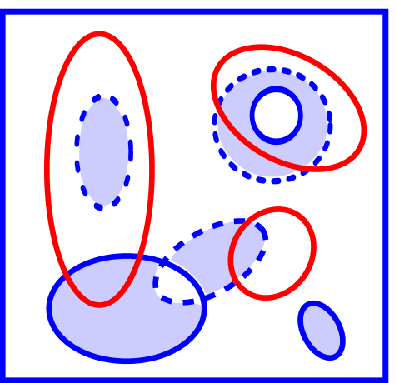}
\end{center}
\caption{A typical picture representing the intersections of an (unlinked) blue cylinder with red and blue membranes. Here we draw the cylinder as a rectangle with top and bottom identified and the left and right being the two ends of the cylinder.
}
\label{typicalpicture1}
\end{figure}

First, we set $f_b(X_b, X_r,S) = 0$ if the membrane configuration $(X_b, X_r)$ violates the closed membrane constraint anywhere in the neighborhood 
of $S$. More precisely, $f_b = 0$ if any link $l, \hat{l}$ that touches or intersects $S$ is adjacent to an odd number of occupied plaquettes. On 
the other hand, if every such link is adjacent to an \emph{even} number of occupied plaquettes, then $f_b(X_b, X_r,S)$ is defined in terms of the two 
intersections $X_b \cap S$ and $X_r \cap S$. The dependence of $f_b$ on $X_b \cap S$ and $X_r \cap S$ is complicated, and in order to explain it, we 
first describe a way to represent $X_b \cap S$ and $X_r \cap S$ as a picture drawn on the surface of the cylinder $S$. The picture we will draw 
consists of a collection of red lines and blue regions with the red lines denoting the places where membranes in $X_r$ intersect $S$, and the blue 
regions denoting the places where membranes in $X_b$ intersect/overlap with $S$. We note that the boundaries of the blue regions correspond to places 
where a membrane in $X_b$ is incident upon the surface $S$. If this membrane is incident upon $S$ from above, we will draw the corresponding boundary 
as a solid line, while if it is incident from below, we will draw the boundary as a dotted line (Fig. \ref{intersection2}). (Note that for this 
step to be well-defined, one needs to specify a convention for what side of $S$ is defined as ``above'' $S$ and what side is defined as ``below'' 
$S$. Equivalently, one needs to choose a normal vector to $S$). Putting this all together, the intersections $X_b \cap S$ and $X_r \cap S$ can be 
represented by a picture, drawn on the cylinder $S$, of the form shown in Fig. \ref{typicalpicture1}.

The value of $f_b(X_b, X_r,S)$ is completely determined by the corresponding picture.  Thus, $f_b$ can be thought of as a complex-valued 
function defined on pictures. All that remains is to specify this function. For reasons that will become clear later on, it is most natural to 
describe $f_b$ implicitly, through local constraint equations, rather than providing an explicit formula. More specifically, we define $f_b$ in 
terms of the constraint equations
\begin{subequations}
	\label{local}
\begin{align}
	f_b\left(\raisebox{-0.12in}{\includegraphics[height=0.3in]{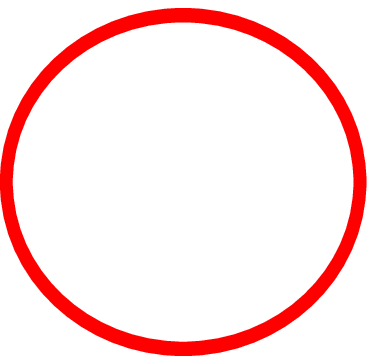}}\right) &= - f_b\left(\cdot\right) \label{r2},\\
	f_b\left(\raisebox{-0.12in}{\includegraphics[height=0.3in]{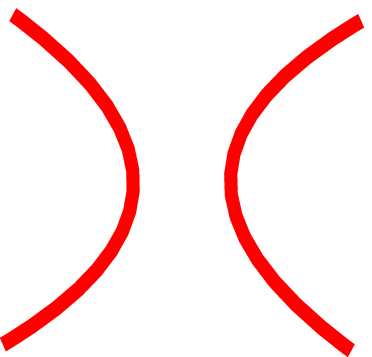}}\right) &= - f_b\left(\raisebox{-0.12in}{\includegraphics[height=0.3in]{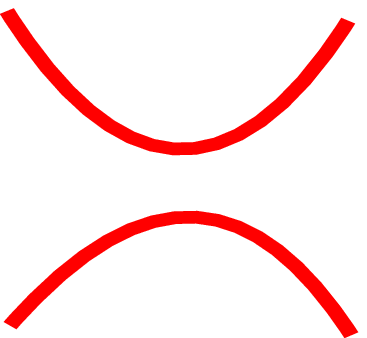}}\right), \label{r3} \\
	f_b\left(\raisebox{-0.12in}{\includegraphics[height=0.3in]{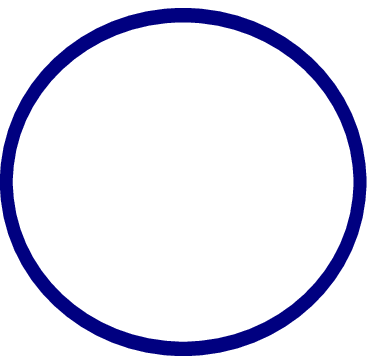}}\right) &= f_b\left(\cdot\right), \label{r4} \\
	f_b\left(\raisebox{-0.12in}{\includegraphics[height=0.3in]{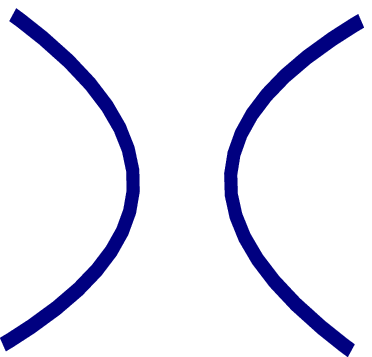}}\right) &= f_b\left(\raisebox{-0.12in}{\includegraphics[height=0.3in]{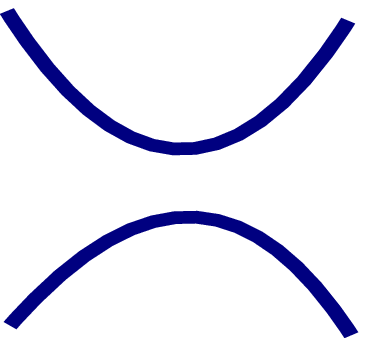}}\right), \label{r5} \\
	f_b\left(\raisebox{-0.12in}{\includegraphics[height=0.3in]{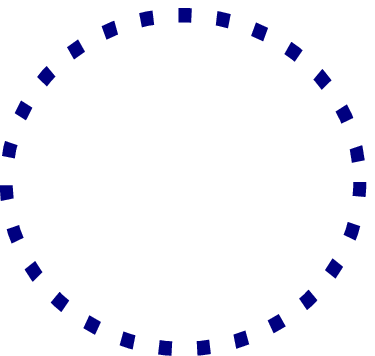}}\right) &= f_b\left(\cdot\right), \label{r6}\\
	f_b\left(\raisebox{-0.12in}{\includegraphics[height=0.3in]{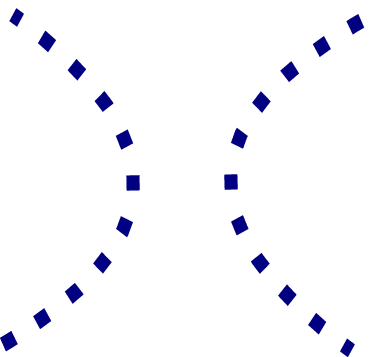}}\right) &= f_b\left(\raisebox{-0.12in}{\includegraphics[height=0.3in]{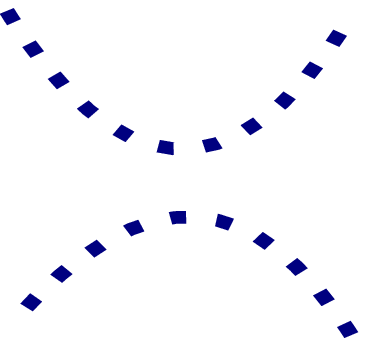}}\right), \label{r7} \\
	f_b\left(\raisebox{-0.12in}{\includegraphics[height=0.3in]{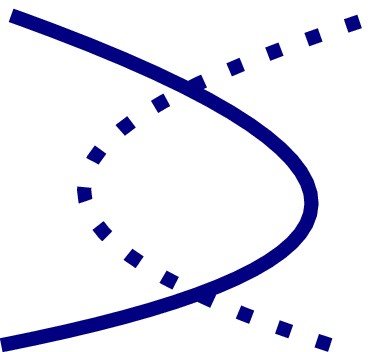}}\right) &=  f_b\left(\raisebox{-0.12in}{\includegraphics[height=0.3in]{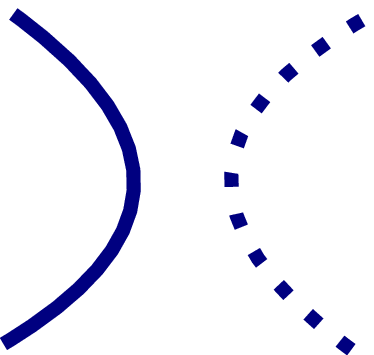}}\right), \label{r8} \\
		f_b\left(\raisebox{-0.12in}{\includegraphics[height=0.3in]{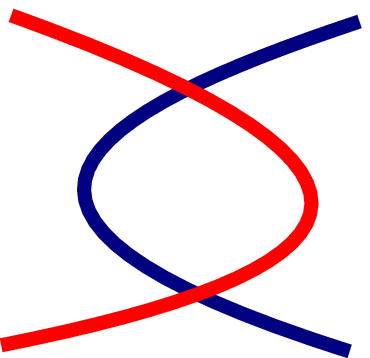}}\right) &= - f_b\left(\raisebox{-0.12in}{\includegraphics[height=0.3in]{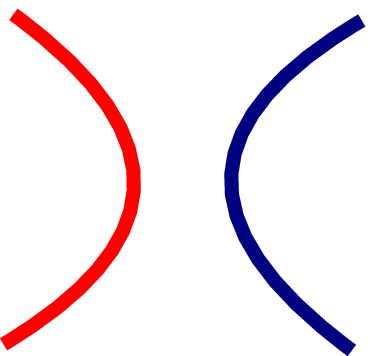}}\right), \label{r10} \\
	f_b\left(\raisebox{-0.12in}{\includegraphics[height=0.3in]{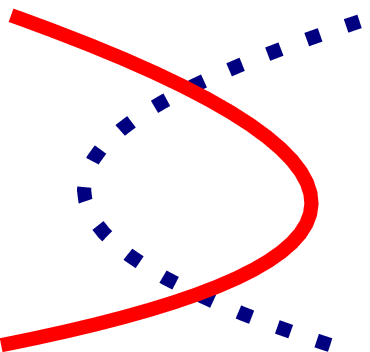}}\right) &= - f_b\left(\raisebox{-0.12in}{\includegraphics[height=0.3in]{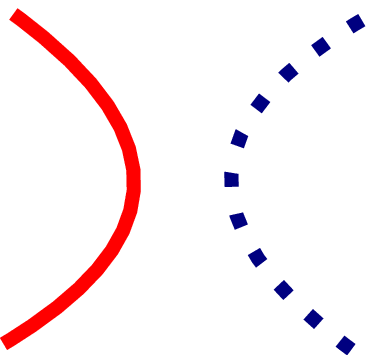}}\right), \label{r9} \\
	f_b\left(\raisebox{-0.12in}{\includegraphics[height=0.3in]{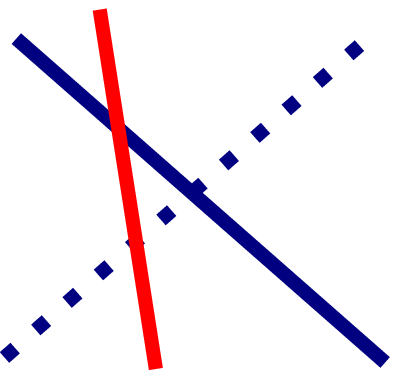}}\right) &= - f_b\left(\raisebox{-0.12in}{\includegraphics[height=0.3in]{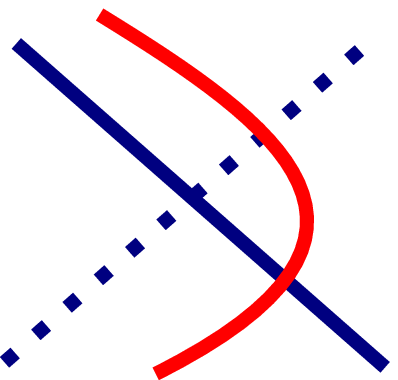}}\right) \label{r11} 
\end{align}
\end{subequations}
where it is understood that the value of $f_b$ depends only on the \emph{topology} of the picture. That is, any two pictures that can be smoothly 
deformed into one another have the same value of $f_b$. The meaning of the above constraint equations is as follows. The first equation (\ref{r2}) 
states that the value of $f_b$ for a picture with a closed red loop is equal to the value of $f_b$ for the same picture, without the loop, up to a 
factor of $-1$. Equation (\ref{r3}) states that two pictures that differ by the recoupling of red curves have values of $f_b$ that differ by a factor 
of $-1$. The remaining equations have a similar meaning.

In the above constraint equations, we have left out the shading of the blue regions. But it should be understood that the blue regions are shaded in 
a consistent way on both sides of the equations. For example, (\ref{r5}) represents two constraints with different shadings
\begin{align*}
f_b\left(\raisebox{-0.12in}{\includegraphics[height=0.3in]{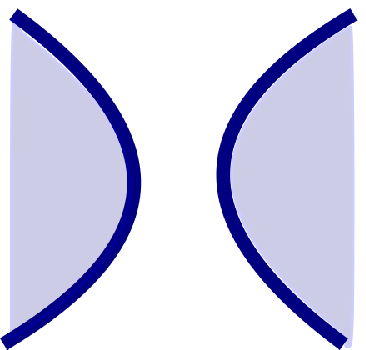}}\right) =  f_b\left(\raisebox{-0.12in}{\includegraphics[height=0.3in]{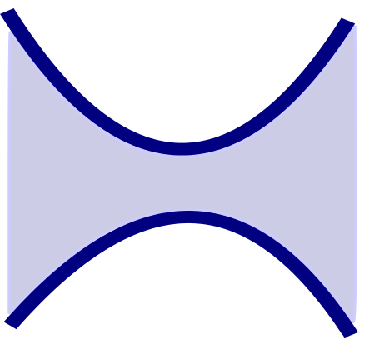}}\right), \\
f_b\left(\raisebox{-0.12in}{\includegraphics[height=0.3in]{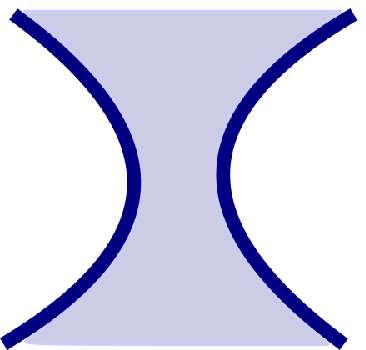}}\right) =  f_b\left(\raisebox{-0.12in}{\includegraphics[height=0.3in]{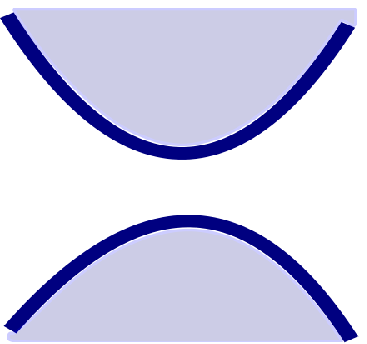}}\right).
	\label{}
\end{align*} 
Similarly, (\ref{r11}) also represents two constraints with different shadings
\begin{align*}
f_b\left(\raisebox{-0.12in}{\includegraphics[height=0.3in]{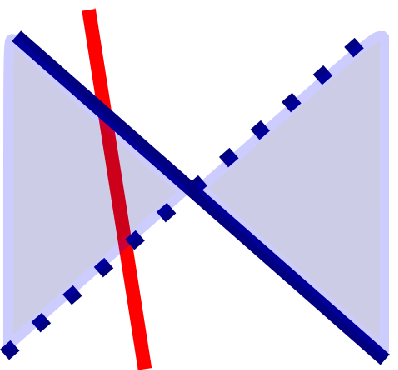}}\right) = - f_b\left(\raisebox{-0.12in}{\includegraphics[height=0.3in]{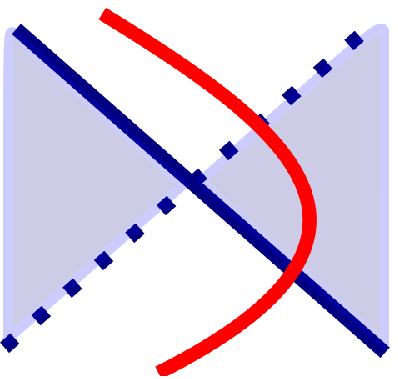}}\right), \\
f_b\left(\raisebox{-0.12in}{\includegraphics[height=0.3in]{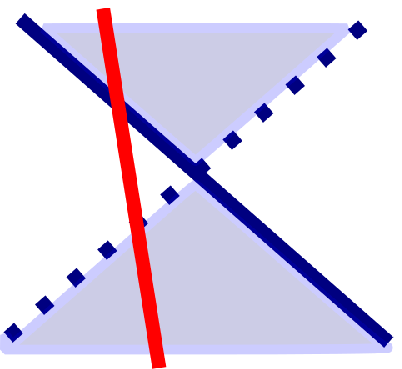}}\right) = - f_b\left(\raisebox{-0.12in}{\includegraphics[height=0.3in]{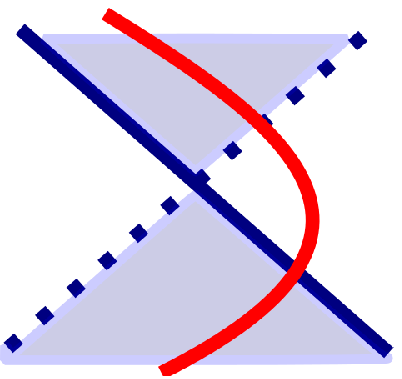}}\right).
	\label{}
\end{align*}

\begin{table}
\begin{tabular}{|c|c|c|c|c|}
\hline
\raisebox{-0.16in}{\includegraphics[height=0.4in]{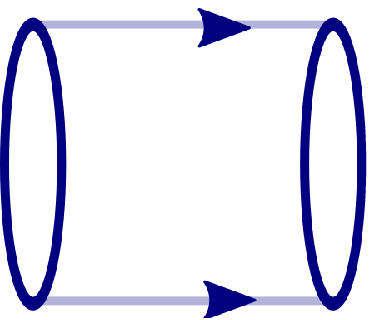}} &
\raisebox{-0.16in}{\includegraphics[height=0.4in]{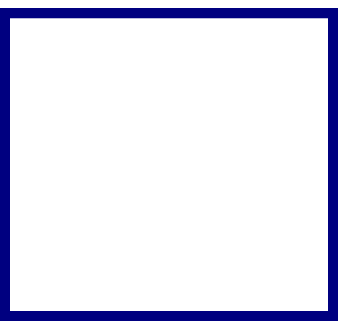}} &
\raisebox{-0.16in}{\includegraphics[height=0.4in]{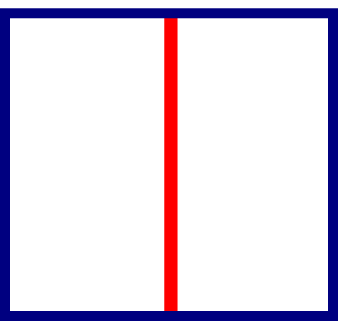}} &
\raisebox{-0.16in}{\includegraphics[height=0.4in]{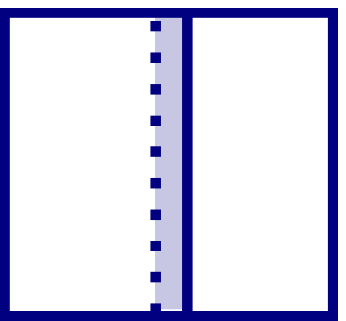}} &
\raisebox{-0.16in}{\includegraphics[height=0.4in]{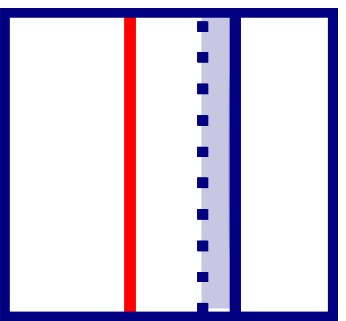}} \\ \hline
$f_b$ & $1$ & $-e^{i\pi q_r}$ & $e^{i\pi q_b}$ & $-e^{i\pi (q_r+q_b)}$\\ \hline
\end{tabular}
\caption{The function $f_b$ that defines the (unlinked) blue cylinder operator is completely determined by its values on four basic pictures that are drawn on the surface of the cylinder. Here the blue cylinder is represented by a rectangle with upper and lower edges identified and with the left and right being the two ends of the cylinder. The corresponding values of $f_b$ are shown below the pictures. The integers $q_r,q_b=0,1$ define four different functions $f_b$ and therefore four different cylinder operators.} 
\label{fig:config1}
\end{table}

To complete the definition, we impose a boundary condition on $f_b$ which states that $f_b = 0$ if any of the red lines or 
blue regions touch the boundaries of $S$. With this boundary condition and the above constraint equations, the value of $f_b$ on any picture can be 
related to one of the four ``basic'' pictures shown in Table \ref{fig:config1}. Thus, once we specify the value of $f_b$ on these basic pictures, we will 
have completely specified $f_b$ and therefore $M_b^1(S)$. 

The values for $f_b$ on the basic pictures are shown in the bottom row of Table \ref{fig:config1}. These values are parameterized by two integers 
$q_b, q_r \in \{0,1\}$. Hence our construction actually defines four different membrane operators. In principle, we should label these 
operators by $M_{(b,q_b,q_r)}^1$ to make the dependence on $q_b, q_r$ explicit. However, this notation is cumbersome so we will denote the membrane 
operators by $M_b^1$ with the understanding that $M_b^1$ is not fully defined until we specify $q_b, q_r$. Similarly to the $H_0$ model, these four 
different blue membrane operators create four different types of blue vortex loops. These vortex loops differ from one another by the amount of 
charge that they carry and we will label them by $(b, q_b, q_r)$. 

At this point, we have fully defined the membrane operator $M_b^1(S)$. All that remains is to show that this membrane operator has the required 
properties. That is, we need to show that $M_b^1(S)$ creates blue vortex loops at its two boundaries and nothing else. We will establish this
fact in section \ref{showpathindep}.

\subsubsection{Cylindrical membrane operators for red vortex loops}
To construct cylindrical membrane operators that create \emph{red} vortex loops, we follow exactly the same recipe as above but with the roles 
of ``red'' and ``blue'' reversed. First, we define
\begin{align}
M_r^1(S) |X_b, X_r\> = f_r(X_b, X_r,S)|X_b, X_r+S\>
\label{mr}
\end{align}
where $f_r(X_b, X_r, S)$ is a complex-valued function. We then define $f_r(X_b, X_r,S)$ in terms of the two intersecting sets $X_b \cap S$ and 
$X_r \cap S$. As in the definition of $f_b$, we represent $X_b \cap S$ and $X_r \cap S$ in terms of a picture drawn on the surface of the cylinder 
$S$, and we think of $f_r$ as a function defined on such pictures. In this case, each picture consists of a collection of \emph{blue} lines and 
\emph{red} regions, where the blue lines denote the intersection $X_b \cap S$, and the red regions denote the intersection $X_r \cap S$. We define 
the value of $f_r$ on each picture using local constraint equations which are identical to Eqs. (\ref{local}), but with the red and blue colors 
reversed. With these constraint equations, the value of $f_r$ on any picture can be related to one of the four basic pictures shown in 
Table \ref{fig:config1} (with the colors reversed). We then define the value of $f_r$ on these basic pictures, just as in Table \ref{fig:config1} 
but with the colors reversed and $q_b$ and $q_r$ exchanged. This procedure completely specifies $f_r$ and therefore $M_r^1$. Note that, like 
$M_b^1$, this construction actually gives four different membrane operators that are parameterized by two integers $q_b, q_r \in \{0,1\}$. These 
membrane operators create four different types of red vortex loops, which we denote by $(r, q_b, q_r)$.

\subsubsection{Cylindrical membrane operators for linked loops}
We now have all the tools we need to create an excited state with a pair of vortex loops: to do this we simply apply one of the above cylindrical 
operators to the ground state $|\Psi_1\>$. But what if we want to build a state with more than two vortex loops? One might try to make such a state 
by applying multiple cylinder operators to the ground state. This approach will work fine if the state we are
trying to build does not contain any linked loops. However, it will fail if any of the loops are linked: the above membrane operators are simply 
incapable of creating excited states with \emph{linked} loops. To see this, imagine we first apply a cylindrical membrane operator $M_b^1(S)$ to 
the ground state, obtaining $M_b^1(S) |\Psi_1\>$. This state contains two blue vortex loop excitations located at the two boundaries of the 
cylinder $S$. Now, suppose we apply another cylinder operator $M_b^1(S')$ where $S'$ is linked with one of these loops. One might hope that the 
result would be a state with linked vortex loops. Unfortunately, however, $M_b^1(S')$ simply annihilates this state: 
$M_b^1(S') M_b^1(S) |\Psi_1\> = 0$. To see this, note that $M_b^1(S) |\Psi_1\>$ is a superposition of membrane states $|X_b, X_r\>$, where all 
the membranes in $X_b$ and $X_r$ form closed surfaces, except at the two boundaries of $S$, where a blue membrane terminates. Since $S'$ is linked 
with one of the boundaries of $S$, it follows that each of the two boundaries of $S'$ must intersect with blue membranes an odd number of times. 
In particular, each boundary must have at least one intersection with a blue membrane. But if we look back at the boundary conditions for $f_b$, 
we see that any state of this kind has $f_b(X_b, X_r,S') = 0$, so any state of this kind is annihilated by $M_b^1(S')$.

We therefore need to find another set of membrane operators to create linked loops. We now define such membrane operators. Before doing this, 
we first need to explain the basic idea behind these operators. In general, we will be interested in excited states containing a collection of 
vortex loops that are all linked with a single ``base loop'' which can be either blue or red. We will build such states with the help of two 
different cylindrical membrane operators --- one for each type of base loop. The operator associated with the blue base is designed to be 
applied to states containing a blue loop that links with the cylinder $S$: when the operator is applied to such a state, it creates two 
vortex loops at the ends of the cylinder $S$ which are linked with the original blue loop. On the other hand, when it is applied to states 
that do not contain a blue loop that links with the cylinder $S$, it simply annihilates them. The operator associated with the red base 
has a similar property. This structure is reminiscent of the cylinder operators constructed in the previous two sections. In fact, the 
previously constructed operators can be thought of as cylindrical membrane operators associated with a \emph{trivial} base: they only 
create loop excitations when applied to states without any loops linked with $S$.

\begin{figure}[ptb]
\begin{center}
\includegraphics[
height=0.9in,
width=1.6in
]{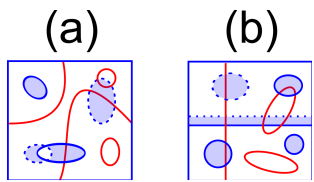}
\end{center}
\caption{ Two typical pictures representing the intersections between (linked) blue cylinders and red and blue membranes. 
Here we draw the cylinder as a rectangle with top and bottom identified. Panel (a) shows a typical picture for a blue cylinder 
linked with a red base loop. Panel (b) shows a typical picture for a blue cylinder linked with a blue base loop.
}
\label{typicalpicture2}
\end{figure}

With this picture in mind, we now construct cylindrical membrane operators for a red base loop and a blue base loop. We begin by defining a 
cylindrical membrane operator that creates blue vortex loops that are linked to a red base loop. This operator is defined in almost exactly 
the same way as the blue membrane operator $M_b^1(S)$ defined above. The only difference is that we change the boundary condition on $f_b$ 
so that it only takes nonzero values on pictures in which a single red line touches the boundary of $S$ at a fixed position $y_0$, and no 
blue regions touch the boundary. (See Fig. \ref{typicalpicture2}a for a typical picture of this type). Given this boundary condition and 
the constraint equations (\ref{local}) we can reduce any such picture to one of the four basic pictures shown in the top panel of 
Table \ref{fig:config2}. We can therefore completely specify the membrane operator once we specify the value of $f_b$ on these four basic 
pictures. The values we choose are shown in Table \ref{fig:config2}.

\begin{table}
\begin{tabular}{|c|c|c|c|c|}
\hline
\raisebox{-0.16in}{\includegraphics[height=0.4in]{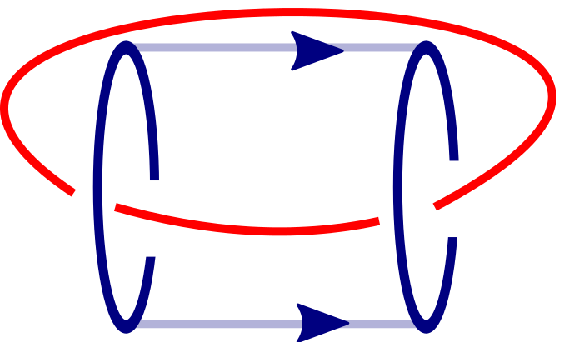}} &
\raisebox{-0.16in}{\includegraphics[height=0.4in]{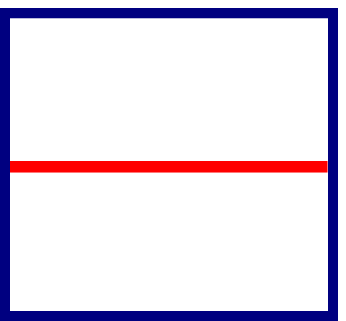}} &
\raisebox{-0.16in}{\includegraphics[height=0.4in]{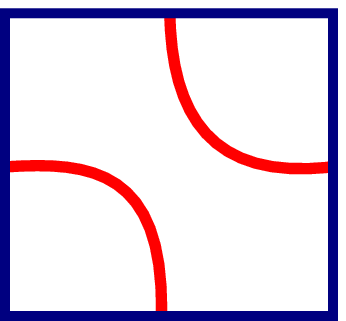}} &
\raisebox{-0.16in}{\includegraphics[height=0.4in]{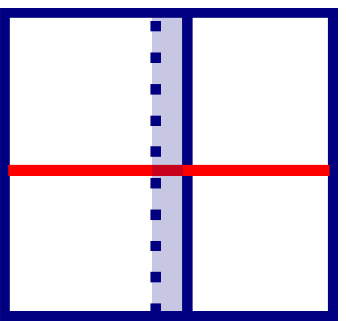}} &
\raisebox{-0.16in}{\includegraphics[height=0.4in]{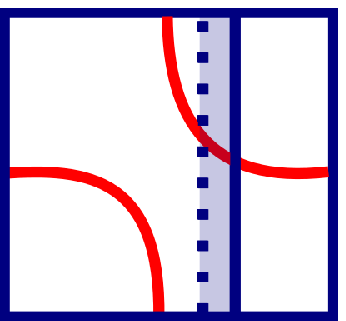}} \\ \hline
$f_b$ & $1$ & $i\cdot e^{i\pi q_r}$ & $i\cdot e^{i\pi q_b}$ & $-e^{i\pi(q_r+q_b)}$\\ 
\hline
\hline
\hline
\raisebox{-0.16in}{\includegraphics[height=0.4in]{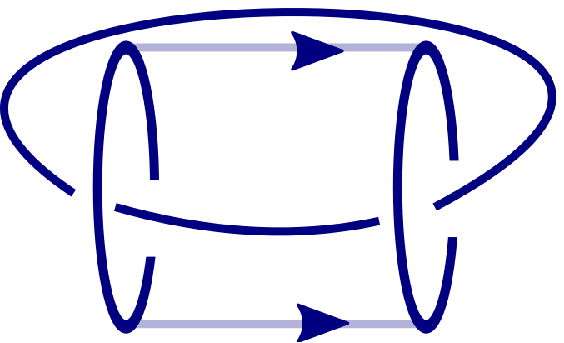}} &
\raisebox{-0.16in}{\includegraphics[height=0.4in]{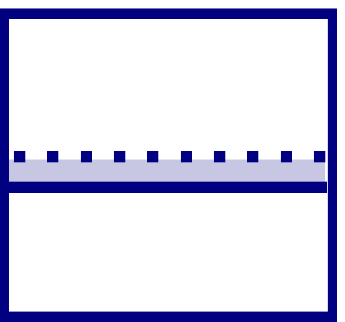}} &
\raisebox{-0.16in}{\includegraphics[height=0.4in]{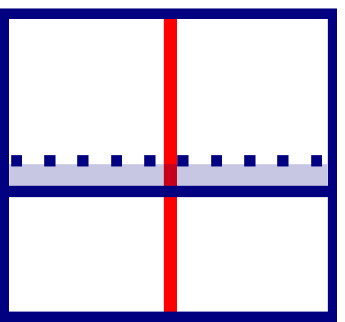}} &
\raisebox{-0.16in}{\includegraphics[height=0.4in]{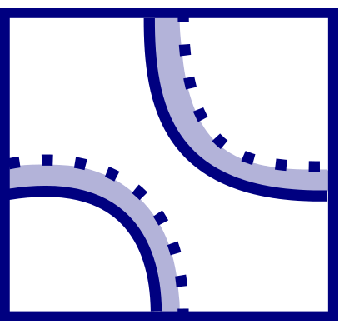}} &
\raisebox{-0.16in}{\includegraphics[height=0.4in]{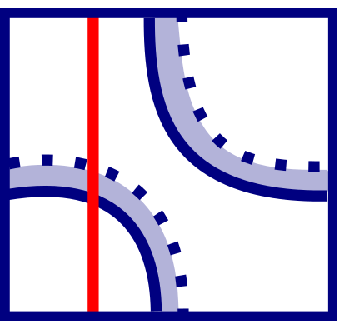}} \\ \hline
$f_b $ & $1$ & $e^{i\pi q_r}$ & $e^{i\pi q_b}$ & $e^{i\pi(q_r+q_b)}$\\ \hline 
\end{tabular}
\caption{The function $f_b$ that defines the linked blue cylinder operator is completely determined by its values on four basic pictures that are drawn on the surface of the cylinder.
The top panel shows the four basic pictures for a blue cylinder linked to a red base loop, while the bottom panel shows the pictures for a cylinder linked to a blue base loop. The corresponding values of $f_b$ are shown below the pictures. Here $q_r,q_b=0,1$ define four different functions $f_b$ and therefore four different cylinder operators for each panel.
}
\label{fig:config2}
\end{table}

We next describe a cylindrical membrane operator which creates a blue vortex loop which is linked to a blue base loop. Here, we choose a boundary 
condition on $f_b$ so that it only takes nonzero values on pictures in which a single thin blue region touches the boundary of $S$ at a fixed 
position $y_0$, and no red lines touch the boundary. (See Fig. \ref{typicalpicture2}b for a typical picture of this type). Given this boundary 
condition and the constraint equations (\ref{local}), we can reduce any such picture to one of the four pictures shown in the bottom panel of 
Table \ref{fig:config2}. The value of $f_b$ on these four pictures is shown in Table \ref{fig:config2}. 

To complete the discussion, we need to explain how to construct cylindrical membrane operators that create \emph{red} vortex loops linked to a 
blue or red base loop. These operators are defined exactly like the two operators described above, but with the roles of ``red'' and ``blue'' reversed. 

Before concluding, we make one comment about our notation: we will use the same symbol $M_b^1(S)$ to denote all of the blue membrane operators 
whether they are associated with a red base (top panel of Table \ref{fig:config2}), a blue base (bottom panel of Table \ref{fig:config2}), or no 
base at all (Table \ref{fig:config1}). Similarly, we will denote all the red membrane operators by $M_r^1(S)$ independent of what base they are 
associated with. This abuse of notation will not introduce confusion since it will always be clear from context which operator we have in mind.

\subsubsection{Toroidal and spherical membrane operators\label{otherM}}
So far we have only discussed cylindrical membrane operators. We now discuss how to construct membrane operators with other topologies --- in 
particular, toroidal and spherical operators. We begin with the toroidal operators. Roughly speaking, these operators can be obtained by connecting 
the two ends of our cylinder operators. More precisely, we define toroidal operators exactly like the cylindrical operators, with only one 
modification: we replace the cylindrical boundary conditions on $f_b$ and $f_r$ with \emph{periodic} boundary conditions in both directions of 
the torus. 

Like cylindrical operators, we can build different torus operators for different base loops. The blue torus operator for a trivial base loop is 
defined by the values of $f_b$ shown in Table \ref{fig:config1} while the blue torus operator for a red base loop is defined by the values shown 
in the top panel of Table \ref{fig:config2}. Finally, the blue torus operator for a blue base loop is defined by the values of $f_b$ shown in 
the bottom panel of Table \ref{fig:config2}. Red torus operators for various base loops can be defined in the same way but with the red and blue 
colors exchanged.

It is important to keep in mind that the two coordinates that describe the torus are \emph{not} equivalent. One coordinate parameterizes the 
movement of the loop in space, while the other coordinate parameterizes the loop itself. Thus to define a torus operator, we not only have to specify 
the torus $S$, but we also have to specify which coordinate has which meaning.

Let us now discuss spherical membrane operators.
We can define blue spherical operators following the same recipe as the blue cylindrical operators. 
The only difference is that in the spherical topology, every picture can be reduced to the vacuum or (empty) picture by application of the 
constraints (\ref{local}) --- in contrast with the cylindrical case where every picture can be reduced to one of the four pictures shown in 
Table \ref{fig:config1} or Table \ref{fig:config2}. Thus, in the spherical case it suffices to define the value of $f_b$ on the vacuum picture. Here 
we define $f_b(\text{vacuum}) = 1$. 
Note that there is only one type of blue spherical operator --- unlike the cylindrical or toroidal case 
where there are four types of operators parameterized by $q_r,q_b \in \{0,1\}$.

In fact, we already encountered spherical membrane operators in the definition of the Hamiltonian $H_1$: the operators $\mathcal{B}^1_{c}$ and 
$\mathcal{B}^1_{\hat{c}}$ (\ref{matrixB}) can be thought of small spherical membrane operators associated with a unit blue cube $S = c$ or unit 
red cube $S = \hat{c}$. To see that $\mathcal{B}^1_c$ is a spherical membrane operator, note that the picture drawn on the surface of the blue unit 
cube has the property that the boundaries of the blue regions are always solid lines ---that is, there are no boundaries that are dotted lines. The 
reason this is so is that all membranes in $X_b$ are incident on the cube from the outside, rather than the inside. Now, since the picture on the 
surface of the cube does not contain any dotted blue lines, the local constraints (\ref{local}) simplify considerably: in fact, we can throw out 
equations (\ref{r6}) - (\ref{r8}) and (\ref{r9}-\ref{r11}). The resulting equations can be solved explicitly, leading to the following formula for $f_b$:
\begin{align}
f_b(X_b, X_r, S) = (-1)&^{ \text{\#\{red loops on $c$\}}}  \nonumber \\
		 &\cdot i^{\text{\#\{blue-red intersections on $c$\}}}.
		 \label{sphere}
\end{align}
We can see that this formula, together with the definition (\ref{wb}) agrees exactly with the definition of $\mathcal{B}^1_{c}$. 

\subsubsection{Showing that the membrane operators have the required properties \label{showpathindep}}
Having defined the membrane operators $M_b^1(S)$ and $M_r^1(S)$, we now show that they have the required properties. We will focus on one case, 
namely blue membrane operators $M_b^1(S)$ that create \emph{unlinked} loop excitations. The arguments for the linked case are similar. 

To begin, we show that if $S$ is a \emph{torus} then $M_b^1(S)$ does not create any excitations at all. That is, we show
\begin{equation}
M_b^1(S) |\Psi_1\> \propto |\Psi_1\>.
\label{closedmem}
\end{equation}
To establish this result, we first rewrite (\ref{closedmem}) in a more convenient form. Multiplying both sides by $\<X_b, X_r|$ gives:
\begin{equation}
\<X_b,X_r| M_b^1(S) |\Psi_1\> \propto \<X_b, X_r|\Psi_1\>.
\end{equation}
Next, using the definition of $M_b^1(S)$ (\ref{wb}), we can rewrite this as
\begin{equation}
f_b^*(X_b,X_r,S) \cdot \Psi_1(X_b+S,X_r) \propto \Psi_1(X_b,X_r).
\label{frat}
\end{equation}
To proceed further, we observe that Eq. (\ref{frat}) is equivalent to the relation
\begin{equation}
\frac{f_b^*(X_b,X_r,S) \cdot \Psi_1(X_b+S,X_r)}{f_b^*(X_b',X_r',S) \cdot \Psi_1(X_b'+S,X_r')} = \frac{\Psi_1(X_b,X_r)}{\Psi_1(X_b',X_r')}
\end{equation}
for any two closed membrane configurations $X_b,X_r$ and $X_b',X_r'$. Finally, we use the fact that $f_b$ is a pure phase to rewrite this equation as
\begin{equation}
	\frac{f_b(X_b',X_r',S)}{f_b(X_b,X_r,S)}   = \frac{\Psi_1(X_b,X_r) \Psi_1(X_b'+S,X_r')}{\Psi_1(X_b',X_r')\Psi_1(X_b+S,X_r)}.
	\label{fratio2}
\end{equation}

Our task is now to prove equation (\ref{fratio2}). First, we claim that it suffices to prove (\ref{fratio2}) for the case where $(X_b',X_r')$ and 
$(X_b, X_r)$ only differ locally --- i.e. only differ in some small region. The reason it is enough to consider this case is that we can get from 
any membrane configuration $(X_b, X_r)$ to any other configuration $(X_b',X_r')$ by a series of local changes: 
\begin{equation*}
(X_b, X_r) \rightarrow (X_{b1}, X_{r1}) \rightarrow (X_{b2}, X_{r2}) \cdots \rightarrow (X_b',X_r').
\end{equation*}
If (\ref{fratio2}) holds for each pair $(X_{bi},X_{ri})$,$(X_{b(i+1)},X_{r(i+1)})$ then once we multiply these relations together, we see that it 
automatically holds for $(X_b',X_r')$ and $(X_b, X_r)$.

There are two cases to consider: the region where $(X_b',X_r')$ and $(X_b, X_r)$ differ may overlap $S$ or not overlap $S$. In the second case, it 
is easy to see that (\ref{fratio2}) holds. Indeed, in this case the left hand side of (\ref{fratio2}) is equal to $1$ since $f_b$ only depends on 
the membrane configuration in the neighborhood of $S$ and hence $f_b(X_b',X_r',S) = f_b(X_b,X_r,S)$. 
Similarly, the right hand side of (\ref{fratio2}) is also equal to $1$ since the ratio $\Psi_1(X_b',X_r')/\Psi_1(X_b,X_r)$ only depends on the 
membrane configuration in the region where $(X_b',X_r')$ and $(X_b, X_r)$ differ from one another.

\begin{figure}[ptb]
\begin{center}
\includegraphics[
height=0.6in,
width=2.4in
]{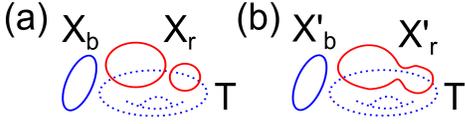}
\end{center}
\caption{A blue torus operator $M_b(S)$ acts on two slightly different membrane configurations (a) $(X_b,X_r)$ and (b) $(X_b',X_r')$. Here, $X_b=X_b'$ is a blue sphere, while $X_r$ consists of two red spheres and $X_r'$ consists of a bigger red sphere obtained by merging the two red spheres in $X_r$.
}
\label{exfb}
\end{figure}

All that remains is the case where $(X_b',X_r')$ and $(X_b, X_r)$ differ in a small region that overlaps $S$. In this case, the ratio on the left 
hand side of (\ref{fratio2}) is directly determined by the local constraint equations (\ref{local}) for $f_b$. The key point is that these constraint 
equations were chosen specifically so that a solution to these equations will automatically obey (\ref{fratio2}). 
This is easiest to see by example. Consider a blue torus operator $M_b^1(S)$ acting on the two configurations $(X_b,X_r)$ and $(X_b',X_r')$ shown in 
Fig. \ref{exfb}. Here $X_b = X_b'$ consists of a single blue sphere, while $X_r$ consists of two red spheres and $X_r'$ consists of single red sphere 
obtained by merging together the two spheres in $X_r$. In this case 
\begin{align*}
f_b(X_b,X_r,S)&=f_b\left(\raisebox{-0.115in}{\includegraphics[height=0.3in]{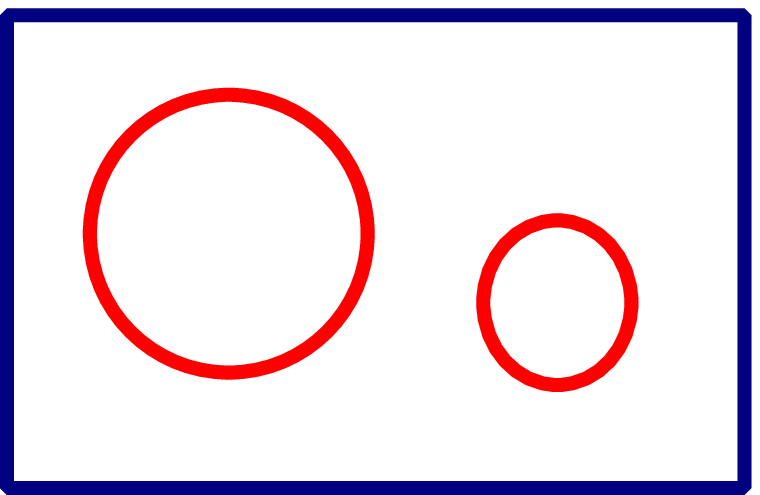}}\right), \\
f_b(X_b',X_r',S)&=f_b\left(\raisebox{-0.115in}{\includegraphics[height=0.3in]{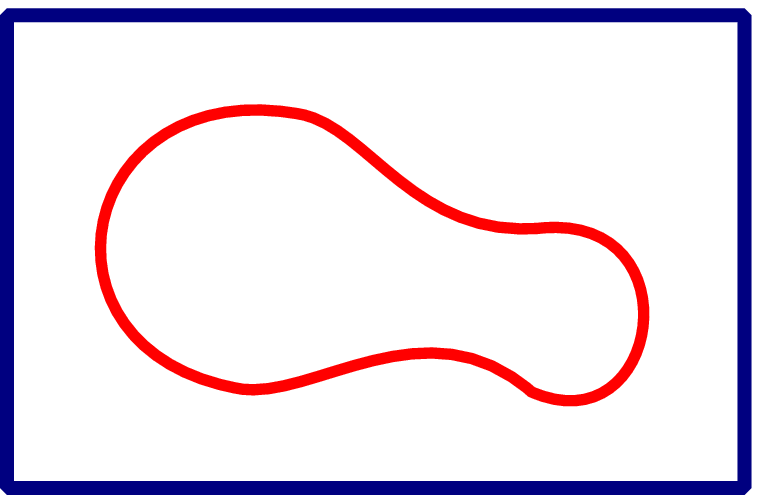}}\right)
\end{align*}
so that 
\begin{eqnarray}
\frac{f_b(X_b',X_r',S)}{f_b(X_b,X_r,S)} &=& \frac{f_b\left(\raisebox{-0.115in}{\includegraphics[height=0.3in]{exfb2.eps}}\right)}{f_b\left(\raisebox{-0.115in}{\includegraphics[height=0.3in]{exfb1.eps}}\right)}=-1
\end{eqnarray}
according to (\ref{r3}). On the other hand, it is easy to see that
\begin{align*}
\frac{\Psi_1(X_b,X_r)}{\Psi_1(X_b',X_r')} =1, \quad \frac{\Psi_1(X_b+S,X_r)}{\Psi_1(X_b'+S,X_r')} =-1 
\end{align*}
using the explicit formula for $\Psi_1$ (\ref{psi1}). We conclude that
\begin{equation}
\frac{\Psi_1(X_b,X_r) \Psi_1(X_b'+S,X_r')}{\Psi_1(X_b',X_r')\Psi_1(X_b+S,X_r)} = -1
\end{equation}
so that equation (\ref{fratio2}) is satisfied in this case. 

The above example serves two purposes. First it demonstrates, in at least one case, that $f_b$ obeys equation (\ref{fratio2}). Second, it reveals 
where the constraint equation (\ref{r3}) comes from: it should be clear that we chose the factor of $-1$ in this equation specifically to ensure 
that (\ref{fratio2}) was satisfied. Similarly, one can check that the other constraint equations (\ref{local}) ensure that $f_b$ obeys 
(\ref{fratio2}) in other cases.

This concludes our argument showing that the torus operator $M_b^1(S)$ does not create any excitations when applied to the ground state. Next, we 
need to show that the \emph{cylinder} operator $M_b^1(S)$ does not create any excitations except at the two boundaries of $S$. This claim can be 
established using a similar argument to the torus case, but we will not repeat the derivation here.
Instead we simply observe that the cylinder and torus operators look identical except near the two boundaries of the cylinder. Therefore it is 
intuitively clear that since the torus operator does not create any excitations, the same must be true of the cylinder operator away from its 
boundaries.

At this point, we have argued that the cylinder operator $M_b^1(S)$ creates blue loop excitations at its two boundaries and no other 
excitations anywhere else. However, we are not quite finished: we still need to verify one more property of the cylinder operator $M_b^1(S)$. Recall 
that $H_1$ supports four topologically distinct types of blue loop excitations $(b,q_b,q_r)$ which differ from one another by attaching red and blue 
charges (see section \ref{memH0}). We need to check that $M_b^1(S)$ creates exactly one of these excitations, and not a linear superposition of 
different types of excitations. In other words, we need to check that $M_b^1(S)$ creates loop excitations that are \emph{eigenstates} of braiding 
measurements. 

To see that $M_b^1$ creates braiding eigenstates, we make use of a result from appendix \ref{simpleop}. In that appendix, we show that the cylinder 
operators $M_b^1(S)$ are guaranteed to create braiding measurement eigenstates provided that $f_b$ is multiplicative in the sense that
\begin{equation}
	f_b(X_b,X_r,S \cup S')= f_b(X_b,X_r,S) \cdot f_b(X_b, X_r, S')
	\label{piecewise}
\end{equation}
for any two cylinders $S$ and $S'$ that share a common boundary and any membrane configuration $(X_b, X_r)$ whose intersection with the common 
boundary obeys the appropriate cylinder operator boundary condition. 

In view of this result, we only have to show that $f_b$ obeys condition (\ref{piecewise}). 
We go through this calculation in appendix \ref{simpleop}, and we show that $f_b$ does indeed 
obey (\ref{piecewise}) provided that the values of $f_b$ on the four basic pictures are those shown in Table \ref{fig:config1}. In fact, this is 
why we picked the particular values shown in that table: we chose those values to ensure that $f_b$ obeys equation (\ref{piecewise}).

To summarize, we have shown that the cylinder operators $M_b^1(S)$ create blue loop excitations at their two boundaries and no other excitations 
anywhere else. We have also shown that these blue loop excitations are eigenstates of braiding measurements. This establishes that the cylinder 
operators have all the required properties.

\subsection{Labeling scheme for loop excitations}
In this section, we discuss some subtleties related to the labeling of loop excitations. As we have emphasized previously, both the $H_0$ and $H_1$ 
models support four different kinds of blue loop excitations and four different kinds of red loop excitations. We label the former by 
$(b,q_b,q_r)$ and the latter by $(r,q_b,q_r)$ where $q_b,q_r = 0,1$. In practice, we first assign labels to the membrane operators; we then 
assign labels to the loop excitations according to which membrane operator creates them.

Now, an important question is how we choose which loop excitations are labeled by $(b,0,0)$ and $(r,0,0)$. We will call these excitations 
\emph{neutral} loops. Once we decide which blue and red loops should called neutral, the labeling of all the other excitations is naturally fixed: 
for example, we assign the label $(b,q_b, q_r)$ to the loop that is obtained by attaching $q_b$ blue charge and $q_r$ red charge to the neutral 
loop, $(b,0,0)$. 

For the case of \emph{unlinked} loops, there is a natural choice for which loop should be called neutral: the neutral loop is the unique loop that 
can be shrunk down to a point and annihilated by a local operator. Thus, for the case of unlinked loops there is a canonical labeling scheme. This 
is the labeling scheme we use here. Indeed, in the case of $H_0$, we assign $(b,0,0)$ to the loop created by the operator (\ref{membrane0}), and one can readily 
verify that this loop is the unique blue loop that can be annihilated locally. Similarly, in the case of $H_1$, we assign $(b,0,0)$ to the loop
created by the membrane operator defined in Table \ref{fig:config1} with $q_b = q_r = 0$ and one can also check that this loop can be annihilated 
locally. (One way to see this is to note that if we shrink the two ends of the cylinder to make a sphere, the operator defined by 
Table \ref{fig:config1} reduces to the spherical membrane operator defined by (\ref{sphere})).

In contrast, for \emph{linked} loops, there is no canonical way to define which loops are neutral. Therefore, we make this assignment 
\emph{arbitrarily}. In other words, for each base loop, we arbitrarily choose one of the blue linked loops and one of the red linked loops, and we 
call them neutral. In the case of $H_1$, this arbitrary choice enters in how we parameterize the values in Table \ref{fig:config2} which define the 
membrane operators and hence define the loop excitations. As we show in appendix \ref{simpleop}, the values in Table \ref{fig:config2} are obtained by solving 
certain algebraic equations. These equations have four different solutions, which define four different membrane operators. All four are on an equal footing, but we 
simply picked one and labeled it by $q_b = 0, q_r = 0$. The labeling of the other solutions is then completely fixed. We could replace 
$q_b \rightarrow q_b + 1$ or $q_r \rightarrow q_r + 1$, for either the top or bottom panel of Table \ref{fig:config2}, and the resulting table would 
define an equally valid labeling convention.

The arbitrariness of our labeling scheme has an important consequence: because we picked the labels in an arbitrary way, there is no sense in which 
an unlinked loop of type $(b,q_b,q_r)$ is the ``same'' type as a linked loop of type $(b,q_b,q_r)$. More generally, we cannot sensibly compare loops 
that are linked with different base loops. Each base loop effectively defines its own universe of excitations.

\section{Braiding statistics of excitations \label{section:statistics}}
In the previous section, we constructed operators that create charges and vortex loops for the $H_0, H_1$ models. In this section, we will use these 
operators to compute the braiding statistics of these excitations. We consider four types of processes: (1) processes involving two charges, 
(2) processes involving a charge and a vortex loop, (3) processes involving two vortex loops, and finally (4)
``three-loop processes'' involving two vortex loops that are both linked with a third loop. We find that the two models have the same braiding 
statistics except for the last process; when we investigate the three-loop braiding process we find a distinction between the two models which 
implies that they belong to distinct topological phases. 

\subsection{Braiding two charges}

\begin{figure}[ptb]
\begin{center}
\includegraphics[
height=0.8in,
width=1.0in
]{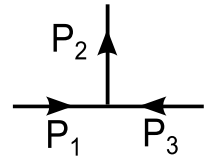}
\end{center}
\caption{The exchange statistics of the charge excitations can be computed from the commutation algebra (\ref{hoppingalg}) of three string operators acting on three paths $P_1, P_2, P_3$ that share a common endpoint.}
\label{hopping}
\end{figure}

To compute the statistics of the charge excitations, we use the ``hopping operator algebra'' derived in Ref. \onlinecite{LevinWenHop}. This algebra 
relates the exchange statistics of particle excitations to the commutation properties of the string operators that create these particles. To see how 
this works, let us consider the blue charge excitations in the $H_0$ model. According to the hopping operator algebra, the exchange statistics of
these excitations can be read off from the phase factor in the commutation relation 
\begin{eqnarray}
W_b^0(P_1) W_b^0(P_2) &W_b^0(P_3)&|\Psi_0\> = \nonumber \\
e^{i \theta} \cdot &W_b^0(P_3)& W_b^0(P_2) W_b^0(P_1)|\Psi_0\> 
\label{hoppingalg}
\end{eqnarray}
where $P_1, P_2, P_3$ are three paths arranged in the geometry of Fig. \ref{hopping} and $|\Psi_0\>$ is the ground state of $H_0$. That is, if 
$\theta = 0$, then the blue charges are bosons, while if $\theta = \pi$, the blue charges are fermions. (Other values of $\theta$ are not possible 
in 3D). If we examine the form of the string operators, we can see that $\theta = 0$ since the string operators 
$W_b^0(P_1), W_b^0(P_2), W_b^0(P_3)$ (\ref{string0}) all commute with one another. We conclude that the blue charges are bosons in the $H_0$ model. 
Using identical reasoning, we can see that the blue and red charges are bosons in both models.
 
\subsection{Braiding a charge and a vortex loop}

\begin{figure}[ptb]
\begin{center}
\includegraphics[
height=0.8in,
width=2.2in
]{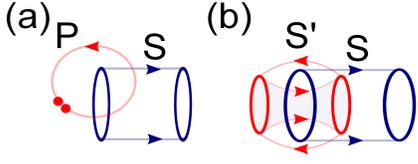}
\end{center}
\caption{(a) The statistical phase associated with braiding a charge around a loop can be computed from the commutation algebra (\ref{wmalg}) of a string operator acting along a path $P$, and a membrane operator acting along a cylinder $S$. 
	(b) The phase associated with braiding two loops around one another can be computed from the commutation algebra (\ref{mmalg0}) of two membrane operators, one acting on a cylinder $S$, and the other acting on a torus $S'$.}
\label{braidingloop}
\end{figure}

Next, we compute the statistical Berry phase associated with braiding a charge around a vortex loop. More specifically, let us consider the 
$H_0$ model and imagine braiding a blue charge around a blue vortex loop. The statistical phase $\theta$ for such a process can be read off from the 
commutation algebra of the corresponding string operator and membrane operator 
\begin{equation}
	M_b^0(S)W_b^0(P)|\Psi_0\>=e^{i\theta} \cdot W_b^0(P)M_b^0(S)|\Psi_0\>
	\label{wmalg}
\end{equation}
where $P$ is a closed path and $S$ is a cylindrical surface, arranged as in Fig. \ref{braidingloop}a. To see where this relation comes from, note 
that the operator $M_b^0(S)$ describes a process in which two vortex loops are created and moved to the ends of the cylinder $S$, while the operator 
$W_b^0(P)$ describes a three-step process in which two charges are created out of the vacuum, one of them is braided around the path $P$, and then 
the two annihilated with one another. With this interpretation, the right hand side of (\ref{wmalg}) describes a process in which two vortex loops 
are created and then a charge is braided around $P$, while the left hand side describes a process in which a charge is braided around $P$ 
first, and then two vortex loops are created. Clearly the phase difference between these two processes is the statistical phase associated with 
braiding a charge around a vortex loop.

Remembering the definition of $M_b^0(S)$ and $W_b^0(P)$, we can see that these operators anticommute with each other. We conclude that braiding a 
blue charge around a blue vortex loop results in a statistical phase of $\pi$. Similarly, we can see that $M_r^0(S)$ and $W_r^0(P)$ anticommute with 
each other so braiding a red charge around a red vortex loop gives a phase of $\pi$. On the other hand, if one braids a red charge around a blue 
vortex loop or a blue charge around a red vortex loop, the statistical phase vanishes since these operators commute with one another. 

A similar calculation for the $H_1$ model gives identical results. Thus the two models share the same statistics between charges and vortex loops. 
We note that these statistical phases agree with the Aharonov-Bohm phases associated with braiding a charge around a vortex loop in 
$\mathbb{Z}_2 \times \mathbb{Z}_2$ gauge theory. This is not a coincidence: as we mentioned in the introduction, the two models can be viewed as two 
different types of  $\mathbb{Z}_2 \times \mathbb{Z}_2$ gauge theories.

\subsection{Braiding two vortex loops}
\subsubsection{\texorpdfstring{$H_0$}{H0}}
We now consider, for the $H_0$ model, the statistical phase associated with braiding a vortex loop around another vortex loop (Fig. \ref{loopbraiding}a). For example, let us 
consider braiding a blue vortex loop around another blue vortex loop. Similarly to Eq. (\ref{wmalg}), the statistical phase $\theta^0_{bb}$ can be 
computed from the commutation relation 
\begin{equation}
	M_{b}^0(S') M_{b}^0(S)|\Psi_0\> = e^{i\theta^0_{bb}} \cdot M_{b}^0(S) M_{b}^0(S') |\Psi_0\>
	\label{mmalg0}
\end{equation}
where $S$ is a cylinder and $S'$ is a torus, arranged as in Fig. \ref{braidingloop}b.
Examining the definition of the membrane operators (\ref{membrane0}), we can see that they commute with one another so 
that $\theta^0_{bb} = 0$. Likewise, we can see that there is no statistical phase associated with braiding a 
red loop around a red loop or a red loop around a blue loop since the corresponding membrane operators all commute with one another.

The above results apply to the blue vortex loop and red vortex loops labeled by $(b,0,0)$ and $(r,0,0)$. One might also wonder
about the braiding statistics of more general vortex loops $(b, q_b, q_r)$ and $(r, q_b, q_r)$. The braiding statistics of
these more general vortex loops can be computed using the same approach as above. The only difference is that the
membrane creation operators are slightly different in this case: they are obtained by multiplying $M_b^0$ (or $M_r^0$) by one of the string 
operators $W_b^0(P)$, $W_r^0(P')$
or $W_b^0(P) \cdot W_r^0(P')$, where the paths $P, P'$ run along the length of the cylinder. Substituting these modified membrane operators into 
Eq. (\ref{mmalg0}), a simple calculation shows that the phase associated with braiding a blue vortex loop $(b, q_{b}, q_{r})$ around another
blue vortex loop $(b, q_{b}', q_{r}')$ is given by
\begin{equation}
\theta^0_{bb} = \pi (q_{b} + q_{b}').
\label{mmalgbb}
\end{equation}
Similarly, the phase associated with braiding a red vortex loop $(r, q_{b}, q_{r})$ around another red vortex loop $(r, q_{b}', q_{r}')$ is
\begin{equation}
\theta^0_{rr} = \pi (q_{r} + q_{r}')
\label{mmalgrr}
\end{equation}
while the phase associated with braiding a red vortex loop $(r, q_{b}, q_{r})$ around a blue vortex loop $(b, q_{b}', q_{r}')$ is
\begin{equation}
\theta^0_{rb} = \pi (q_{b} + q_{r}').
\label{mmalgrb}
\end{equation}
These expressions have a natural interpretation in terms of Aharonov-Bohm phases: the first term in each formula is the statistical phase 
associated with braiding the charge on the first vortex loop around the flux of the second loop while the second term is the phase 
associated with braiding the flux of the first vortex loop around the charge on the second loop.

\subsubsection{\texorpdfstring{$H_1$}{H1}}
\label{h1twoloopsect}
Similarly to the $H_0$ case, the statistical phase $\theta^1_{bb}$ associated with braiding a blue vortex loop around a blue 
vortex loop in the $H_1$ model can be read off from the commutation algebra of the corresponding membrane operators:
\begin{equation}
	M_{b}^1(S') M_{b}^1(S)|\Psi_1\> = e^{i\theta^1_{bb}} \cdot M_{b}^1(S) M_{b}^1(S') |\Psi_1\>.
	\label{mmalg1}
\end{equation}
However the membrane operators $M_b^1$ are more complicated than their counterparts in the $H_0$ model, so it is more difficult to apply 
Eq. (\ref{mmalg1}). To deal with this issue, we 
now derive a simpler version of Eq. (\ref{mmalg1}) which is more convenient for our purposes. The first step is to rewrite Eq. (\ref{mmalg1}) as
\begin{align*}
	\sum_X M_b^1(S')M_b^1(S) & |X\>\<X|\Psi_{1}\> = \\
	&e^{i\theta^1_{bb}}\sum_X M_b^1(S)M_b^1(S')|X\>\<X|\Psi_{1}\>.
\end{align*}
where $|X\> \equiv |X_b, X_r\>$.
Next, we use the fact that $M_b^1(S') M_b^1(S)|X_b, X_r\> \propto |X_b + S + S', X_r\>$, which implies that all of the states 
$M_b^1(S') M_b^1(S)|X\>$ that appear in the
above sum are linearly independent from one another. It follows that the above equality must hold for each term \emph{separately}. Hence, we must have 
\begin{equation}
M_b^1(S') M_b^1(S) |X\> = e^{i\theta^1_{bb}} \cdot M_b^1(S)M_b^1(S')|X\>
\label{mmalgsimp1}
\end{equation}
for every membrane state $|X\>$ that has a nonzero amplitude in the state $|\Psi_1\>$. 

We will now use (\ref{mmalgsimp1}) to compute the statistics of the loops in the $H_1$ model. To this end, we set $|X\> = |0,0\>$, the 
``no-membrane'' state. We then define $|Y\> = |S,0\>$, $|Y'\> = |S',0\>$, and $|Z\> = |S+S',0\>$. With this notation, the left 
side of (\ref{mmalgsimp1}) can be computed as
\begin{eqnarray}
M_b^1(S') M_b^1(S) |X\> &=& M_b^1(S') \cdot f_b(X,S) |Y\> \nonumber \\
&=& f_b(Y,S') f_b(X,S) |Z\>.
\end{eqnarray}
Similarly, the term on the right side is given by
\begin{eqnarray}
M_b^1(S) M_b^1(S') |X\> &=& M_b^1(S) \cdot f_b(X,S') |Y'\> \nonumber \\
&=& f_b(Y', S) f_b(X,S') |Z\>.
\end{eqnarray}
Comparing these two expressions with (\ref{mmalgsimp1}), we derive
\begin{equation}
e^{i\theta^1_{bb}} = \frac{f_b(Y,S') f_b(X,S)}{f_b(Y', S) f_b(X,S')}.
\label{thetabbform}
\end{equation}
To complete the calculation, we compute the value of $f_b$ for each of these configurations
\begin{align}
f_b(X,S) &= f_b\left( \raisebox{-0.12in}{\includegraphics[height=0.3in]{p1a.eps}}\right) = 1, \nonumber \\
f_b(X,S') &= f_b\left( \raisebox{-0.12in}{\includegraphics[height=0.3in]{p1a.eps}} \right) = 1, \nonumber \\
f_b(Y,S') &= f_b\left( \raisebox{-0.12in}{\includegraphics[height=0.3in]{p1c.eps}} \right) = e^{i\pi q_{b'}}, \nonumber \\
f_b(Y',S) &= f_b\left( \raisebox{-0.12in}{\includegraphics[height=0.3in]{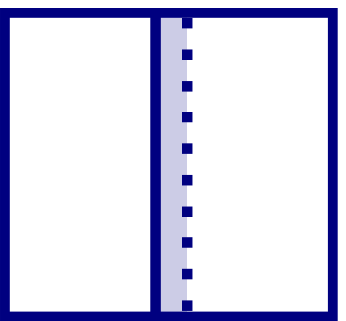}} \right) = e^{i \pi q_{b}}. 
\end{align}
Substituting these values into (\ref{thetabbform}), we find that the statistical phase associated with braiding a loop $(b, q_{b}, q_{b})$ 
around another loop $(b,q_{b}', q_{b}')$ is
\begin{equation}
\theta^1_{bb} = \pi (q_{b} + q_{b}').
\end{equation}
We note that this result is identical to the statistical phase $\theta^0_{bb}$ (\ref{mmalgbb}) in the $H_0$ model. Similarly, it is easy to 
check that the phase associated with braiding two red loops
around one another agrees with (\ref{mmalgrr}) while the phase associated with braiding a red loop around a blue loop agrees with (\ref{mmalgrb}). 
Thus, the two models
share the same two-loop braiding statistics. In fact the agreement between the two models is not surprising, since the two models are equivalent to 
$\mathbb{Z}_2 \times \mathbb{Z}_2$ gauge theories, 
and in such systems the two-loop braiding statistics always has an Aharonov-Bohm form, as explained in Ref. \onlinecite{WangLevin}. 

\subsection{Three-loop braiding \label{threeloop}}
Finally, we discuss the three-loop braiding statistics in the two models. Specifically, we consider a braiding process in which a loop is braided 
around another loop, while both are linked to a third ``base'' loop (Fig. \ref{loopbraiding}b). Unlike the other processes we have considered until now, we will see that 
the two models can be distinguished by their three-loop statistics.

\subsubsection{\texorpdfstring{$H_0$}{H0}}

\begin{figure}[ptb]
\begin{center}
\includegraphics[
height=0.8in,
width=1.2in
]{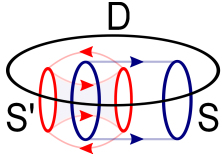}
\end{center}
\caption{The statistical phase associated with a three-loop braiding process can be computed from the commutation algebra of two membrane operators, one acting along the cylinder $S$, and the other acting along the torus $S'$. Here both $S$ and $S'$ are linked with a base loop which lies along the boundary of the disk $D$.}
\label{braiding}
\end{figure}

First we compute the statistics in the $H_0$ model. As in the two-loop case, the three-loop braiding statistics can be read off from the 
commutation algebra for the membrane 
operators. For example, the statistical phase $\theta^0_{bb,r}$ associated with braiding a blue vortex loop around a blue vortex 
loop, while both are linked to a red base loop can be obtained from
\begin{equation}
	M_{b}^0(S') M_{b}^0(S)|\Psi_{ex}\> = e^{i\theta^0_{bb,r}} \cdot M_{b}^0(S) M_{b}^0(S') |\Psi_{ex}\>
\end{equation}
where $S$ is a cylinder, $S'$ is a torus, and $|\Psi_{ex}\>$ is an excited state with a red vortex loop that links with both $S$ 
and $S'$ (Fig. \ref{braiding}). Remembering that the membrane operators $M_b^0$ all commute with each other, we deduce that 
$\theta^0_{bb,r} = 0$. In the same way, we can see that $\theta^0_{bb,b} = 0$.

In the above calculation we implicitly assumed loops of the form $(b,0,0)$. If we instead consider a process in which a general vortex
loop $(b, q_{b}, q_{r})$ is braided around another loop $(b,q_{b}',q_{r}')$ with either a red base loop or a blue base loop, 
then, just as in the two-loop case (\ref{mmalgbb}), one finds
\begin{equation}
\theta^0_{bb,r} = \theta^0_{bb,b} = \pi (q_{b} + q_{b}'). 
\label{thetabbr0}
\end{equation}
Similarly, if we braid a red loop $(r,q_{b}, q_{r})$ around another red loop $(r,q_{b}',q_{r}')$ with either base, the
statistical phase is
\begin{equation}
\theta^0_{rr,r} = \theta^0_{rr,b} = \pi (q_{r} + q_{r}'). 
\label{thetarrr0}
\end{equation}
Finally, braiding a red loop $(r,q_b,q_r)$ around a blue loop $(b,q_b',q_r')$ with either base gives the phase
\begin{equation}
\theta^0_{rb,r} = \theta^0_{rb,b} = \pi (q_b + q_r').
\label{thetarbr0}
\end{equation}  
The agreement with the two-loop statistics (\ref{mmalgbb}-\ref{mmalgrb}) is to be expected
since the membrane operators $M_b^0(S)$,$M_b^0(S')$ obey the same commutation algebra independent of whether they act on the
ground state $|\Psi_0\>$ (as in the two loop case) or an excited state $|\Psi_{ex}\>$ (as in the three loop case).

\subsubsection{\texorpdfstring{$H_1$}{H1}}
We begin by computing the statistical phase associated with braiding a blue vortex loop around another blue vortex loop, while
both are linked with a red base loop. As in the $H_0$ case, the statistical phase $\theta^1_{bb,r}$ is given by
\begin{equation}
        M_{b}^1(S') M_{b}^1(S)|\Psi_{ex}\> = e^{i\theta^1_{bb,r}} \cdot M_{b}^1(S) M_{b}^1(S') |\Psi_{ex}\>
\label{mmalgpsiex}
\end{equation}
where $S$ is a cylinder, $S'$ is a torus, and $|\Psi_{ex}\>$ is an excited state with a red vortex loop that links with both $S$
and $S'$ (Fig. \ref{braiding}). Following the same logic as in section \ref{h1twoloopsect}, it is straightforward to deduce an alternative and more
convenient form of equation (\ref{mmalgpsiex}):
\begin{equation}
M_b^1(S') M_b^1(S) |X\> = e^{i\theta^1_{bb,r}} M_b^1(S)M_b^1(S')|X\>
\label{mmalgpsiexsimp}
\end{equation}
where $|X\> \equiv |X_b,X_r\>$ is any membrane state that has nonzero overlap with $|\Psi_{ex}\>$. To compute $\theta^1_{bb,r}$, we set $|X\> = |0,D\>$
where $D$ is a disk with a boundary along the red vortex loop. We then define $|Y\> = |S,D\>$, $|Y'\> = |S',D\>$, and
$|Z\> = |S+S',D\>$. With this notation, the left hand side of (\ref{mmalgpsiexsimp}) can be computed as
\begin{equation}
M_b^1(S') M_b^1(S) |X\> = f_b(Y,S') f_b(X,S) |Z\>
\end{equation}
while the expression on the right hand side can be written as
\begin{equation}
M_{b}^1(S) M_b^1(S') |X\> = f_b(Y',S) f_b(X,S') |Z\>.
\end{equation}
The three-loop statistics is then given by
\begin{equation}
 e^{i\theta^1_{bb,r}}= \frac{f_b(Y,S') f_b(X,S)}{ f_b(Y',S) f_b(X,S')}.
\label{thetabbrform}
\end{equation}
To complete the calculation, we need to compute the values of $f_b$ on each of these configurations. We have
\begin{align}
f_b(X,S) &= f_b\left(\raisebox{-0.12in}{\includegraphics[height=0.3in]{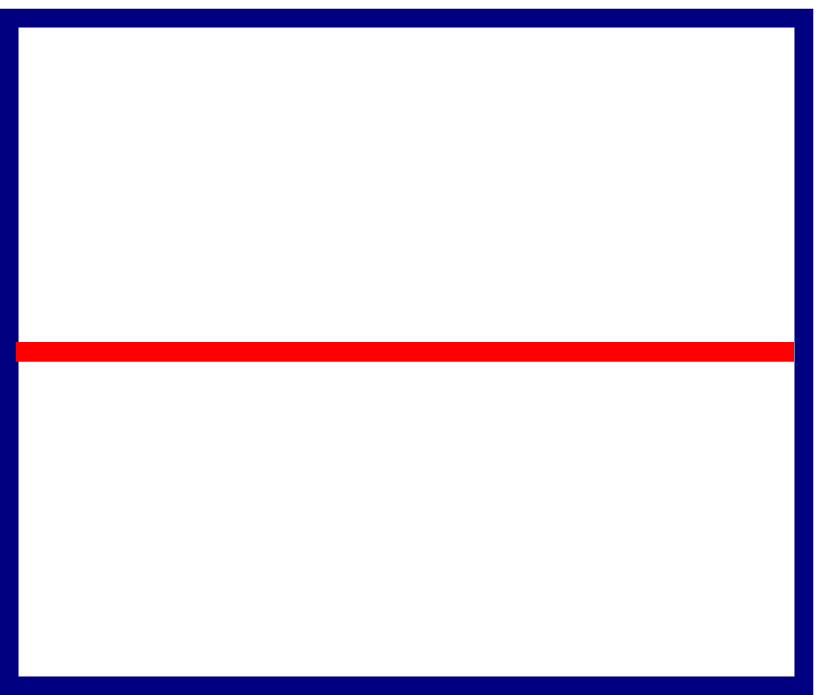}}\right) = 1, \nonumber \\
f_b(X,S') &= f_b\left(\raisebox{-0.12in}{\includegraphics[height=0.3in]{RRB1.eps}}\right) = 1, \nonumber \\
f_b(Y,S') &= f_b\left(\raisebox{-0.12in}{\includegraphics[height=0.3in]{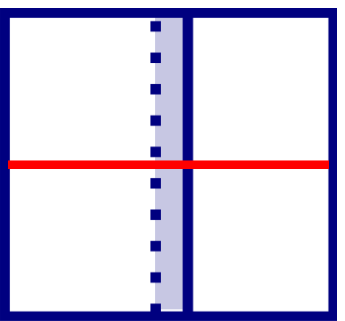}}\right) = i \cdot e^{i\pi q_b'}, \nonumber \\
f_b(Y',S) &= f_b\left(\raisebox{-0.12in}{\includegraphics[height=0.3in]{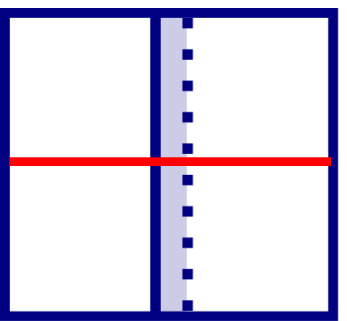}}\right) = -i \cdot e^{i\pi q_b},
\end{align}
where the `$-$' sign in the last equation comes from using the local rules (\ref{local}) to reduce the picture to one of the basic pictures in
Table \ref{fig:config2}.
Substituting these values into (\ref{thetabbrform}), we find that the statistical phase associated with braiding a loop
$(b,q_b,q_r)$ around another loop $(b,q_b',q_r')$ while both are linked to a red base loop is given by:
\begin{equation}
\theta^1_{bb,r} = \pi + \pi (q_b + q_b'). 
\label{thetabbr1}
\end{equation}
In a similar manner, it is easy to show that the phase associated with braiding a red loop $(r,q_b,q_r)$ around another red loop $(r,q_b',q_r')$, while
both are linked to a red base is:
\begin{equation}
\theta^1_{rr,r} = \pi (q_r + q_r').
\label{thetarrr1}
\end{equation}

As another example, let us compute the phase associated with braiding a red loop around a blue loop, with a red base.
The analogue of equation (\ref{mmalgpsiexsimp}) in this case is
\begin{equation}
M_r^1(S') M_b^1(S) |X\> = e^{i\theta^1_{rb,r}} \cdot M_b^1(S)M_r^1(S')|X\>
\end{equation}
where $|X\> \equiv |X_b, X_r\>$ is any state that has nonzero overlap with $|\Psi_{ex}\>$ and $|\Psi_{ex}\>$ is an excited state with a red
vortex loop that links with both $S$ and $S'$. Letting $|X\> = |0,D\>$, $|Y\> = |S,D\>$, $|Y'\> = |0, D+S'\>$, and
$|Z\> = |S,D+S'\>$, we derive
\begin{equation}
e^{i\theta^1_{rb,r}}= \frac{f_r(Y,S') f_b(X,S)}{ f_b(Y',S) f_r(X,S')}.
\label{thetarbrform}
\end{equation}
Proceeding as before, we find
\begin{align}
f_b(X,S) &= f_b\left(\raisebox{-0.12in}{\includegraphics[height=0.3in]{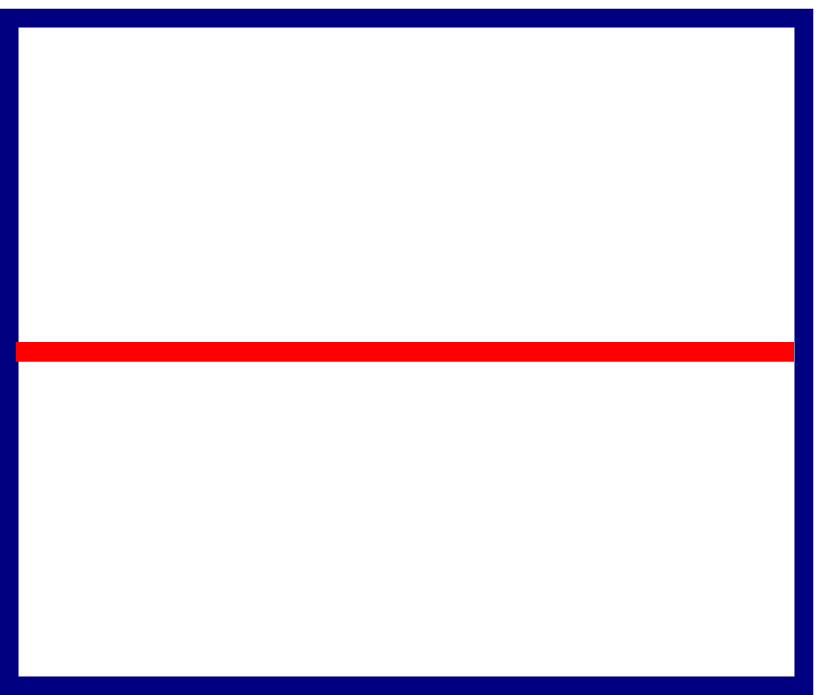}}\right) = 1, \nonumber \\
f_r(X,S') &= f_r\left(\raisebox{-0.12in}{\includegraphics[height=0.3in]{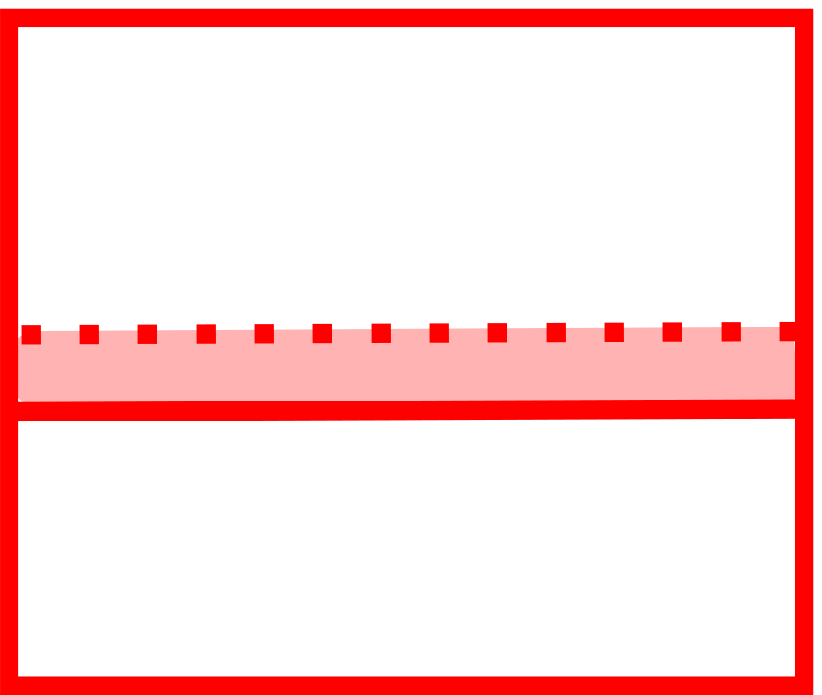}}\right) = 1, \nonumber \\
f_r(Y,S') &= f_r\left(\raisebox{-0.12in}{\includegraphics[height=0.3in]{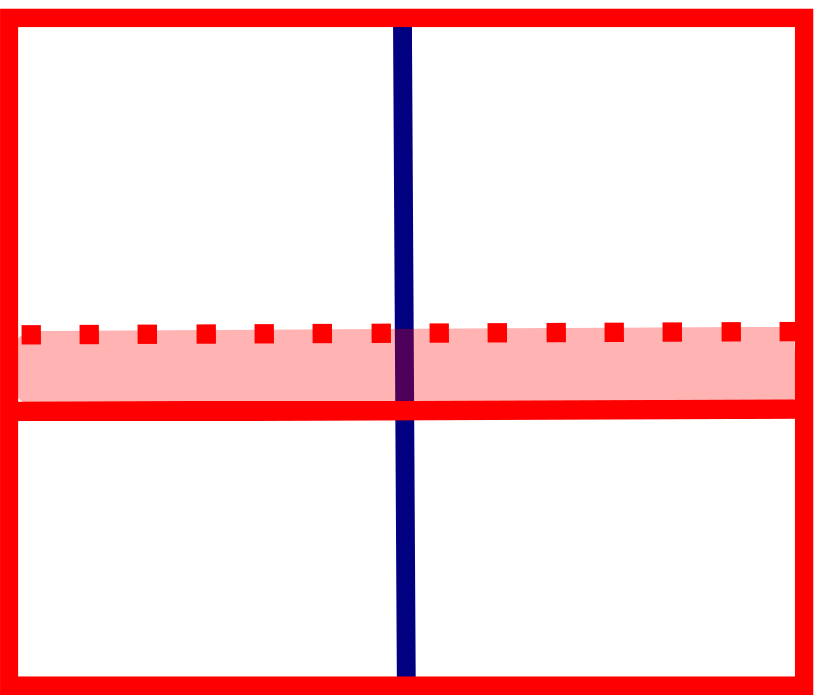}}\right) = e^{i \pi q_b'}, \nonumber \\
f_b(Y',S) &= f_b\left(\raisebox{-0.12in}{\includegraphics[height=0.3in]{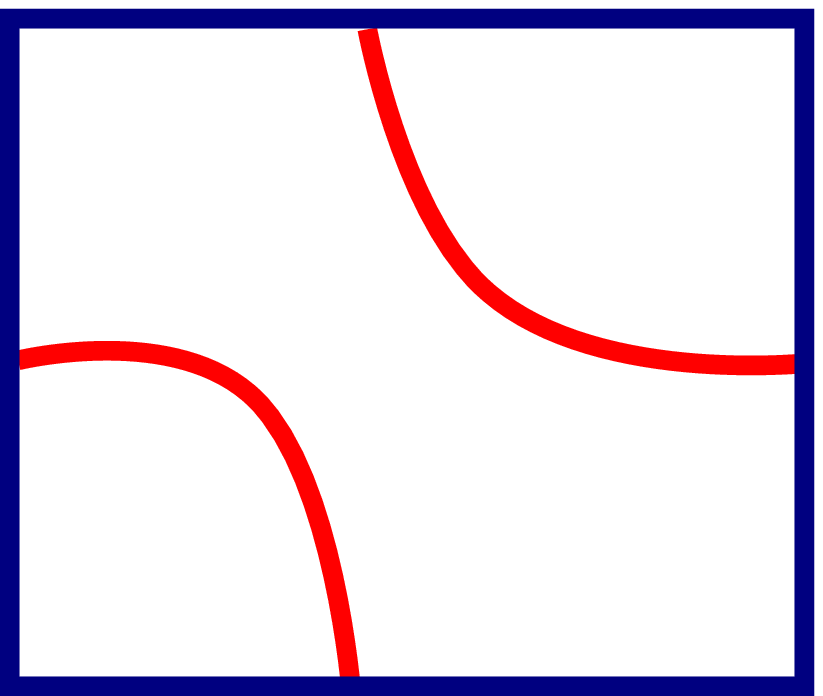}}\right) = i \cdot e^{i \pi q_r}.
\end{align}
Substituting these values into (\ref{thetarbrform}), we conclude that the statistical phase associated with braiding a loop
$(r,q_b,q_r)$ around another loop $(b,q_b',q_r')$ while both are linked to a red base loop is
\begin{equation}
\theta^1_{rb,r}= -\frac{\pi}{2} + \pi(q_b + q_r').
\label{thetarbr1}
\end{equation}

So far we have only discussed braiding processes involving a red base loop. We can find the braiding statistics associated with a \emph{blue} base loop,
by simply switching the roles of ``red'' and ``blue'':
\begin{eqnarray}
\theta^1_{rr,b} &=& \pi + \pi(q_r + q_r'), \nonumber \\
\theta^1_{bb,b} &=& \pi (q_b + q_b'), \nonumber \\
\theta^1_{br,b} &=& -\frac{\pi}{2} + \pi(q_r + q_b'). 
\label{bluebase1}
\end{eqnarray}
Putting together equations (\ref{thetabbr1}),(\ref{thetarrr1}),(\ref{thetarbr1}) and (\ref{bluebase1}), we have found all the three-loop braiding 
statistics of the $H_1$ model. (Actually, one quantity that we have not computed is the \emph{exchange} statistics of the loops. However, this quantity is not necessary for
our purposes, since the mutual statistics computed above is sufficient to distinguish the two models).

Before concluding this section, we would like to mention that the three-loop statistics for $H_1$ appears to be identical to the three-loop statistics of one of the  
$\mathbb{Z}_2 \times \mathbb{Z}_2$ Dijkgraaf-Witten models\cite{DijkgraafWitten,Wan3dmodel}. To see this, we follow the notation of 
Ref. \onlinecite{WangLevin,WangLevin2} and we summarize the three-loop statistics for $H_1$ in terms of the quantity $\Theta_{ij,k} \equiv 2 \theta^1_{ij,k}$ 
where the indices $i,j,k \in \{r,b\}$, and $\Theta_{ij,k}$ is defined modulo $2\pi$. Translating equations (\ref{thetabbr1}),(\ref{thetarrr1}),
(\ref{thetarbr1}) and (\ref{bluebase1}) into the $\Theta$ variables gives: 
\begin{eqnarray}
\Theta_{bb,r} &=& \Theta_{rr,r} = \Theta_{rr,b} = \Theta_{bb,b} = 0 \nonumber \\
\Theta_{rb,r} &=& \Theta_{br,b} = \pi,
\end{eqnarray}
If we compare this data to the results of Ref. \onlinecite{WangLevin}, we can see that it matches the three-loop statistics for 
the $\mathbb{Z}_2 \times \mathbb{Z}_2$ Dijkgraaf-Witten model labeled by $(p_1 = 1, p_2 = 1)$. Based on this fact, we conjecture that $H_1$ belongs to the same phase as this Dijkgraaf-Witten model. \footnote{In fact, if our conjecture is correct then $H_1$ also belongs to the same phase as the Dijkgraaf-Witten models labeled by 
$(1, 0)$ and $(0,1)$, since the three Dijkgraaf-Witten models labeled by $(1,1)$, $(1,0)$ and $(0,1)$ all belong to the same phase, when viewed as spin models rather than gauge theories.}

\subsubsection{Comparing the three-loop statistics in the two models}
With the above results, we now show that the $H_0$ and $H_1$ models have distinct three-loop statistics. To see an example of a difference between the 
two models, consider the formula for $\theta^1_{rb,r}$ (\ref{thetarbr1}). This formula implies that, in the $H_1$ model, if we braid a red loop around 
a blue loop, while both are linked to a red loop the resulting phase is $\pm \pi/2$ depending on what type of red and blue loops are being braided. 
In contrast, in the $H_0$ model, we can see from (\ref{thetabbr0}), (\ref{thetarrr0}), (\ref{thetarbr0}) that if we braid \emph{any} two loops around 
one another, the resulting phase is always $0$ or $\pi$, independent of the choice of base loop or what loops are being braided. 

For another example of a difference, consider the formula for $\theta^1_{bb,r}$ (\ref{thetabbr1}) for the case where the two blue loops 
$(b, q_b, q_r)$, $(b,q_b', q_r')$ are \emph{identical}, i.e. $q_b = q_b'$ and $q_r = q_r'$. In this case, $\theta^1_{bb,r} = \pi$. This means that if 
we braid two identical blue loops around one another while they are both linked to a red loop, the resulting statistical phase is $\pi$ in the $H_1$ 
model. On the other hand, in the $H_0$ model, we can see from (\ref{thetabbr0}), (\ref{thetarrr0}), (\ref{thetarbr0}) that if we braid \emph{any} two 
identical loops around one another the resulting phase is always $0$, independent of which loops or base loops are involved.

From the above examples, it is clear that there is no way to map the loop excitations of $H_0$ onto the loop excitations of $H_1$ in such a way that 
the corresponding loops have the same statistics. Importantly, we can rule out \emph{all} possible mappings between the two sets of loop excitations 
including those that change the ``colors'' or ``charges'' of the loop, e.g. that map the excitation $(b,1,0)$ in $H_0$ onto $(r,0,1)$ in $H_1$. 
(This generality is important because the ``colors'' and ``charges'' of the loops have no physical meaning in this context except as a scheme for 
labeling excitations). We conclude that the two models have physically distinct three-loop statistics, and hence must belong to different phases.

\section{Conclusion}
In this paper, we have presented an explicit computation of the three-loop braiding statistics of two spin models, $H_0, H_1$. The key step in our 
analysis was our construction of membrane operators that create and move loop excitations. With the help of these operators, we were able to 
implement the three-loop braiding process on the lattice and find the associated statistical phase in each of the models. While technically 
complicated, this membrane operator approach has the advantage that it is more direct than previous calculations based on dimensional 
reduction\cite{WangLevin} or modular transformations on a 3D torus.\cite{JiangMesarosRan,WangWen14} An additional feature of our results is that they provide a concrete 
demonstration of the utility of three-loop braiding statistics for distinguishing 3D gapped phases: indeed, we have shown that the two models 
$H_0$ and $H_1$ share the same particle exchange statistics and the same particle-loop and loop-loop braiding statistics. The only way to see that 
these models belong to distinct phases is to examine their three-loop braiding statistics. 

The discussion in this paper raises an important question, namely whether three-loop braiding statistics, together with particle exchange statistics 
and particle-loop braiding statistics, provides enough data to uniquely distinguish all 3D gapped phases. The existence of the 3D cubic code 
model\cite{Haah} suggests that the answer to this question may be `no' in general. Indeed, the cubic code model does not support deconfined 
particle-like and loop-like excitations like the models studied in this paper, so it is not clear whether one can define particle or loop braiding 
statistics for this system. In light of this example, a more natural question may be whether the above braiding statistics data is sufficient to 
distinguish an appropriate subset of 3D gapped phases, such as the ``gapped quantum liquids'' defined in Ref. \onlinecite{ZengWen}. As far as we know, 
this is an open question: we are not aware of any counterexamples or arguments one way or the other.

One direction for future work would be to investigate the implications of our results for symmetry protected topological (SPT) phases. Indeed, as 
discussed in the introduction, both $H_0$ and $H_1$ can be thought of as $\mathbb{Z}_2 \times \mathbb{Z}_2$ gauge theories obtained by gauging the 
$\mathbb{Z}_2 \times \mathbb{Z}_2$ symmetry of two different spin models. The $H_1$ model comes from a spin model\cite{ChenLuVishwanath} belonging to a nontrivial 
SPT phase, while $H_0$ comes from a spin model in a trivial SPT phase. Following the same approach as in Ref. \onlinecite{LevinGu}, it should be 
possible to derive bulk and surface properties of the two SPT phases from the braiding statistics in the associated gauge theories $H_0, H_1$. Results
of this kind will be discussed in a separate publication. \cite{LinLevinprep}

Another direction would be to construct exactly soluble lattice models that can realize more general types of 3D gapped phases. For example, since the 
two models $H_0, H_1$ are built out of two species of intersecting membranes, it may be possible to build more general models by considering more 
species of membranes or by allowing membranes to \emph{branch} --- in analogy with the string-net models
of Ref. \onlinecite{LevinWenstrnet}. Given our suspicion that $H_0, H_1$ belong to the same phase as $\mathbb{Z}_2 \times \mathbb{Z}_2$ 
Dijkgraaf-Witten models, a key question is whether such a construction can realize any phases that cannot be realized by previously known exactly 
soluble models, such as the 3D Dijkgraaf-Witten models\cite{DijkgraafWitten,Wan3dmodel} or 3D string-net models\cite{LevinWenstrnet}.

\section*{Acknowledgements}
We thank Chenjie Wang for useful discussions. This work is supported by the Alfred P. Sloan foundation and NSF DMR-1254741.

\appendix

\section{Proving the identities (\ref{commutingbc},\ref{psi0prop},\ref{psi1prop}) \label{app:show}}
In this section, we prove a few identities involving the operators $A_l, A_{\hat{l}}, B_c^0, B_{\hat{c}}^0, B_c^1, B_{\hat{c}}^1$ and the wave functions 
$|\Psi_0\>$ and $|\Psi_1\>$. We used these identities when we solved the two models, $H_0$ and $H_1$. The first set of identities state that $|\Psi_0\>$ and
$|\Psi_1\>$ are eigenstates of the $A_l, A_{\hat{l}}$ operators:
\begin{align}
	A_l|\Psi_0\> &=A_{\hat{l}}|\Psi_0\>=|\Psi_0\>, \nonumber \\
	A_l|\Psi_1\> & =A_{\hat{l}}|\Psi_1\>=|\Psi_1\>. \label{Qidapp}
\end{align}
Similarly, the second set of identities state that $|\Psi_0\>$ is an eigenstate of $B_c^0, B_{\hat{c}}^0$ and $|\Psi_1\>$ is an eigenstate of $B_c^1, B_{\hat{c}}^1$:
\begin{align}
	B_c^0|\Psi_0\>&=B_{\hat{c}}^0|\Psi_0\>=|\Psi_0\>,  \label{Bidapp1} \\
	B_c^1|\Psi_1\>&=B_{\hat{c}}^1|\Psi_1\>=|\Psi_1\>. \label{Bidapp2}
\end{align}
The last identity states that the $B_c^1, B_{\hat{c}}^1$ operators commute with one another:
\begin{align}
	[B_c^1,B_{c'}^1]&=[B_{\hat{c}}^1,B_{\hat{c}'}^1]= [B_c^1,B_{\hat{c}}^1] = 0. \label{Bcommidapp}
\end{align}

We begin by deriving Eqs. (\ref{Qidapp}). These relations are obvious since $|\Psi_0\>$ and $|\Psi_1\>$ are linear superpositions of closed membrane 
configurations $|X_b, X_r\>$, which by definition obey $A_l |X_b, X_r\> = A_{\hat{l}} |X_b, X_r\> = |X_b, X_r\>$. The relation (\ref{Bidapp1}) is also easy to 
prove: first we note that
\begin{eqnarray}
\mathcal{B}_c^0|\Psi_0\> &=& \sum_{\text{closed } X_b, X_r} \mathcal{B}_c^0|X_b, X_r\> \nonumber \\
&=& \sum_{\text{closed } X_b, X_r} |X_b + c, X_r\> \nonumber \\
&=& \sum_{\text{closed } X_b', X_r} |X_b', X_r\> \nonumber \\
&=& |\Psi_0\>
\end{eqnarray}
where we made the change variables $X_b' = X_b + c$ in the third line. Then using the fact that $B_c^0 = \frac{1}{2}(1+ \mathcal{B}_c^0)$, we 
immediately derive the required relation $B_c^0 |\Psi_0\> = |\Psi_0\>$. The same argument works for $B_{\hat{c}}^0$.

\begin{figure}[ptb]
\begin{center}
\includegraphics[
height=1in,
width=2.5in
]{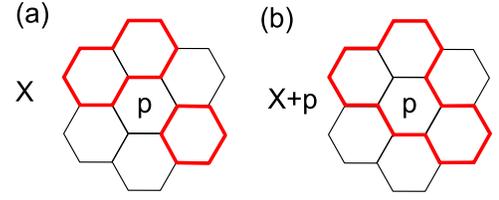}
\end{center}
\caption{An example of the 2D identity (\ref{2dloopid}) for the case of loops living on the honeycomb lattice. Panel (a) shows a loop configuration $X$, while panel (b) shows the corresponding configuration $X+p$. In this example, $N_{\text{loop}}(X)=2$ and $N_{\text{loop}}(X+p) = 1$. Also $n_p(X) = 4$, since four of the six legs adjacent to the plaquette $p$ are occupied by strings. The identity (\ref{2dloopid}) holds for this example, since $(-1)^{1-2} = (-1) \cdot i^4$.}
\label{fig:loopid}
\end{figure}

The relation (\ref{Bidapp2}) requires a little more work. First, we note that
\begin{eqnarray}
\mathcal{B}_c^1 |\Psi_1\> &=& \sum_{\text{closed } X_b, X_r} (-1)^{N_g(X_b,X_r)} \mathcal{B}_c^1 |X_b, X_r\>  \label{bcapp} \\
&=& \sum_{\text{closed } X_b, X_r} (-1)^{N_g(X_b,X_r)} (-1)^{m_c}i^{n_c} |X_b+c, X_r\> \nonumber 
\end{eqnarray}
where $N_g$ is the number of green loops in the configuration $X_b, X_r$ and $m_c,n_c$ are integer valued functions of $X_b, X_r$ which 
are defined in section \ref{section:h1}. To proceed further, we use the following identity:
\begin{equation}
(-1)^{N_g(X_b+c,X_r) - N_g(X_b, X_r)} = (-1)^{m_c(X_b,X_r)} i^{n_c(X_b, X_r)}.
\label{3dloopid}
\end{equation}
Here, we explicitly show the dependence of $m_c$ and $n_c$ on $X_b, X_r$, for clarity. Equation (\ref{3dloopid}) says that when we add a blue 
cube, i.e. $X_b \rightarrow X_b +c$, the resulting change in the parity of the number of green loops $N_g$ can be computed in terms of $m_c$ 
and $n_c$. This relation is nontrivial because $(-1)^{N_g}$ depends on the \emph{global} properties of the membrane configuration $X_b, X_r$, 
while $m_c$ and $n_c$ only depend on \emph{local} properties of $X_b, X_r$ in the neighborhood of the cube $c$. 

To derive the above identity (\ref{3dloopid}), we first recall an analogous, but simpler identity involving 2D loop models. The 2D identity 
applies to 2D loop configurations $X$ that live on the links of a 2D planar trivalent lattice. It states that if we add a loop around the 
boundary of a single plaquette $p$ to form a new loop configuration $X+p$, then 
\begin{equation}
(-1)^{N_{\text{loop}}(X+p) - N_{\text{loop}}(X)} = (-1) \cdot i^{n_p(X)}
\label{2dloopid}
\end{equation}
where $N_{\text{loop}}(X)$ and $N_{\text{loop}}(X+p)$ are the number of loops in the configurations $X$ and $X+p$, and where $n_p(X)$ is the 
number of occupied legs of the plaquette $p$. (See Fig. \ref{fig:loopid} for an example). The 2D identity (\ref{2dloopid}) is closely related 
to the 2D exactly soluble double semion model of Ref. \onlinecite{LevinWenstrnet}, and can be established using the local rules that define 
that model. 

To derive the 3D identity (\ref{3dloopid}) from its 2D cousin (\ref{2dloopid}), it is helpful to think about the configuration of green loops in 
$X_b, X_r$, for a fixed red 
membrane configuration $X_r$. These green loops lie on the surfaces of the red membranes $X_r$. When we add a blue cube, this effectively adds 
a collection of green loops along all the red membranes that intersect the cube. Adding these green loops is similar to adding the loop $p$ in 
the 2D identity (\ref{2dloopid}), so when we add these loops, the change in the parity of the number of green loops can be calculated by 
multiplying the factors on the right hand side of (\ref{2dloopid}) over all the additional green loops. Taking the product of these factors gives 
the 3D identity (\ref{3dloopid}).

Now that we have established the identity (\ref{3dloopid}), we can complete our derivation of (\ref{Bidapp2}): substituting (\ref{3dloopid}) into (\ref{bcapp}), we obtain
\begin{eqnarray}
\mathcal{B}_c^1 |\Psi_1\> &=&  \sum_{\text{closed } X_b, X_r} (-1)^{N_g(X_b+c,X_r)} |X_b+c, X_r\>  \nonumber \\
&=&  \sum_{\text{closed } X_b', X_r} (-1)^{N_g(X_b',X_r)} |X_b', X_r\> \nonumber \\
&=& |\Psi_1\>
\end{eqnarray}
where we made the change of variables $X_b' = X_b + c$ in the second line. Finally, using the fact that $B_c^1 = \frac{1}{2}(1+ \mathcal{B}_c^1) P_c$ 
and $P_c |\Psi_1\> = |\Psi_1\>$, we derive the required relation $B_c^1 |\Psi_1\> = |\Psi_1\>$. The same argument works for $B_{\hat{c}}^1$.

All that remains is the third relation (\ref{Bcommidapp}). In principle, one could establish (\ref{Bcommidapp}) by writing out the explicit form of 
$B_c^1,B_{\hat{c}}^1$ and calculating their commutators. However such a calculation would be tedious and not particularly illuminating. Therefore we 
use a different approach to show (\ref{Bcommidapp}). 
We make use of two properties of the operators $\mathcal{B}_c^1$. The first property is the relation
\begin{equation}
\mathcal{B}_c^1 |\Psi_1\> = |\Psi_1\>
\label{prop1}
\end{equation}
which we just established above. The second property is that, for any closed membrane configuration $(X_b, X_r)$,
\begin{equation}
\mathcal{B}_c^1 |X_b, X_r\> = e^{i \theta(X_b,X_r,c)} |X_b+c,X_r\>
\label{prop2}
\end{equation}
where $e^{i \theta(X_b,X_r,c)}$ is a phase factor that depends on $X_b, X_r, c$. The latter property follows immediately from
the definition of $\mathcal{B}_c^1$.

Using the above two properties (\ref{prop1}) and (\ref{prop2}), we will now show that the two operators $\mathcal{B}_c^1, \mathcal{B}_{c'}^1$ commute with one another when acting on closed membrane states. The argument is simple. From property (\ref{prop2}), we compute
\begin{align}
&\mathcal{B}_{c'}^1\mathcal{B}_c^1 |X_b, X_r\> = \mathcal{B}_{c'}^1 \cdot e^{i \theta(X_b,X_r,c)} |X_b+c,X_r\>  \label{border1} \\
&= e^{i \theta(X_b+c,X_r,c')} \cdot e^{i \theta(X_b,X_r,c)} |X_b+c+c',X_r\> \nonumber 
\end{align}
On the other hand, if we reverse the order we find
\begin{eqnarray}
\mathcal{B}_{c}^1\mathcal{B}_{c'}^1 |X_b, X_r\> &=& \nonumber \\
 e^{i \theta(X_b+c',X_r,c)} &\cdot& e^{i \theta(X_b,X_r,c')} |X_b+c+c',X_r\>
\label{border2}
\end{eqnarray}
We need to show that the expressions on the right hand sides of (\ref{border1}) and (\ref{border2}) are equal. To this end, we note that
property (\ref{prop1}) implies that
\begin{equation}
\<X_b, X_r| \mathcal{B}_c^1| \Psi_1\> = \<X_b, X_r | \Psi_1\>.
\end{equation}
Then, using property (\ref{prop2}), we can write this as
\begin{equation}
e^{-i \theta(X_b,X_r,c)}\<X_b+c, X_r| \Psi_1\> = \<X_b, X_r | \Psi_1\>
\end{equation}
so that
\begin{equation}
e^{i\theta(X_b,X_r,c)} = \frac{\<X_b+c, X_r| \Psi_1\>}{\<X_b, X_r | \Psi_1\>}.
\label{thetaform}
\end{equation}
We now substitute (\ref{thetaform}) into equation (\ref{border1}). The result is
\begin{equation*}
\mathcal{B}_{c'}^1\mathcal{B}_c^1 |X_b, X_r\> = |X_b+c+c',X_r\> \cdot \frac{\<X_b+c+c', X_r| \Psi_1\>}{\<X_b, X_r | \Psi_1\>}.
\end{equation*}
Likewise, if we substitute (\ref{thetaform}) into equation (\ref{border2}), we obtain the same expression:
\begin{equation*}
\mathcal{B}_{c}^1\mathcal{B}_{c'}^1 |X_b, X_r\> = |X_b+c+c',X_r\> \cdot \frac{\<X_b+c+c', X_r| \Psi_1\>}{\<X_b, X_r | \Psi_1\>},
\end{equation*}
implying that $\mathcal{B}_c^1$ and $\mathcal{B}_{c'}^1$ commute when acting on closed membrane states.

Now that we know that $\mathcal{B}_c^1$ and $\mathcal{B}_{c'}^1$ commute when acting on closed membrane states, we can immediately deduce that the 
combinations $\mathcal{B}_c^1 P_c $ and $\mathcal{B}_{c'}^1 P_{c'}$ commute when acting on \emph{arbitrary} membrane states, since the operators 
$P_c$ and $P_{c'}$ project onto states that are closed in the neighborhood of $c$ and $c'$. It then follows that the operators $B_c^1$ and 
$B_{c'}^1$ commute with one another, since $B_c^1$ is a linear combination of $\mathcal{B}_c^1 P_c$
and $P_c$. 

We have shown that $[B_c^1, B_{c'}^1] = 0$ for any cubes $c,c'$ in the cubic lattice. This establishes the first identity in Eq. (\ref{Bcommidapp}). 
The other two identities in
Eq. (\ref{Bcommidapp}) can be proven in exactly the same way.

\section{Showing that the cylinder operators create braiding eigenstates \label{simpleop}}
In this section we show that the blue cylinder operator $M_b^1(S)$ creates loop excitations which are eigenstates of braiding measurements. Our argument consists of two parts. In the first part, we show that as long as $f_b$ obeys condition (\ref{piecewise}), then the loop excitations are guaranteed to be eigenstates of braiding measurements. In the second part, we check that $f_b$ does in fact obey (\ref{piecewise}). The second part can also be thought of as a derivation of Table \ref{fig:config1} and Table \ref{fig:config2}, since we will see that the values in those tables are completely fixed by the requirement that $f_b$ obey equation (\ref{piecewise}).

\subsection{Connection between equation (\ref{piecewise}) and braiding eigenstates}
In this section we show that if $f_b$ obeys condition (\ref{piecewise}) then the blue cylinder operator is guaranteed to create loop excitations which are braiding eigenstates. We reprint 
equation (\ref{piecewise}) below for convenience:
\begin{equation}
	f_b(X_b,X_r,S_1 \cup S_2)= f_b(X_b,X_r,S_1) \cdot f_b(X_b, X_r, S_2).
	\label{piecewiseapp}
\end{equation}
Here $S_1$ and $S_2$ can be any two cylinders that share a common boundary while $(X_b, X_r)$ can be any membrane configuration whose intersection with the common boundary obeys the relevant cylinder operator boundary condition.

To begin, it is helpful to consider an example. Let $|\Psi_{ex}\>$ be an excited state with a single red loop. We will call this red loop the ``base loop.'' Let $M_b^1(S)$ be a blue cylinder operator that is linked with the base loop. Consider the state $M_b^1(S) |\Psi_{ex}\>$. This state contains two blue loops at the two ends of the cylinder $S$, both of which are linked with the red base loop. Now, what we want to show is that these blue loops are eigenstates of braiding measurements. That is, we want to show that if we braid any other loop around one of these loops, the system returns to its original state multiplied by a complex number (in fact, a phase factor). We now translate this claim into a mathematical equation.

Consider, for example, braiding another blue loop around one of these loops. This braiding process can be implemented by applying a blue torus operator $M_b^1(S')$ where the torus $S'$ encircles the loop that we wish to braid around (see Fig. \ref{braiding} for a similar geometry). The fact that the blue loop is an eigenstate of this braiding process is equivalent to the equation
\begin{equation}
M_b^1(S') \cdot (M_b^1(S) |\Psi_{ex}\>) = \text{(const.)} \cdot M_b^1(S) |\Psi_{ex}\>.
\label{braideigen}
\end{equation}

Our task is to show that if $f_b$ obeys (\ref{piecewiseapp}) then the operators $M_b^1$ obey equation (\ref{braideigen}). To establish this result, we first rewrite (\ref{braideigen}) in a more convenient form. We begin by recalling that since $S'$ is a torus operator, it does not create any excitations --- in other words, $M_b^1(S') |\Psi_{ex}\> \propto |\Psi_{ex}\>$. Using this fact, the above equation can be written as
\begin{equation*}
M_b^1(S') M_b^1(S) |\Psi_{ex}\> = \text{(const.)} \cdot M_b^1(S) M_b^1(S') |\Psi_{ex}\>.
\end{equation*}
Next, multiplying both sides by $\<X_b, X_r|$ gives
\begin{align*}
\<X_b, X_r| M_b^1(S') & M_b^1(S) |\Psi_{ex}\> = \\
\text{(const.)} &\cdot \<X_b, X_r| M_b^1(S) M_b^1(S') |\Psi_{ex}\>.
\end{align*}
Using the definition of $M_b^1$, we rewrite the matrix elements on the left and right hand sides as:
\begin{align*}
\<X_b, X_r| &M_b^1(S') M_b^1(S) |\Psi_{ex}\> = f_b^*(X_b,X_r,S') \\
& \cdot f_b^*(X_b+S',X_r,S) \cdot \Psi_{ex}(X_b+S+S',X_r),  \nonumber \\
\<X_b, X_r| &M_b^1(S) M_b^1(S') |\Psi_{ex}\> = f_b^*(X_b,X_r,S) \\
& \cdot f_b^*(X_b+S,X_r,S') \cdot \Psi_{ex}(X_b+S+S',X_r).  \nonumber
\end{align*}
Therefore, what we need to prove is
\begin{align*}
f_b(X_b,X_r,S') &\cdot f_b(X_b+S',X_r,S) = \nonumber \\
\text{(const.)} &\cdot f_b(X_b,X_r,S) \cdot f_b(X_b+S,X_r,S') 
\end{align*}
or equivalently
\begin{equation}
\frac{f_b(X_b+S',X_r,S)}{f_b(X_b,X_r,S)} \cdot \frac{f_b(X_b,X_r,S')}{f_b(X_b+S,X_r,S')} = \text{(const.)}.
\label{braideigen2}
\end{equation}

We now prove (\ref{braideigen2}). Focusing on the first ratio in (\ref{braideigen2}), we note that the numerator and denominator can be represented graphically as
\begin{eqnarray*}
f_b(X_b+S',X_r,S) &=& f_b\left(\raisebox{-0.115in}{\includegraphics[height=0.3in]{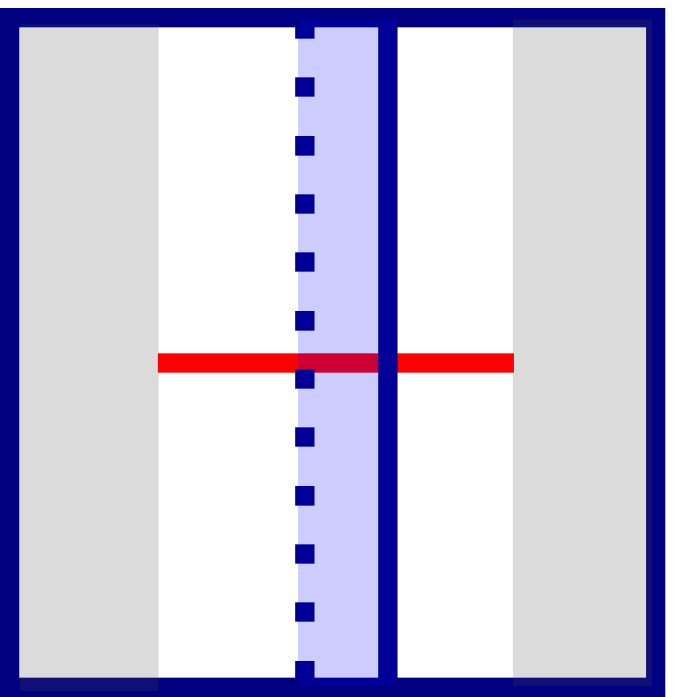}}\right), \nonumber \\
f_b(X_b,X_r,S) &=& f_b\left(\raisebox{-0.115in}{\includegraphics[height=0.3in]{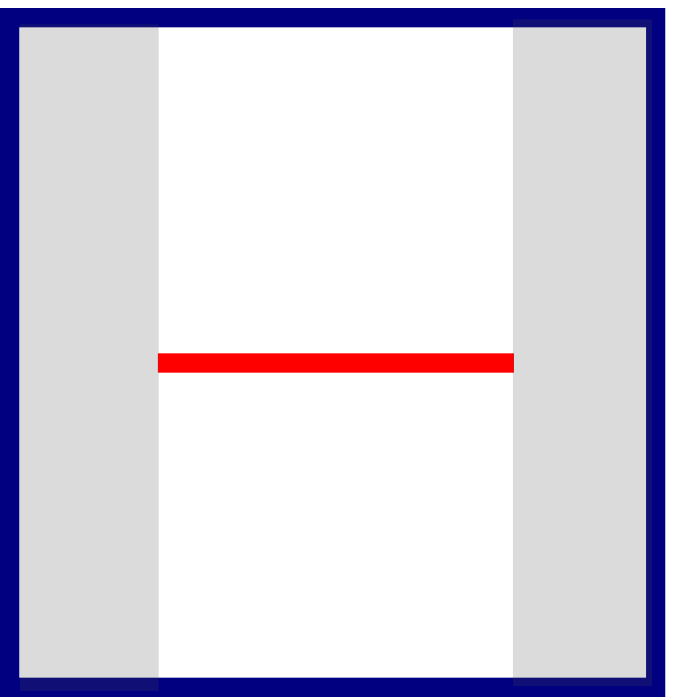}}\right).
\end{eqnarray*}
Here the thin vertical blue region corresponds to the intersection $S' \cap S$, while the gray areas on the left and on the right are meant to denote some arbitrary pictures corresponding to $X_b \cap S$ and $X_r \cap S$. (We can assume without loss of generality that the pictures are of the above simple form, since we can use the path independence property, discussed below Eqs. (\ref{local}), to deform $S$ and $S'$ as we wish). Next we use equation (\ref{piecewiseapp}) twice to break up the cylinder $S$ into three smaller cylinders, giving us
\begin{eqnarray*}
f_b\left(\raisebox{-0.115in}{\includegraphics[height=0.3in]{exfig1.eps}}\right) &=& f_b\left(\raisebox{-0.115in}{\includegraphics[height=0.3in]{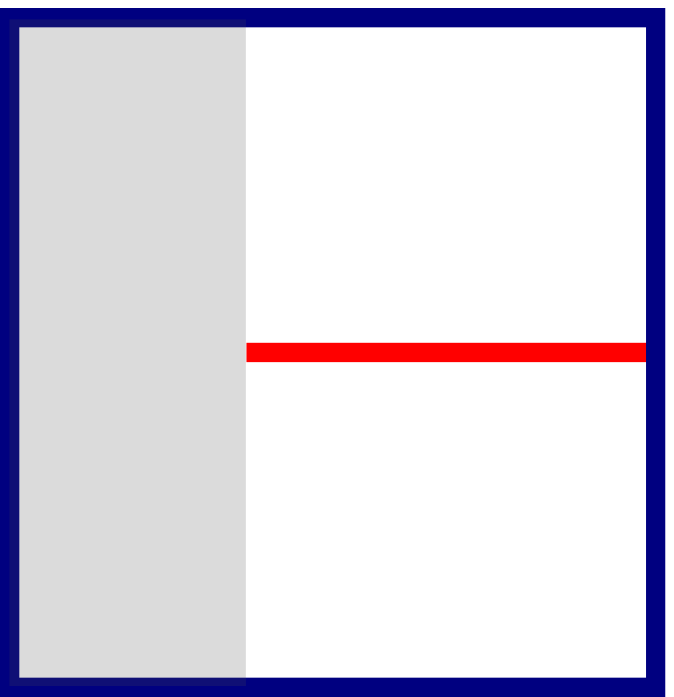}}\right) \cdot f_b\left(\raisebox{-0.115in}{\includegraphics[height=0.3in]{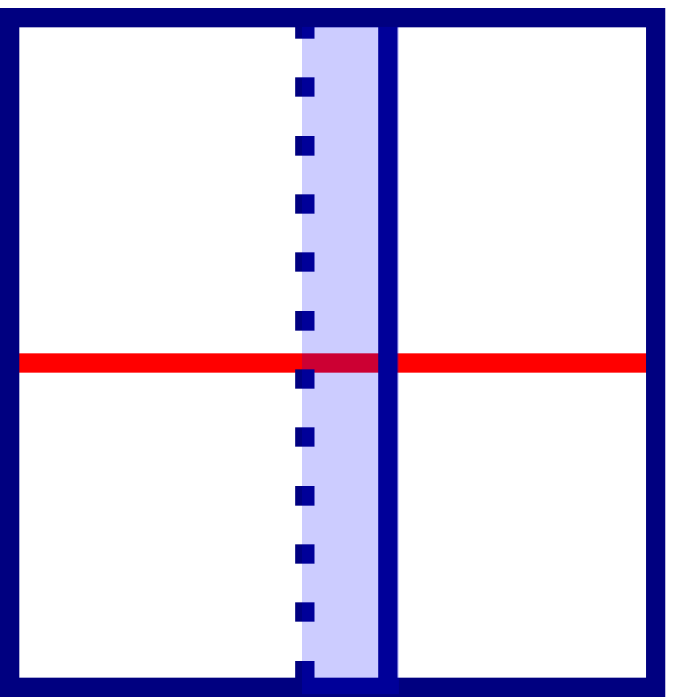}}\right) \cdot f_b\left(\raisebox{-0.115in}{\includegraphics[height=0.3in]{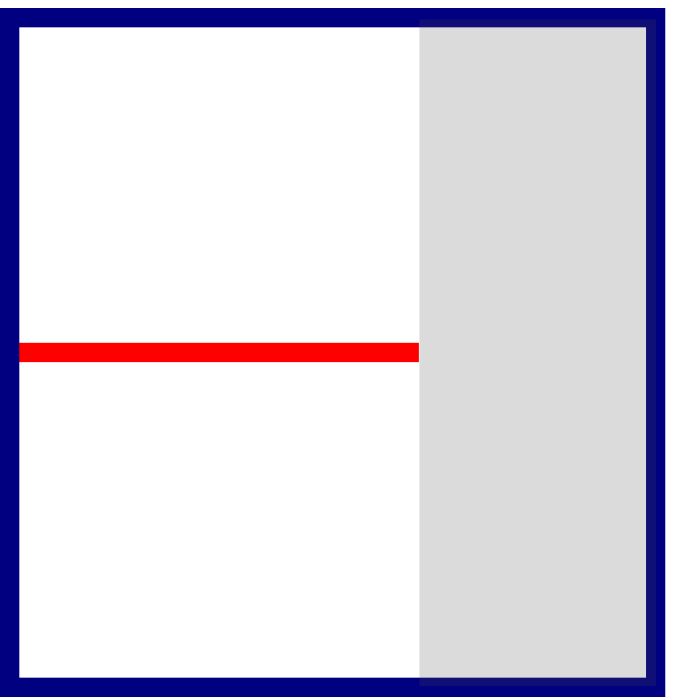}}\right), \nonumber \\
f_b\left(\raisebox{-0.115in}{\includegraphics[height=0.3in]{exfig2.eps}}\right) &=& f_b\left(\raisebox{-0.115in}{\includegraphics[height=0.3in]{exfig1a.eps}}\right) \cdot f_b\left(\raisebox{-0.115in}{\includegraphics[height=0.3in]{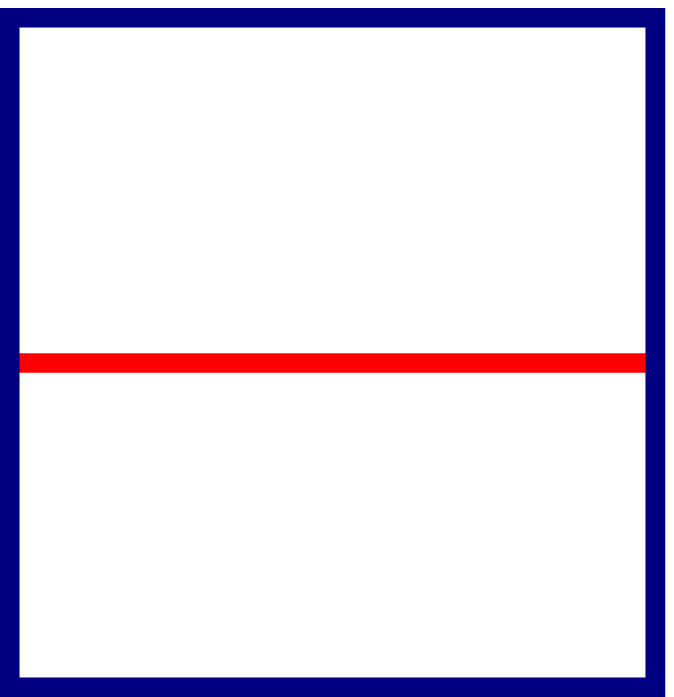}}\right) \cdot f_b\left(\raisebox{-0.115in}{\includegraphics[height=0.3in]{exfig1c.eps}}\right). 
\end{eqnarray*}
Taking the ratio of these expressions gives
\begin{equation*}
\frac{f_b(X_b+S',X_r,S)}{f_b(X_b,X_r,S)} = \frac{f_b\left(\raisebox{-0.115in}{\includegraphics[height=0.3in]{exfig1b.eps}}\right)}{f_b\left(\raisebox{-0.115in}{\includegraphics[height=0.3in]{exfig2b.eps}}\right)}.
\end{equation*}
In particular, we see that $\frac{f_b(X_b+S',X_r,S)}{f_b(X_b,X_r,S)}$ is \emph{independent} of $X_b,X_r$. In exactly the same way, one can show that $\frac{f_b(X_b,X_r,S')}{f_b(X_b+S,X_r,S')}$ is independent of $X_b, X_r$. This proves equation (\ref{braideigen2}): both terms on the left hand side are independent of $X_b, X_r$, so their product must be independent of $X_b, X_r$ as well.

So far, we have shown that (\ref{piecewiseapp}) implies the braiding eigenstate property in \emph{one} case. The case we considered was a blue loop linked to a red base loop. We showed that the blue loop is a braiding eigenstate with respect to braiding another blue loop around it. To prove the general claim, we need to establish the same braiding eigenstate property for all other cases, that is all other combinations of red and blue loops. We will not discuss the other cases here, but it should be clear that the above argument is not specific to the case considered above and applies equally well to the other cases.

\subsection{Checking that \texorpdfstring{$f_b$}{fb} obeys equation (\ref{piecewise})}

In this section, we check that $f_b$ obeys equation (\ref{piecewise}) --- or equivalently (\ref{piecewiseapp}). To begin, we note that it suffices to check (\ref{piecewiseapp}) for the case where the intersections $X_b \cap S_i$ and $X_r \cap S_i$ look like one of the four basic pictures shown in Table \ref{fig:config1} and Table \ref{fig:config2}. The reason that we only need to consider this case is that the constraint equations (\ref{local}) that define $f_b$ are \emph{local}, so if (\ref{piecewiseapp}) holds for basic pictures, then it automatically holds for general $X_b, X_r$. 

Let's start with the simplest case: blue cylinder operators that create \emph{unlinked} loops. In this case, the four basic pictures are those shown in Table \ref{fig:config1}. Specializing to these pictures, equation (\ref{piecewiseapp}) reduces to $4^2 = 16$ different relations that we need to check. Many of these equations are redundant, so we only write the four independent ones below:
\begin{align*}
	& f_b \left(\raisebox{-0.115in}{\includegraphics[height=0.3in]{p1a.eps}} \right) \cdot f_b \left(\raisebox{-0.115in}{\includegraphics[height=0.3in]{p1a.eps}} \right) 
	= f_b \left(\raisebox{-0.115in}{\includegraphics[height=0.3in]{p1a.eps}} \right), \\
	& f_b \left(\raisebox{-0.115in}{\includegraphics[height=0.3in]{p1b.eps}} \right) \cdot f_b \left(\raisebox{-0.115in}{\includegraphics[height=0.3in]{p1b.eps}} \right) 
	= f_b \left(\raisebox{-0.115in}{\includegraphics[height=0.3in]{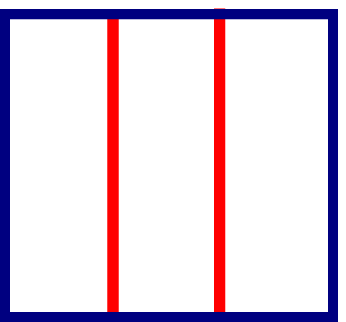}} \right),  \\
	& f_b \left(\raisebox{-0.115in}{\includegraphics[height=0.3in]{p1c.eps}} \right) \cdot f_b \left(\raisebox{-0.115in}{\includegraphics[height=0.3in]{p1c.eps}} \right) 
	= f_b \left(\raisebox{-0.115in}{\includegraphics[height=0.3in]{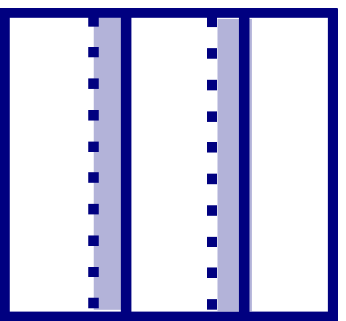}} \right),  \\ 
	& f_b \left(\raisebox{-0.115in}{\includegraphics[height=0.3in]{p1b.eps}} \right) \cdot f_b \left(\raisebox{-0.115in}{\includegraphics[height=0.3in]{p1c.eps}} \right) 
	= f_b \left(\raisebox{-0.115in}{\includegraphics[height=0.3in]{p1d.eps}} \right).
\end{align*}
Next, we use the constraint equations (\ref{local}) to reduce the pictures in the second and third lines to basic pictures:
\begin{align*}
f_b \left(\raisebox{-0.115in}{\includegraphics[height=0.3in]{2p1b.eps}} \right) &= f_b \left(\raisebox{-0.115in}{\includegraphics[height=0.3in]{p1a.eps}} \right), \\
f_b \left(\raisebox{-0.115in}{\includegraphics[height=0.3in]{2p1c.eps}} \right) &= f_b \left(\raisebox{-0.115in}{\includegraphics[height=0.3in]{p1a.eps}} \right).
\end{align*}
Putting this together, we have
\begin{align*}
	& f_b \left(\raisebox{-0.115in}{\includegraphics[height=0.3in]{p1a.eps}} \right) \cdot f_b \left(\raisebox{-0.115in}{\includegraphics[height=0.3in]{p1a.eps}} \right) 
	= f_b \left(\raisebox{-0.115in}{\includegraphics[height=0.3in]{p1a.eps}} \right), \nonumber \\
	& f_b \left(\raisebox{-0.115in}{\includegraphics[height=0.3in]{p1b.eps}} \right) \cdot f_b \left(\raisebox{-0.115in}{\includegraphics[height=0.3in]{p1b.eps}} \right) 
	= f_b \left(\raisebox{-0.115in}{\includegraphics[height=0.3in]{p1a.eps}} \right),  \nonumber \\
	& f_b \left(\raisebox{-0.115in}{\includegraphics[height=0.3in]{p1c.eps}} \right) \cdot f_b \left(\raisebox{-0.115in}{\includegraphics[height=0.3in]{p1c.eps}} \right) 
	= f_b \left(\raisebox{-0.115in}{\includegraphics[height=0.3in]{p1a.eps}} \right),  \nonumber \\ 
	& f_b \left(\raisebox{-0.115in}{\includegraphics[height=0.3in]{p1b.eps}} \right) \cdot f_b \left(\raisebox{-0.115in}{\includegraphics[height=0.3in]{p1c.eps}} \right) 
	= f_b \left(\raisebox{-0.115in}{\includegraphics[height=0.3in]{p1d.eps}} \right).
\end{align*}
Letting
\begin{align*}
	f_b \left(\raisebox{-0.115in}{\includegraphics[height=0.3in]{p1a.eps}} \right) = A, \quad
	f_b \left(\raisebox{-0.115in}{\includegraphics[height=0.3in]{p1b.eps}} \right) = B,\\
	f_b \left(\raisebox{-0.115in}{\includegraphics[height=0.3in]{p1c.eps}} \right) = C, \quad
	f_b \left(\raisebox{-0.115in}{\includegraphics[height=0.3in]{p1d.eps}} \right) = D ,
\end{align*}
we can rewrite these equations as
\begin{align}
	A^2 = A, \nonumber \\ 
	B^2 = A, \nonumber \\
	C^2 = A, \nonumber\\
	B \cdot C = D . \label{set1}
\end{align}
Now let us compare with the values for $A,B,C,D$ given in Table \ref{fig:config1}:
\begin{equation}
	A = 1, \quad B= - e^{i \pi q_r},\quad C=e^{i \pi q_b},\quad D=- e^{i \pi (q_r+q_b)}
	\label{set1sol}
\end{equation}
where $q_b, q_r = 0$ or $1$. First, we can see that the above expressions (\ref{set1sol}) do in fact obey equations (\ref{set1}). Thus, we have successfully verified (\ref{piecewiseapp}) for the case of an unlinked blue cylinder operator. In fact, we can see that (\ref{set1sol}) is actually the most general possible solution to the equations (\ref{set1}). Thus, the above calculation can also be regarded as a derivation of Table \ref{fig:config1}.

Next, we consider the case of a blue cylinder operator linked with a red base loop. We proceed in exactly the same way as in the unlinked case. First, we note that equation (\ref{piecewiseapp}) reduces to $4^2 = 16$ different relations, of which $4$ are independent:
\begin{align*}
	& f_b \left(\raisebox{-0.115in}{\includegraphics[height=0.3in]{p2a.eps}} \right) \cdot f_b \left(\raisebox{-0.115in}{\includegraphics[height=0.3in]{p2a.eps}} \right) 
	= f_b \left(\raisebox{-0.115in}{\includegraphics[height=0.3in]{p2a.eps}} \right), \\
	& f_b \left(\raisebox{-0.115in}{\includegraphics[height=0.3in]{p2b.eps}} \right) \cdot f_b \left(\raisebox{-0.115in}{\includegraphics[height=0.3in]{p2b.eps}} \right) 
	= f_b \left(\raisebox{-0.115in}{\includegraphics[height=0.3in]{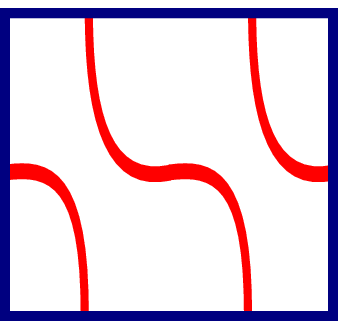}} \right),  \\
	& f_b \left(\raisebox{-0.115in}{\includegraphics[height=0.3in]{p2c.eps}} \right) \cdot f_b \left(\raisebox{-0.115in}{\includegraphics[height=0.3in]{p2c.eps}} \right) 
	= f_b \left(\raisebox{-0.115in}{\includegraphics[height=0.3in]{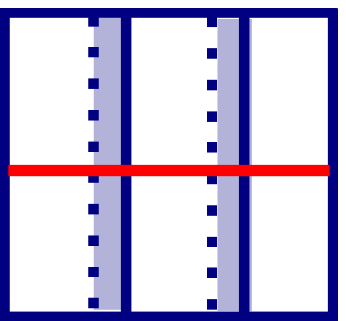}} \right),  \\ 
	& f_b \left(\raisebox{-0.115in}{\includegraphics[height=0.3in]{p2b.eps}} \right) \cdot f_b \left(\raisebox{-0.115in}{\includegraphics[height=0.3in]{p2c.eps}} \right) 
	= f_b \left(\raisebox{-0.115in}{\includegraphics[height=0.3in]{p2d.eps}} \right).
\end{align*}
Next, we use the constraint equations (\ref{local}) to derive:
\begin{align*}
	f_b \left(\raisebox{-0.115in}{\includegraphics[height=0.3in]{2p2b.eps}} \right) &= -f_b \left(\raisebox{-0.115in}{\includegraphics[height=0.3in]{p2a.eps}} \right),\\
	f_b \left(\raisebox{-0.115in}{\includegraphics[height=0.3in]{2p2c.eps}} \right) &= -f_b \left(\raisebox{-0.115in}{\includegraphics[height=0.3in]{p2a.eps}} \right).
\end{align*}
Putting this together gives
\begin{align*}
	& f_b \left(\raisebox{-0.115in}{\includegraphics[height=0.3in]{p2a.eps}} \right) \cdot f_b \left(\raisebox{-0.115in}{\includegraphics[height=0.3in]{p2a.eps}} \right) 
	= f_b \left(\raisebox{-0.115in}{\includegraphics[height=0.3in]{p2a.eps}} \right), \\
	& f_b \left(\raisebox{-0.115in}{\includegraphics[height=0.3in]{p2b.eps}} \right) \cdot f_b \left(\raisebox{-0.115in}{\includegraphics[height=0.3in]{p2b.eps}} \right) 
	= -f_b \left(\raisebox{-0.115in}{\includegraphics[height=0.3in]{p2a.eps}} \right),  \\
	& f_b \left(\raisebox{-0.115in}{\includegraphics[height=0.3in]{p2c.eps}} \right) \cdot f_b \left(\raisebox{-0.115in}{\includegraphics[height=0.3in]{p2c.eps}} \right) 
	= -f_b \left(\raisebox{-0.115in}{\includegraphics[height=0.3in]{p2a.eps}} \right),  \\ 
	& f_b \left(\raisebox{-0.115in}{\includegraphics[height=0.3in]{p2b.eps}} \right) \cdot f_b \left(\raisebox{-0.115in}{\includegraphics[height=0.3in]{p2c.eps}} \right) 
	= f_b \left(\raisebox{-0.115in}{\includegraphics[height=0.3in]{p2d.eps}} \right).
\end{align*}
Letting
\begin{align*}
	f_b \left(\raisebox{-0.115in}{\includegraphics[height=0.3in]{p2a.eps}} \right) = A, \quad
	f_b \left(\raisebox{-0.115in}{\includegraphics[height=0.3in]{p2b.eps}} \right) = B, \\
	f_b \left(\raisebox{-0.115in}{\includegraphics[height=0.3in]{p2c.eps}} \right)  = C, \quad
	f_b \left(\raisebox{-0.115in}{\includegraphics[height=0.3in]{p2d.eps}} \right) = D,
\end{align*}
we arrive at the following algebraic equations:
\begin{align}
	A^2=A, \nonumber \\
	B^2=-A, \nonumber \\
	C^2=-A, \nonumber \\
	B\cdot C=D.
	\label{set2}
\end{align}
Let us compare with the values of $A,B,C,D$ given in the top panel of Table \ref{fig:config2}:
\begin{equation}
	A = 1, \quad B=i \cdot e^{i \pi q_r},\quad C=i \cdot e^{i \pi q_b},\quad D=- e^{i \pi (q_r+q_b)}.
	\label{set2sol}
\end{equation}
Again, we can see that the above expressions (\ref{set2sol}) obey (\ref{set2}). Thus, we have proven equation (\ref{piecewiseapp}) for the case of a blue cylinder operator linked to a red base loop.
We can also see that the above expressions are the most general solutions to these equations. Thus, our calculation can also be thought of as a derivation of the top panel of Table \ref{fig:config2}.

Finally, we consider the case of a blue cylinder operator linked with a blue base loop. Proceeding in the same away, we set
\begin{align*}
	f_b \left(\raisebox{-0.115in}{\includegraphics[height=0.3in]{p3a.eps}} \right) = A, \quad
	f_b \left(\raisebox{-0.115in}{\includegraphics[height=0.3in]{p3b.eps}} \right) = B, \\
	f_b \left(\raisebox{-0.115in}{\includegraphics[height=0.3in]{p3c.eps}} \right) = C, \quad
	f_b \left(\raisebox{-0.115in}{\includegraphics[height=0.3in]{p3d.eps}} \right) = D.
\end{align*}
and we derive the algebraic equations
\begin{align}
	A^2 = A, \nonumber \\ 
	B^2 = A, \nonumber \\
	C^2 = A, \nonumber\\
	B \cdot C = D. \label{set3}
\end{align}
Again, it is easy to check that the values of $A,B,C,D$ given in the bottom panel of Table \ref{fig:config2} obey equations (\ref{set3}). This establishes equation (\ref{piecewiseapp}) for the case of a blue cylinder operator linked with a blue base loop.

\bibliography{reference}

\newcommand{\noopsort}[1]{} \newcommand{\printfirst}[2]{#1}
  \newcommand{\singleletter}[1]{#1} \newcommand{\switchargs}[2]{#2#1}
\begin{thebibliography}{43}%
\makeatletter
\providecommand \@ifxundefined [1]{%
 \@ifx{#1\undefined}
}%
\providecommand \@ifnum [1]{%
 \ifnum #1\expandafter \@firstoftwo
 \else \expandafter \@secondoftwo
 \fi
}%
\providecommand \@ifx [1]{%
 \ifx #1\expandafter \@firstoftwo
 \else \expandafter \@secondoftwo
 \fi
}%
\providecommand \natexlab [1]{#1}%
\providecommand \enquote  [1]{``#1''}%
\providecommand \bibnamefont  [1]{#1}%
\providecommand \bibfnamefont [1]{#1}%
\providecommand \citenamefont [1]{#1}%
\providecommand \href@noop [0]{\@secondoftwo}%
\providecommand \href [0]{\begingroup \@sanitize@url \@href}%
\providecommand \@href[1]{\@@startlink{#1}\@@href}%
\providecommand \@@href[1]{\endgroup#1\@@endlink}%
\providecommand \@sanitize@url [0]{\catcode `\\12\catcode `\$12\catcode
  `\&12\catcode `\#12\catcode `\^12\catcode `\_12\catcode `\%12\relax}%
\providecommand \@@startlink[1]{}%
\providecommand \@@endlink[0]{}%
\providecommand \url  [0]{\begingroup\@sanitize@url \@url }%
\providecommand \@url [1]{\endgroup\@href {#1}{\urlprefix }}%
\providecommand \urlprefix  [0]{URL }%
\providecommand \Eprint [0]{\href }%
\providecommand \doibase [0]{http://dx.doi.org/}%
\providecommand \selectlanguage [0]{\@gobble}%
\providecommand \bibinfo  [0]{\@secondoftwo}%
\providecommand \bibfield  [0]{\@secondoftwo}%
\providecommand \translation [1]{[#1]}%
\providecommand \BibitemOpen [0]{}%
\providecommand \bibitemStop [0]{}%
\providecommand \bibitemNoStop [0]{.\EOS\space}%
\providecommand \EOS [0]{\spacefactor3000\relax}%
\providecommand \BibitemShut  [1]{\csname bibitem#1\endcsname}%
\let\auto@bib@innerbib\@empty
\bibitem [{\citenamefont {Hasan}\ and\ \citenamefont
  {Kane}(2010)}]{HasanKaneRMP}%
  \BibitemOpen
  \bibfield  {author} {\bibinfo {author} {\bibfnamefont {M.~Z.}\ \bibnamefont
  {Hasan}}\ and\ \bibinfo {author} {\bibfnamefont {C.~L.}\ \bibnamefont
  {Kane}},\ }\href@noop {} {\bibfield  {journal} {\bibinfo  {journal} {Rev.
  Mod. Phys.}\ }\textbf {\bibinfo {volume} {82}},\ \bibinfo {pages} {3045}
  (\bibinfo {year} {2010})}\BibitemShut {NoStop}%
\bibitem [{\citenamefont {Qi}\ and\ \citenamefont {Zhang}(2011)}]{QiZhangRMP}%
  \BibitemOpen
  \bibfield  {author} {\bibinfo {author} {\bibfnamefont {X.-L.}\ \bibnamefont
  {Qi}}\ and\ \bibinfo {author} {\bibfnamefont {S.-C.}\ \bibnamefont {Zhang}},\
  }\href@noop {} {\bibfield  {journal} {\bibinfo  {journal} {Rev. Mod. Phys.}\
  }\textbf {\bibinfo {volume} {83}},\ \bibinfo {pages} {1057} (\bibinfo {year}
  {2011})}\BibitemShut {NoStop}%
\bibitem [{Note1()}]{Note1}%
  \BibitemOpen
  \bibinfo {note} {The 1D case is less interesting, since it is known that in
  the absence of symmetry all 1D bosonic systems belong to the same phase\cite
  {PollmannSPT2,XieSPT1,XieSPT2,NorbertSPT,LukaszfSPT2,PollmannSPT1} while all
  1D fermionic systems belong to one of two phases.\cite
  {LukaszfSPT2}}\BibitemShut {NoStop}%
\bibitem [{Note2()}]{Note2}%
  \BibitemOpen
  \bibinfo {note} {Here, when we say braiding statistics, we mean the complete
  set of algebraic data for anyon systems, including quantum dimensions and
  fusion rules. For more details, see e.g. Appendix E of Ref. \protect
  \rev@citealpnum {KitaevHoneycomb}.}\BibitemShut {Stop}%
\bibitem [{\citenamefont {Kane}\ and\ \citenamefont
  {Fisher}(1997)}]{KaneFisherThermal}%
  \BibitemOpen
  \bibfield  {author} {\bibinfo {author} {\bibfnamefont {C.~L.}\ \bibnamefont
  {Kane}}\ and\ \bibinfo {author} {\bibfnamefont {M.~P.~A.}\ \bibnamefont
  {Fisher}},\ }\href {\doibase 10.1103/PhysRevB.55.15832} {\bibfield  {journal}
  {\bibinfo  {journal} {Phys. Rev. B}\ }\textbf {\bibinfo {volume} {55}},\
  \bibinfo {pages} {15832} (\bibinfo {year} {1997})}\BibitemShut {NoStop}%
\bibitem [{Note3()}]{Note3}%
  \BibitemOpen
  \bibinfo {note} {Here, we implicitly exclude 3D layered systems, like a stack
  of fractional quantum Hall states, from our discussion.}\BibitemShut {Stop}%
\bibitem [{\citenamefont {Aharonov}\ and\ \citenamefont
  {Bohm}(1959)}]{AharonovBohm}%
  \BibitemOpen
  \bibfield  {author} {\bibinfo {author} {\bibfnamefont {Y.}~\bibnamefont
  {Aharonov}}\ and\ \bibinfo {author} {\bibfnamefont {D.}~\bibnamefont
  {Bohm}},\ }\href {\doibase 10.1103/PhysRev.115.485} {\bibfield  {journal}
  {\bibinfo  {journal} {Phys. Rev.}\ }\textbf {\bibinfo {volume} {115}},\
  \bibinfo {pages} {485} (\bibinfo {year} {1959})}\BibitemShut {NoStop}%
\bibitem [{\citenamefont {Alford}\ and\ \citenamefont
  {Wilczek}(1989)}]{AlfordWilczek}%
  \BibitemOpen
  \bibfield  {author} {\bibinfo {author} {\bibfnamefont {M.~G.}\ \bibnamefont
  {Alford}}\ and\ \bibinfo {author} {\bibfnamefont {F.}~\bibnamefont
  {Wilczek}},\ }\href {\doibase 10.1103/PhysRevLett.62.1071} {\bibfield
  {journal} {\bibinfo  {journal} {Phys. Rev. Lett}\ }\textbf {\bibinfo {volume}
  {62}},\ \bibinfo {pages} {1071} (\bibinfo {year} {1989})}\BibitemShut
  {NoStop}%
\bibitem [{\citenamefont {Krauss}\ and\ \citenamefont
  {Wilczek}(1989)}]{KraussWilczek}%
  \BibitemOpen
  \bibfield  {author} {\bibinfo {author} {\bibfnamefont {L.~M.}\ \bibnamefont
  {Krauss}}\ and\ \bibinfo {author} {\bibfnamefont {F.}~\bibnamefont
  {Wilczek}},\ }\href {\doibase 0.1103/PhysRevLett.62.1221} {\bibfield
  {journal} {\bibinfo  {journal} {Phys. Rev. Lett.}\ }\textbf {\bibinfo
  {volume} {62}},\ \bibinfo {pages} {1221} (\bibinfo {year}
  {1989})}\BibitemShut {NoStop}%
\bibitem [{\citenamefont {Preskill}\ and\ \citenamefont
  {Krauss}(1990)}]{PreskillKrauss}%
  \BibitemOpen
  \bibfield  {author} {\bibinfo {author} {\bibfnamefont {J.}~\bibnamefont
  {Preskill}}\ and\ \bibinfo {author} {\bibfnamefont {L.~M.}\ \bibnamefont
  {Krauss}},\ }\href {\doibase 10.1016/0550-3213(90)90262-C} {\bibfield
  {journal} {\bibinfo  {journal} {Nucl. Phys. B}\ }\textbf {\bibinfo {volume}
  {341}},\ \bibinfo {pages} {50} (\bibinfo {year} {1990})}\BibitemShut
  {NoStop}%
\bibitem [{\citenamefont {Aneziris}\ \emph {et~al.}(1991)\citenamefont
  {Aneziris}, \citenamefont {Balachandran}, \citenamefont {Kauffman},\ and\
  \citenamefont {Srivastava}}]{Anezirisloop}%
  \BibitemOpen
  \bibfield  {author} {\bibinfo {author} {\bibfnamefont {C.}~\bibnamefont
  {Aneziris}}, \bibinfo {author} {\bibfnamefont {A.~P.}\ \bibnamefont
  {Balachandran}}, \bibinfo {author} {\bibfnamefont {L.}~\bibnamefont
  {Kauffman}}, \ and\ \bibinfo {author} {\bibfnamefont {A.~M.}\ \bibnamefont
  {Srivastava}},\ }\href {\doibase 10.1142/S0217751X91001210} {\bibfield
  {journal} {\bibinfo  {journal} {Int. J. Mod. Phys. A}\ }\textbf {\bibinfo
  {volume} {06}},\ \bibinfo {pages} {2519} (\bibinfo {year}
  {1991})}\BibitemShut {NoStop}%
\bibitem [{\citenamefont {Alford}\ \emph {et~al.}(1992)\citenamefont {Alford},
  \citenamefont {Lee}, \citenamefont {March-Russell},\ and\ \citenamefont
  {Preskill}}]{Alfordloop}%
  \BibitemOpen
  \bibfield  {author} {\bibinfo {author} {\bibfnamefont {M.~G.}\ \bibnamefont
  {Alford}}, \bibinfo {author} {\bibfnamefont {K.-M.}\ \bibnamefont {Lee}},
  \bibinfo {author} {\bibfnamefont {J.}~\bibnamefont {March-Russell}}, \ and\
  \bibinfo {author} {\bibfnamefont {J.}~\bibnamefont {Preskill}},\ }\href
  {\doibase 10.1016/0550-3213(92)90468-Q} {\bibfield  {journal} {\bibinfo
  {journal} {Nucl. Phys. B}\ }\textbf {\bibinfo {volume} {384}},\ \bibinfo
  {pages} {251} (\bibinfo {year} {1992})}\BibitemShut {NoStop}%
\bibitem [{\citenamefont {Baez}\ \emph {et~al.}(2007)\citenamefont {Baez},
  \citenamefont {Wise},\ and\ \citenamefont {Crans}}]{Baezloop}%
  \BibitemOpen
  \bibfield  {author} {\bibinfo {author} {\bibfnamefont {J.~C.}\ \bibnamefont
  {Baez}}, \bibinfo {author} {\bibfnamefont {D.~K.}\ \bibnamefont {Wise}}, \
  and\ \bibinfo {author} {\bibfnamefont {A.~S.}\ \bibnamefont {Crans}},\ }\href
  {\doibase 10.4310/ATMP.2007.v11.n5.a1} {\bibfield  {journal} {\bibinfo
  {journal} {Adv. Theor. Math. Phys.}\ }\textbf {\bibinfo {volume} {11}},\
  \bibinfo {pages} {707} (\bibinfo {year} {2007})}\BibitemShut {NoStop}%
\bibitem [{\citenamefont {Wang}\ and\ \citenamefont
  {Levin}(2014{\natexlab{a}})}]{WangLevin}%
  \BibitemOpen
  \bibfield  {author} {\bibinfo {author} {\bibfnamefont {C.}~\bibnamefont
  {Wang}}\ and\ \bibinfo {author} {\bibfnamefont {M.}~\bibnamefont {Levin}},\
  }\href {\doibase 10.1103/PhysRevLett.113.080403} {\bibfield  {journal}
  {\bibinfo  {journal} {Phys. Rev. Lett.}\ }\textbf {\bibinfo {volume} {113}},\
  \bibinfo {pages} {080403} (\bibinfo {year} {2014}{\natexlab{a}})}\BibitemShut
  {NoStop}%
\bibitem [{\citenamefont {Jiang}\ \emph {et~al.}(2014)\citenamefont {Jiang},
  \citenamefont {Mesaros},\ and\ \citenamefont {Ran}}]{JiangMesarosRan}%
  \BibitemOpen
  \bibfield  {author} {\bibinfo {author} {\bibfnamefont {S.}~\bibnamefont
  {Jiang}}, \bibinfo {author} {\bibfnamefont {A.}~\bibnamefont {Mesaros}}, \
  and\ \bibinfo {author} {\bibfnamefont {Y.}~\bibnamefont {Ran}},\ }\href
  {\doibase 10.1103/PhysRevX.4.031048} {\bibfield  {journal} {\bibinfo
  {journal} {Phys. Rev. X}\ }\textbf {\bibinfo {volume} {4}},\ \bibinfo {pages}
  {031048} (\bibinfo {year} {2014})}\BibitemShut {NoStop}%
\bibitem [{\citenamefont {Wang}\ and\ \citenamefont {Wen}(2014)}]{WangWen14}%
  \BibitemOpen
  \bibfield  {author} {\bibinfo {author} {\bibfnamefont {J.}~\bibnamefont
  {Wang}}\ and\ \bibinfo {author} {\bibfnamefont {X.-G.}\ \bibnamefont {Wen}},\
  }\href@noop {} {\bibfield  {journal} {\bibinfo  {journal} {arXiv:1404.7854}\
  } (\bibinfo {year} {2014})}\BibitemShut {NoStop}%
\bibitem [{\citenamefont {Jian}\ and\ \citenamefont {Qi}(2014)}]{JianQi}%
  \BibitemOpen
  \bibfield  {author} {\bibinfo {author} {\bibfnamefont {C.-M.}\ \bibnamefont
  {Jian}}\ and\ \bibinfo {author} {\bibfnamefont {X.-L.}\ \bibnamefont {Qi}},\
  }\href {\doibase 10.1103/PhysRevX.4.041043} {\bibfield  {journal} {\bibinfo
  {journal} {Phys. Rev. X}\ }\textbf {\bibinfo {volume} {4}},\ \bibinfo {pages}
  {041043} (\bibinfo {year} {2014})}\BibitemShut {NoStop}%
\bibitem [{\citenamefont {Wang}\ and\ \citenamefont
  {Levin}(2014{\natexlab{b}})}]{WangLevin2}%
  \BibitemOpen
  \bibfield  {author} {\bibinfo {author} {\bibfnamefont {C.}~\bibnamefont
  {Wang}}\ and\ \bibinfo {author} {\bibfnamefont {M.}~\bibnamefont {Levin}},\
  }\href@noop {} {\bibfield  {journal} {\bibinfo  {journal} {arxiv:1412.1781}\
  } (\bibinfo {year} {2014}{\natexlab{b}})}\BibitemShut {NoStop}%
\bibitem [{\citenamefont {Chen}\ \emph {et~al.}(2015)\citenamefont {Chen},
  \citenamefont {Burnell}, \citenamefont {Vishwanath},\ and\ \citenamefont
  {Fidkowski}}]{SurfaceTOChen}%
  \BibitemOpen
  \bibfield  {author} {\bibinfo {author} {\bibfnamefont {X.}~\bibnamefont
  {Chen}}, \bibinfo {author} {\bibfnamefont {F.~J.}\ \bibnamefont {Burnell}},
  \bibinfo {author} {\bibfnamefont {A.}~\bibnamefont {Vishwanath}}, \ and\
  \bibinfo {author} {\bibfnamefont {L.}~\bibnamefont {Fidkowski}},\ }\href@noop
  {} {\bibfield  {journal} {\bibinfo  {journal} {arxiv:1403.6491}\ } (\bibinfo
  {year} {2015})}\BibitemShut {NoStop}%
\bibitem [{\citenamefont {Dijkgraaf}\ and\ \citenamefont
  {Witten}(1990)}]{DijkgraafWitten}%
  \BibitemOpen
  \bibfield  {author} {\bibinfo {author} {\bibfnamefont {R.}~\bibnamefont
  {Dijkgraaf}}\ and\ \bibinfo {author} {\bibfnamefont {E.}~\bibnamefont
  {Witten}},\ }\href@noop {} {\bibfield  {journal} {\bibinfo  {journal} {Comm.
  Math. Phys.}\ }\textbf {\bibinfo {volume} {129}},\ \bibinfo {pages} {393}
  (\bibinfo {year} {1990})}\BibitemShut {NoStop}%
\bibitem [{\citenamefont {Kitaev}(2003)}]{KitaevToric}%
  \BibitemOpen
  \bibfield  {author} {\bibinfo {author} {\bibfnamefont {A.~Y.}\ \bibnamefont
  {Kitaev}},\ }\href@noop {} {\bibfield  {journal} {\bibinfo  {journal} {Annals
  of Physics}\ }\textbf {\bibinfo {volume} {303}},\ \bibinfo {pages} {2}
  (\bibinfo {year} {2003})}\BibitemShut {NoStop}%
\bibitem [{\citenamefont {Levin}\ and\ \citenamefont
  {Wen}(2003)}]{LevinWenHop}%
  \BibitemOpen
  \bibfield  {author} {\bibinfo {author} {\bibfnamefont {M.}~\bibnamefont
  {Levin}}\ and\ \bibinfo {author} {\bibfnamefont {X.-G.}\ \bibnamefont
  {Wen}},\ }\href {\doibase 10.1103/PhysRevB.67.245316} {\bibfield  {journal}
  {\bibinfo  {journal} {Phys. Rev. B}\ }\textbf {\bibinfo {volume} {67}},\
  \bibinfo {pages} {245316} (\bibinfo {year} {2003})}\BibitemShut {NoStop}%
\bibitem [{\citenamefont {Levin}\ and\ \citenamefont
  {Wen}(2005)}]{LevinWenstrnet}%
  \BibitemOpen
  \bibfield  {author} {\bibinfo {author} {\bibfnamefont {M.~A.}\ \bibnamefont
  {Levin}}\ and\ \bibinfo {author} {\bibfnamefont {X.-G.}\ \bibnamefont
  {Wen}},\ }\href {\doibase 10.1103/PhysRevB.71.045110} {\bibfield  {journal}
  {\bibinfo  {journal} {Phys. Rev. B}\ }\textbf {\bibinfo {volume} {71}},\
  \bibinfo {pages} {045110} (\bibinfo {year} {2005})}\BibitemShut {NoStop}%
\bibitem [{\citenamefont {Hamma}\ \emph {et~al.}(2005)\citenamefont {Hamma},
  \citenamefont {Zanardi},\ and\ \citenamefont {Wen}}]{HammaZanardiWen}%
  \BibitemOpen
  \bibfield  {author} {\bibinfo {author} {\bibfnamefont {A.}~\bibnamefont
  {Hamma}}, \bibinfo {author} {\bibfnamefont {P.}~\bibnamefont {Zanardi}}, \
  and\ \bibinfo {author} {\bibfnamefont {X.-G.}\ \bibnamefont {Wen}},\ }\href
  {\doibase 10.1103/PhysRevB.72.035307} {\bibfield  {journal} {\bibinfo
  {journal} {Phys. Rev. B}\ }\textbf {\bibinfo {volume} {72}},\ \bibinfo
  {pages} {035307} (\bibinfo {year} {2005})}\BibitemShut {NoStop}%
\bibitem [{\citenamefont {Castelnovo}\ and\ \citenamefont
  {Chamon}(2007)}]{CastelnovoChamon}%
  \BibitemOpen
  \bibfield  {author} {\bibinfo {author} {\bibfnamefont {C.}~\bibnamefont
  {Castelnovo}}\ and\ \bibinfo {author} {\bibfnamefont {C.}~\bibnamefont
  {Chamon}},\ }\href {\doibase 10.1103/PhysRevB.76.184442} {\bibfield
  {journal} {\bibinfo  {journal} {Phys. Rev. B}\ }\textbf {\bibinfo {volume}
  {76}},\ \bibinfo {pages} {184442} (\bibinfo {year} {2007})}\BibitemShut
  {NoStop}%
\bibitem [{\citenamefont {Kogut}(1979)}]{Kogut}%
  \BibitemOpen
  \bibfield  {author} {\bibinfo {author} {\bibfnamefont {J.~B.}\ \bibnamefont
  {Kogut}},\ }\href {\doibase 10.1103/RevModPhys.51.659} {\bibfield  {journal}
  {\bibinfo  {journal} {Rev. Mod. Phys.}\ }\textbf {\bibinfo {volume} {51}},\
  \bibinfo {pages} {659} (\bibinfo {year} {1979})}\BibitemShut {NoStop}%
\bibitem [{\citenamefont {Chen}\ \emph {et~al.}(2014)\citenamefont {Chen},
  \citenamefont {Lu},\ and\ \citenamefont {Vishwanath}}]{ChenLuVishwanath}%
  \BibitemOpen
  \bibfield  {author} {\bibinfo {author} {\bibfnamefont {X.}~\bibnamefont
  {Chen}}, \bibinfo {author} {\bibfnamefont {Y.-M.}\ \bibnamefont {Lu}}, \ and\
  \bibinfo {author} {\bibfnamefont {A.}~\bibnamefont {Vishwanath}},\ }\href
  {\doibase 10.1038/ncomms4507} {\bibfield  {journal} {\bibinfo  {journal}
  {Nature Communications}\ }\textbf {\bibinfo {volume} {5}},\ \bibinfo {pages}
  {3507} (\bibinfo {year} {2014})}\BibitemShut {NoStop}%
\bibitem [{\citenamefont {Gu}\ and\ \citenamefont {Wen}(2009)}]{GuSPT}%
  \BibitemOpen
  \bibfield  {author} {\bibinfo {author} {\bibfnamefont {Z.-C.}\ \bibnamefont
  {Gu}}\ and\ \bibinfo {author} {\bibfnamefont {X.-G.}\ \bibnamefont {Wen}},\
  }\href {\doibase 10.1103/PhysRevB.80.155131} {\bibfield  {journal} {\bibinfo
  {journal} {Phys. Rev. B}\ }\textbf {\bibinfo {volume} {80}},\ \bibinfo
  {pages} {155131} (\bibinfo {year} {2009})}\BibitemShut {NoStop}%
\bibitem [{\citenamefont {Pollmann}\ \emph {et~al.}(2010)\citenamefont
  {Pollmann}, \citenamefont {Turner}, \citenamefont {Berg},\ and\ \citenamefont
  {Oshikawa}}]{PollmannSPT2}%
  \BibitemOpen
  \bibfield  {author} {\bibinfo {author} {\bibfnamefont {F.}~\bibnamefont
  {Pollmann}}, \bibinfo {author} {\bibfnamefont {A.~M.}\ \bibnamefont
  {Turner}}, \bibinfo {author} {\bibfnamefont {E.}~\bibnamefont {Berg}}, \ and\
  \bibinfo {author} {\bibfnamefont {M.}~\bibnamefont {Oshikawa}},\ }\href
  {\doibase 10.1103/PhysRevB.81.064439} {\bibfield  {journal} {\bibinfo
  {journal} {Phys. Rev. B}\ }\textbf {\bibinfo {volume} {81}},\ \bibinfo
  {pages} {064439} (\bibinfo {year} {2010})}\BibitemShut {NoStop}%
\bibitem [{\citenamefont {Fidkowski}\ and\ \citenamefont
  {Kitaev}(2010)}]{LukaszfSPT2}%
  \BibitemOpen
  \bibfield  {author} {\bibinfo {author} {\bibfnamefont {L.}~\bibnamefont
  {Fidkowski}}\ and\ \bibinfo {author} {\bibfnamefont {A.}~\bibnamefont
  {Kitaev}},\ }\href {\doibase 10.1103/PhysRevB.81.134509} {\bibfield
  {journal} {\bibinfo  {journal} {Phys. Rev. B}\ }\textbf {\bibinfo {volume}
  {81}},\ \bibinfo {pages} {134509} (\bibinfo {year} {2010})}\BibitemShut
  {NoStop}%
\bibitem [{\citenamefont {Chen}\ \emph
  {et~al.}(2011{\natexlab{a}})\citenamefont {Chen}, \citenamefont {Gu},\ and\
  \citenamefont {Wen}}]{XieSPT1}%
  \BibitemOpen
  \bibfield  {author} {\bibinfo {author} {\bibfnamefont {X.}~\bibnamefont
  {Chen}}, \bibinfo {author} {\bibfnamefont {Z.-C.}\ \bibnamefont {Gu}}, \ and\
  \bibinfo {author} {\bibfnamefont {X.-G.}\ \bibnamefont {Wen}},\ }\href
  {\doibase 10.1103/PhysRevB.83.035107} {\bibfield  {journal} {\bibinfo
  {journal} {Phys. Rev. B}\ }\textbf {\bibinfo {volume} {83}},\ \bibinfo
  {pages} {035107} (\bibinfo {year} {2011}{\natexlab{a}})}\BibitemShut
  {NoStop}%
\bibitem [{\citenamefont {Chen}\ \emph
  {et~al.}(2011{\natexlab{b}})\citenamefont {Chen}, \citenamefont {Gu},\ and\
  \citenamefont {Wen}}]{XieSPT2}%
  \BibitemOpen
  \bibfield  {author} {\bibinfo {author} {\bibfnamefont {X.}~\bibnamefont
  {Chen}}, \bibinfo {author} {\bibfnamefont {Z.-C.}\ \bibnamefont {Gu}}, \ and\
  \bibinfo {author} {\bibfnamefont {X.-G.}\ \bibnamefont {Wen}},\ }\href
  {\doibase 10.1103/PhysRevB.84.235128} {\bibfield  {journal} {\bibinfo
  {journal} {Phys. Rev. B}\ }\textbf {\bibinfo {volume} {84}},\ \bibinfo
  {pages} {235128} (\bibinfo {year} {2011}{\natexlab{b}})}\BibitemShut
  {NoStop}%
\bibitem [{\citenamefont {Schuch}\ \emph {et~al.}(2011)\citenamefont {Schuch},
  \citenamefont {Perez-Garcia},\ and\ \citenamefont {Cirac}}]{NorbertSPT}%
  \BibitemOpen
  \bibfield  {author} {\bibinfo {author} {\bibfnamefont {N.}~\bibnamefont
  {Schuch}}, \bibinfo {author} {\bibfnamefont {D.}~\bibnamefont
  {Perez-Garcia}}, \ and\ \bibinfo {author} {\bibfnamefont {I.}~\bibnamefont
  {Cirac}},\ }\href {\doibase 10.1103/PhysRevB.84.165139} {\bibfield  {journal}
  {\bibinfo  {journal} {Phys. Rev. B}\ }\textbf {\bibinfo {volume} {84}},\
  \bibinfo {pages} {165139} (\bibinfo {year} {2011})}\BibitemShut {NoStop}%
\bibitem [{\citenamefont {Chen}\ \emph {et~al.}(2013)\citenamefont {Chen},
  \citenamefont {Gu}, \citenamefont {Liu},\ and\ \citenamefont
  {Wen}}]{ChenGuWenSPT}%
  \BibitemOpen
  \bibfield  {author} {\bibinfo {author} {\bibfnamefont {X.}~\bibnamefont
  {Chen}}, \bibinfo {author} {\bibfnamefont {Z.-C.}\ \bibnamefont {Gu}},
  \bibinfo {author} {\bibfnamefont {Z.-X.}\ \bibnamefont {Liu}}, \ and\
  \bibinfo {author} {\bibfnamefont {X.-G.}\ \bibnamefont {Wen}},\ }\href
  {\doibase 10.1103/PhysRevB.87.155114} {\bibfield  {journal} {\bibinfo
  {journal} {Phys. Rev. B}\ }\textbf {\bibinfo {volume} {87}},\ \bibinfo
  {pages} {155114} (\bibinfo {year} {2013})}\BibitemShut {NoStop}%
\bibitem [{\citenamefont {Lin}\ and\ \citenamefont {Levin}()}]{LinLevinprep}%
  \BibitemOpen
  \bibfield  {author} {\bibinfo {author} {\bibfnamefont {C.-H.}\ \bibnamefont
  {Lin}}\ and\ \bibinfo {author} {\bibfnamefont {M.}~\bibnamefont {Levin}},\
  }\href@noop {} {\bibinfo  {journal} {in preparation.}\ }\BibitemShut
  {NoStop}%
\bibitem [{\citenamefont {Wan}\ \emph {et~al.}(2014)\citenamefont {Wan},
  \citenamefont {Wang},\ and\ \citenamefont {He}}]{Wan3dmodel}%
  \BibitemOpen
\bibfield  {journal} {  }\bibfield  {author} {\bibinfo {author} {\bibfnamefont
  {Y.}~\bibnamefont {Wan}}, \bibinfo {author} {\bibfnamefont {J.}~\bibnamefont
  {Wang}}, \ and\ \bibinfo {author} {\bibfnamefont {H.}~\bibnamefont {He}},\
  }\href@noop {} {\bibfield  {journal} {\bibinfo  {journal} {arxiv:1409.3216}\
  } (\bibinfo {year} {2014})}\BibitemShut {NoStop}%
\bibitem [{Note4()}]{Note4}%
  \BibitemOpen
  \bibinfo {note} {For example, toroidal operators can be decorated by string
  operators that encircle the torus while spherical membrane operators cannot
  be decorated in this way. See section \ref {otherM} for more
  details.}\BibitemShut {Stop}%
\bibitem [{Note5()}]{Note5}%
  \BibitemOpen
  \bibinfo {note} {In fact, if our conjecture is correct then $H_1$ also
  belongs to the same phase as the Dijkgraaf-Witten models labeled by $(1, 0)$
  and $(0,1)$, since the three Dijkgraaf-Witten models labeled by $(1,1)$,
  $(1,0)$ and $(0,1)$ all belong to the same phase, when viewed as spin models
  rather than gauge theories.}\BibitemShut {Stop}%
\bibitem [{\citenamefont {Haah}(2011)}]{Haah}%
  \BibitemOpen
  \bibfield  {author} {\bibinfo {author} {\bibfnamefont {J.}~\bibnamefont
  {Haah}},\ }\href {\doibase 10.1103/PhysRevA.83.042330} {\bibfield  {journal}
  {\bibinfo  {journal} {Phys. Rev. A}\ }\textbf {\bibinfo {volume} {83}},\
  \bibinfo {pages} {042330} (\bibinfo {year} {2011})}\BibitemShut {NoStop}%
\bibitem [{\citenamefont {Zeng}\ and\ \citenamefont {Wen}(2014)}]{ZengWen}%
  \BibitemOpen
  \bibfield  {author} {\bibinfo {author} {\bibfnamefont {B.}~\bibnamefont
  {Zeng}}\ and\ \bibinfo {author} {\bibfnamefont {X.-G.}\ \bibnamefont {Wen}},\
  }\href@noop {} {\bibfield  {journal} {\bibinfo  {journal} {arxiv:1406.5090}\
  } (\bibinfo {year} {2014})}\BibitemShut {NoStop}%
\bibitem [{\citenamefont {Levin}\ and\ \citenamefont {Gu}(2012)}]{LevinGu}%
  \BibitemOpen
  \bibfield  {author} {\bibinfo {author} {\bibfnamefont {M.}~\bibnamefont
  {Levin}}\ and\ \bibinfo {author} {\bibfnamefont {Z.-C.}\ \bibnamefont {Gu}},\
  }\href {\doibase 10.1103/PhysRevB.86.115109} {\bibfield  {journal} {\bibinfo
  {journal} {Phys. Rev. B}\ }\textbf {\bibinfo {volume} {86}},\ \bibinfo
  {pages} {115109} (\bibinfo {year} {2012})}\BibitemShut {NoStop}%
\bibitem [{\citenamefont {Pollmann}\ \emph {et~al.}(2012)\citenamefont
  {Pollmann}, \citenamefont {Berg}, \citenamefont {Turner},\ and\ \citenamefont
  {Oshikawa}}]{PollmannSPT1}%
  \BibitemOpen
  \bibfield  {author} {\bibinfo {author} {\bibfnamefont {F.}~\bibnamefont
  {Pollmann}}, \bibinfo {author} {\bibfnamefont {E.}~\bibnamefont {Berg}},
  \bibinfo {author} {\bibfnamefont {A.~M.}\ \bibnamefont {Turner}}, \ and\
  \bibinfo {author} {\bibfnamefont {M.}~\bibnamefont {Oshikawa}},\ }\href
  {\doibase 10.1103/PhysRevB.85.075125} {\bibfield  {journal} {\bibinfo
  {journal} {Phys. Rev. B}\ }\textbf {\bibinfo {volume} {85}},\ \bibinfo
  {pages} {075125} (\bibinfo {year} {2012})}\BibitemShut {NoStop}%
\bibitem [{\citenamefont {Kitaev}(2006)}]{KitaevHoneycomb}%
  \BibitemOpen
  \bibfield  {author} {\bibinfo {author} {\bibfnamefont {A.}~\bibnamefont
  {Kitaev}},\ }\href@noop {} {\bibfield  {journal} {\bibinfo  {journal} {Ann.
  Phys.}\ }\textbf {\bibinfo {volume} {321}},\ \bibinfo {pages} {2} (\bibinfo
  {year} {2006})}\BibitemShut {NoStop}%
\end{thebibliography}%

\end{document}